\documentstyle[12pt,axodraw]{article}
\begin{document}
\baselineskip 0.7cm

\newcommand{\gsim}{ \mathop{}_{\textstyle \sim}^{\textstyle >} }
\newcommand{\lsim}{ \mathop{}_{\textstyle \sim}^{\textstyle <} }
\newcommand{\vev}[1]{ \left\langle {#1} \right\rangle }
\newcommand{\bra}[1]{ \langle {#1} | }
\newcommand{\ket}[1]{ | {#1} \rangle }
\newcommand{\EV}{ {\rm eV} }
\newcommand{\KEV}{ {\rm keV} }
\newcommand{\MEV}{ {\rm MeV} }
\newcommand{\GEV}{ {\rm GeV} }
\newcommand{\TEV}{ {\rm TeV} }
\newcommand{\mex}{m_{\psi}}
\newcommand{\EX}{R}
\newcommand{\HS}{\tilde{F}}
\def\tr{\mathop{\rm tr}\nolimits}
\def\Tr{\mathop{\rm Tr}\nolimits}
\def\Re{\mathop{\rm Re}\nolimits}
\def\Im{\mathop{\rm Im}\nolimits}
\setcounter{footnote}{1}

\begin{titlepage}
\begin{flushright}
KEK-TH-676 \\
YITP-00-12 \\
UT-879 \\
\end{flushright}

\vskip 1cm

\begin{center}
{\large \bf  Natural Effective Supersymmetry}
\vskip 1.2cm
Junji Hisano$^1$, Kiichi Kurosawa$^{2,3}$ and Yasunori Nomura$^3$

\vskip 0.4cm

$^{1}$ {\it Theory Group, KEK, Tsukuba, Ibaraki 305-0801, Japan}\\
$^{2}$ {\it YITP, Kyoto University, Kyoto 606-8502, Japan}\\
$^{3}$ {\it Department of Physics, University of Tokyo, 
            Tokyo 113-0033, Japan}

\vskip 1.5cm

\abstract{Much heavier sfermions of the first-two generations than the
 other superparticles provide a natural explanation for the flavor and
 CP problems in the supersymmetric standard model (SUSY SM). 
 However, the heavy sfermions may drive the mass squareds for the light
 third generation sfermions to be negative through two-loop
 renormalization group (RG) equations, breaking color and charge.
 Introducing extra matters to the SUSY SM, it is possible to construct 
 models where the sfermion masses are RG invariant at the two-loop level
 in the limit of vanishing gaugino-mass and Yukawa-coupling
 contributions.
 We calculate the finite corrections to the light sfermion masses at the 
 two-loop level in the models. 
 We find that the finite corrections to the light-squark mass squareds
 are negative and can be less than $(0.3-1)\%$ of the heavy-squark mass
 squareds, depending on the number and the parameters of the extra
 matters.
 We also discuss whether such models realized by the U(1)$_X$ gauge
 interaction at the GUT scale can satisfy the constraints from 
 $\Delta m_K$ and $\epsilon_K$ naturally.
 When both the left- and right-handed down-type squarks of the first-two
 generations have common U(1)$_X$ charges, the supersymmetric
 contributions to $\Delta m_K$ and $\epsilon_K$ are sufficiently
 suppressed without spoiling naturalness, even if the flavor-violating
 supergravity contributions to the sfermion mass matrices are included. 
 When only the right-handed squarks of the first-two generations have a
 common U(1)$_X$ charge, we can still satisfy the constraint from 
 $\Delta m_K$ naturally, but evading the bound from $\epsilon_K$
 requires the CP phase smaller than $10^{-2}$. }
\end{center}
\end{titlepage}

\renewcommand{\thefootnote}{\arabic{footnote}}
\setcounter{footnote}{0}
\baselineskip 0.5cm

%
%
%
%

\section{Introduction}

The Standard Model (SM) has enjoyed remarkable successes in describing
physics down to a scale of $10^{-18}$ m, the weak scale. However,
the quadratically divergent radiative correction to the Higgs boson mass
leads to the well-known naturalness problem. Supersymmetry (SUSY)
provides a solution to this problem by extending the chiral symmetry to
the scalar partners. At present, the minimal supersymmetric standard
model (MSSM) is considered to be the most promising candidate beyond the
SM \cite{MSSM}.

The MSSM is severely constrained by flavor physics. Introduction of
SUSY-breaking terms provides new sources of the flavor-changing neutral
currents (FCNC's) and CP violation. For instance, the experimental value
of the mass difference between $K_L$ and $K_S$, $\Delta m_K$, leads to
the upper bound on the squark contribution written as
\begin{eqnarray}
  \sin^2  \theta_{\tilde{d}}
  \left(\frac{\Delta m_{\tilde{d}}^2}{\bar{m}_{\tilde{d}}^2}\right)^2
  \left(\frac{30~\TEV}{\bar{m}_{\tilde{d}}}\right)^2
  \lsim 1,
\end{eqnarray}
in a basis where the quark mass matrices are diagonal
\cite{SUSY-flavor}. Here, $\bar{m}_{\tilde{d}}^2$ is the averaged squark
mass, and $\Delta m_{\tilde{d}}^2$ and $\sin^2\theta_{\tilde{d}}$ are
the mass-squared difference and the mixing angle between the down-type
squarks of the first-two generations. In order to satisfy this
constraint, {\it i}) $\Delta m_{\tilde{d}}^2 \simeq 0$ (degeneracy
\cite{degeneracy}), {\it ii}) $\sin \theta_{\tilde{d}} \simeq 0$
(alignment \cite{alignment}), {\it iii}) ${\bar{m}_{\tilde{d}}} \gsim
30~\TEV$ (decoupling \cite{decoupling}), or the hybrid of them is
required.

In this article, we discuss the third possibility, the decoupling
scenario.  In this scenario, the masses for the third generation
squarks, Higgsinos, and gauginos are of the order of the weak scale
while the other squarks and sleptons are heavy enough to suppress the
FCNC and CP-violating processes.  This is still natural at the one-loop
level, since the squarks and sleptons of the first-two generations are
not strongly coupled with the Higgs boson. This mass spectrum is
sometimes called as effective supersymmetry or the more minimal
supersymmetric standard model. It is realized in the anomalous U(1)
SUSY-breaking models \cite{anom-U1_1, anom-U1_2}, U(1)$^\prime$ models
\cite{U(1)'}, composite models of the first-two generation particles
\cite{composite}, or radiatively-driven models with specific boundary
conditions \cite{radiative}.

However, this scenario generically suffers from a problem that the third
generation sfermions obtain the vacuum expectation value (VEV), breaking
color and charge \cite{AM,AG}. At the two-loop level the heavy first-two
generation sfermions contribute to the the renormalization group (RG)
equations for the third generation sfermion masses through the gauge
interactions. The RG equations for the SUSY-breaking sfermion masses
$\tilde{m}^2_{r}$ through the SM gauge interactions are given by 
\cite{2loop-RGE}
\begin{eqnarray}
  \mu\frac{d \tilde{m}^2_{r}}{d \mu} &=& 
    \sum_{A} 8 \left(\frac{\alpha_A}{4\pi}\right)^2 C^A_{r} 
    \sum_s T^A_{s} \tilde{m}^2_{s}
\nonumber\\
&& 
  + 2 \left(\frac{\alpha_Y}{4\pi}\right) 
       Y_{r} \sum_s Y_{s} \tilde{m}^2_{s}
  + \sum_A 8 \left(\frac{\alpha_Y}{4\pi}\right)
    \left(\frac{\alpha_A}{4\pi}\right) Y_{r} 
    \sum_s Y_{s} C_{s}^{A} \tilde{m}^2_{s},
\label{rge2sf}
\end{eqnarray}
at the two-loop level in the $\overline{\rm DR}^{\prime}$ scheme
\cite{DR_Prime}.
Here, we have taken a limit where the gaugino masses vanish. An index
$A$ ($=1-3$) represents the SM gauge groups, and we have adopted the
SU(5)$_{\rm GUT}$ normalization for the U(1)$_Y$ gauge coupling
($\alpha_1 \equiv ({5/3})\, \alpha_Y$). $T^A_{r}$, $C^A_{r}$ and $Y_{r}$ 
are the Dynkin index, the quadratic Casimir coefficient and the
hypercharge for the sfermion $r$, respectively. We find that all
sfermion mass squareds contribute to the RG equations at the two-loop
level. Thus, the contribution from the heavy sfermions may destroy the
mass spectrum by driving the light-sfermion mass squareds to be negative
at low energies. This fact makes it difficult to construct models where
such a hierarchical sfermion mass spectrum is generated at much higher
energy scale than the weak scale.

However, the hierarchical mass spectrum can be stabilized against the RG
evolution if the following relations among the SUSY-breaking mass
squareds for the heavy sfermions are imposed \cite{HKN}:
\begin{eqnarray}
  \sum_r T_r^{A} \tilde{m}^2_{r} &=&0, 
\label{mass_af}\\
  \sum_r Y_{r}  \tilde{m}^2_{r} &=&0,
\label{mass_u1d1}\\
  \sum_r Y_{r} C_{r}^{A} \tilde{m}^2_{r} &=&0,
\label{mass_u1d2}
\end{eqnarray}
for $A=1-3$. These relations cannot be satisfied by the heavy MSSM
sfermions alone, since all of them have positive SUSY-breaking mass
squareds.  However, the relations can be satisfied if we introduce extra
fields with negative SUSY-breaking mass squareds which transform
nontrivially under the SM gauge groups, since the sum is taken over the
extra fields as well as the heavy MSSM sfermions.  The extra fields 
should have the supersymmetric masses to avoid the breaking of the
SM gauge groups caused by their condensations. Thus, the
extra fields have to be in vector-like representations of the SM gauge
groups, and we call them {\it extra matters}.  The supersymmetric masses
for the extra matters should not be much larger than the SUSY-breaking
scalar masses. Otherwise, the large radiative correction is generated
below the energy scale where the extra matters are decoupled.  We refer
this extension of the MSSM to the Natural Effective SUSY SM (NESSM),
hereafter.

The NESSM can be realized by assuming that the SUSY-breaking masses for
the heavy sfermions and extra matters are generated by a $D$-term VEV of
some U(1)$_X$ gauge interaction such that $\tilde{m}^2_{r}= Q^{X}_{r}
\vev{-D_X}$. 
The U(1)$_X$ should be broken in order to give the supersymmetric masses
for the extra matters.
In terms of the U(1)$_X$ charges, Eqs.~(\ref{mass_af}, \ref{mass_u1d1},
\ref{mass_u1d2}) are written as
\begin{eqnarray}
  \sum_r T_r^{A} Q^{X}_{r} &=&0, 
\label{U1X_af}\\
  \sum_r Y_{r} Q^{X}_{r} &=&0,
\label{U1X_u1d1}\\
  \sum_r Y_{r} C_{r}^{A} Q^{X}_{r} &=&0.
\label{U1X_u1d2}
\end{eqnarray}
The first equation corresponds to vanishing mixed anomalies between
U(1)$_X$ and the SM gauge groups. The second and third ones mean
that U(1)$_X$ has no kinetic mixing with U(1)$_Y$ at the one- and
two-loop levels. If U(1)$_X$ is anomaly-free and all fields charged
under U(1)$_X$ are embedded in the SU(5)$_{\rm GUT}$ multiplets,
Eqs.~(\ref{U1X_af}, \ref{U1X_u1d1}) are satisfied automatically.
In this case, we have only to choose the U(1)$_X$ charges for the extra
matters to satisfy Eq.~(\ref{U1X_u1d2}).  Even if U(1)$_X$ is
anomalous, however, we can still construct a model in which some of the
fields are decoupled at the U(1)$_X$ breaking scale and
Eqs.~(\ref{U1X_af}, \ref{U1X_u1d1}, \ref{U1X_u1d2}) are satisfied at
low energies.

In this article, we calculate the finite corrections to the
light-sfermion mass squareds at the two-loop level in the NESSM,
assuming that all fields are embedded in the SU(5)$_{\rm GUT}$
multiplets at the GUT scale.  
In the NESSM, while the dangerous contributions which lead to color and
charge breaking do not exist in the RG equations for the SUSY-breaking
scalar masses, the finite corrections from the heavy sfermions and the
extra matters at the two-loop level may still drive the light-sfermion
mass squareds to be negative.
We find that the finite corrections to the light-squark mass squareds
depend on the number and the parameters of the extra matters and they
are less than $(0.3-1)\%$ of the heavy-squark mass squareds.

We also discuss whether the models in which the NESSM is realized by
the U(1)$_X$ gauge interaction at the GUT scale ($M_{\rm GUT} \simeq
2\times 10^{16}~\GEV$) are viable or not in the light of the
experimental constraints.  
The anomalous U(1) SUSY-breaking model given in Ref.~\cite{HKN} is an
explicit example for such models.  On this setup, the
supergravity contributions to the SUSY-breaking terms, which are
generically non-universal in flavor space, are suppressed compared with
the U(1)$_X$ $D$-term contribution. Moreover, the breaking of the
U(1)$_X$ symmetry can naturally explain the hierarchical structure of
the quark and lepton masses by the Froggatt-Nielsen mechanism
\cite{FN}. 
In this paper, we consider the constraints from $\Delta m_K$ and
$\epsilon_K$, which lead to the severest bound on the heavy-sfermion
mass scale.
We find that, when both the left- and right-handed down-type
squarks (only the right-handed squarks) of the first-two generations
have common U(1)$_X$ charges, the contributions to $\Delta m_K$ and
$\epsilon_K$ ($\Delta m_K$) are sufficiently suppressed without spoiling
naturalness even if the flavor-violating supergravity contributions to
the sfermion mass matrices are included. 

This paper is organized as follows.  In the next section, we show our
setup and the U(1)$_X$ charge assignments. We explain that the squark
and slepton masses are inversely related to the quark and lepton
masses through their U(1)$_X$ charges.  In Section 3, we
derive the formula for the finite corrections to the light sfermion
masses at the two-loop level in the NESSM, and evaluate them
numerically.  In Section 4, using the constraints from $\Delta m_K$
and $\epsilon_K$ the lower bounds on the third generation sfermion
masses at the GUT scale are derived. 
We neglect the effects of the Yukawa interactions in Section 3 and 4,
since they are model-dependent. 
The effects are evaluated in Section 5.  We also discuss the three-loop
RG contributions to the light sfermion masses there. Section 6 is
devoted to the conclusions and discussion.  In Appendix~\ref{Ap_model},
we review the anomalous U(1) SUSY-breaking model given in
Ref.~\cite{HKN} as one of the explicit realizations of the NESSM. 
In Appendix~\ref{Ap_finite}, we show the
complete formulae for the contributions to the light sfermion masses
from the heavy ones at the two-loop level. The formulae for $\Delta
m_K$ and $\epsilon_K$ are given in Appendix~\ref{Ap_KK}.

\section{Setup} 

In this section, we show the setup of the NESSM on which we perform our
analyses in later sections.  In the NESSM, the SUSY-breaking scalar
masses for the heavy sfermions and extra scalars are given by the
U(1)$_X$ $D$-term.
We here assume that the nonzero U(1)$_X$ $D$-term is generated
associated with the breaking of the U(1)$_X$ gauge symmetry caused by
the VEV of a $\Phi$ field which has a U(1)$_X$ charge of $-1$ 
($Q_{\Phi}^X=-1$). 
In this case, the U(1)$_X$ $D$-term is related to the $F$-term of the
$\Phi$ field at a (local) minimum of the potential as follows:
\begin{eqnarray}
  -D_X = \frac{g_X^2}{M_X^2} |-F_{\Phi}|^2,
\end{eqnarray}
where $M_X$ ($=g_X |\vev{\Phi}|$) is the U(1)$_X$ gauge boson mass.
Here, we have assumed that the VEV's and $F$-terms of the other
fields are negligibly small for simplicity.
Then, the relation between the supergravity contribution $m_0$ and
the U(1)$_X$ $D$-term ($m_D\equiv \sqrt{|-D_X|}$) is given by
\begin{eqnarray}
 m_0 \simeq \frac{|-F_\Phi|}{M_{\rm pl}} 
 = \frac{|\vev{\Phi}|}{M_{\rm pl}} m_D,
\label{m0-mD-rel}
\end{eqnarray}
where $M_{\rm pl}$ is the reduced Planck scale 
($M_{\rm pl} \simeq 2 \times 10^{18}~\GEV$).
This shows that the supergravity contributions to the sfermion masses
are, in general, suppressed compared with the $D$-term contribution,
$m_D$, as long as $|\vev{\Phi}|$ is smaller than $M_{\rm pl}$.
Then, the U(1)$_X$ charged sfermions have large masses of the order of
the U(1)$_X$ $D$-term, $m_D$, while the neutral sfermions only receive
smaller masses of order $m_0$ from the supergravity contributions.
In order to obtain desired mass spectrum, we set $m_0$ to be of the
order of the weak scale and $|\vev{\Phi}|/M_{\rm pl} \simeq 
(10^{-1} \sim 10^{-2})$.

Next, we consider the U(1)$_X$ charge assignment.
We assume that all the fields of the NESSM are embedded in SU(5)$_{\rm
GUT}$ multiplets at the GUT scale. It not only maintains the successful
gauge coupling unification but also guarantees vanishing U(1)$_{Y}$
$D$-term contribution at the one-loop level (Eq.~(\ref{U1X_u1d1})). 
We also assume that one pair of the extra matters are introduced in
(${\bf 5+5^\star}$) representation of the SU(5)$_{\rm GUT}$ for
simplicity. 
The extension to the other cases is straightforward. 
Then, the U(1)$_X$ charges for the ${\bf 5}$ and ${\bf 5^\star}$ extra
matters, $Q^{X}_{{\bf 5}_{\rm ex}}$ and 
$Q^{X}_{{\bf 5^\star}_{\rm ex}}$, are determined from those for the
quarks and leptons in the SM to satisfy Eqs.~(\ref{U1X_af},
\ref{U1X_u1d2}) as
\begin{eqnarray}
  Q^{X}_{{\bf 5}_{\rm ex}} &=& - 2 \sum_{i=1}^{3} Q^{X}_{{\bf 10}_i},
\label{Q_5_ex}\\
  Q^{X}_{{\bf 5^\star}_{\rm ex}} &=& - \sum_{i=1}^{3}
  (Q^{X}_{{\bf 10}_i}+ Q^{X}_{{\bf 5^\star}_i}),
\label{Q_5_star_ex}
\end{eqnarray}
where $Q^{X}_{{\bf 10}_i}$ and $Q^{X}_{{\bf 5^\star}_i}$ $(i=1-3)$ are
the U(1)$_X$ charges for the quarks and leptons embedded in the 
${\bf 10}$ and ${\bf 5^\star}$ representations of the SU(5)$_{\rm GUT}$, 
respectively.  An index $i$ represents the generation. 

The U(1)$_X$ charges for the SM matter multiplets are determined by the
following considerations.
The naturalness argument tells us that the superparticles
which are strongly coupled with the Higgs boson are necessary to have
masses around the weak scale and thus should have zero U(1)$_X$ charges.
This requires that $Q^X_{{\bf 10}_3}=0$, since the top squarks and
left-handed bottom squark are coupled with the Higgs boson through large 
top Yukawa coupling.
On the other hand, the superpartners of the light quarks and leptons
have to be heavy enough to sufficiently suppress the FCNC processes, so
that they should have positive U(1)$_X$ charges. 
This explains the hierarchy of the quark and lepton masses naturally by
the Froggatt-Nielsen mechanism, since the Yukawa matrices are generated
through the VEV of the $\Phi$ field suppressed by suitable powers of
$\vev{\Phi}/M_{\rm pl}$ \cite{FN} as
\begin{eqnarray}
  W &=& 
    \sum_{i,j=1}^{3} (f_u)_{ij}
      \left( \frac{\vev{\Phi}}{M_{\rm pl}} 
      \right)^{Q^X_{{\bf 10}_i}+Q^X_{{\bf 10}_j}}
      \Psi_{{\bf 10}_i} \Psi_{{\bf 10}_j} H_u \nonumber\\
  &&+\sum_{i,j=1}^{3} (f_d)_{ij}
      \left( \frac{\vev{\Phi}}{M_{\rm pl}} 
      \right)^{Q^X_{{\bf 10}_i}+Q^X_{{\bf 5^\star}_j}}
      \Psi_{{\bf 10}_i} \Psi_{{\bf 5^\star}_j} H_d,
\label{hierarchy}
\end{eqnarray}
where $H_u$ and $H_d$ are the Higgs doublets for which we have assumed
the vanishing U(1)$_X$ charges.
Thus, it is possible to reproduce the observed quark and lepton mass
matrices with $(f_u)_{ij}, (f_d)_{ij} = O(1)$ if we appropriately
choose the U(1)$_X$ charges $Q^X_{{\bf 10}_i}$ and 
$Q^X_{{\bf 5^\star}_i}$ \cite{flavor_U1X}.

According to the above arguments, we consider the following U(1)$_X$
charge assignments:
\begin{center}
\begin{tabular}{|c||c|c|c|c|}
\hline
Model                            &   I&  II& III&  IV\\\hline\hline
$Q^{X}_{{\bf 10}_1}$ 		 & $1$& $2$& $2$& $2$\\ 
$Q^{X}_{{\bf 10}_2}$		 & $1$& $1$& $1$& $1$\\ 
$Q^{X}_{{\bf 10}_3}$		 & $0$& $0$& $0$& $0$\\ \hline
$Q^{X}_{{\bf 5^\star}_1}$	 & $1$& $1$& $1$& $2$\\ 
$Q^{X}_{{\bf 5^\star}_2}$	 & $1$& $1$& $1$& $1$\\ 
$Q^{X}_{{\bf 5^\star}_3}$	 & $0$& $1$& $0$& $1$\\ \hline
$Q^{X}_{{\bf 5}_{\rm ex}}$	 &$-4$&$-6$&$-6$&$-6$\\ 
$Q^{X}_{{\bf 5^\star}_{\rm ex}}$ &$-4$&$-6$&$-5$&$-7$\\ \hline 
\end{tabular}
\end{center}
Here, the U(1)$_X$ charges for the Higgs multiplets are zero in all the
models and the charges for the extra-matter multiplets were determined
by Eqs.~(\ref{Q_5_ex}, \ref{Q_5_star_ex}).
Model (I) is the simplest possibility realizing the decoupling
scenario. In this model, the FCNC's are strongly suppressed due to the
degeneracy of the SUSY-breaking masses between the first-two
generation sfermions, while the hierarchy among the quark and lepton
masses cannot be explained completely. Models (II-IV) are motivated
to explain the fermion mass hierarchy by the Froggatt-Nielsen
mechanism.  In particular, in Models (II, IV) the second and third
generation doublet leptons have the same U(1)$_X$ charges,
$Q^{X}_{{\bf 5^\star}_2} = Q^{X}_{{\bf 5^\star}_3}$, so that the
observed large mixing between $\nu_{\mu}$ and $\nu_{\tau}$
\cite{SuperK} is naturally explained.\footnote{
Models (II) and (IV) prefer large and small angle MSW solutions to the
solar neutrino problem, respectively \cite{HMW, YS-R}.}
In these models, the FCNC processes are less suppressed than in 
Model (I). Models (II, III) have
non-degenerate SUSY-breaking masses for the left-handed down-type
squarks in the first-two generations.  Thus, $K^0$-$\bar{K}^0$
oscillation has dominant contribution from the off-diagonal elements
in the left-handed squark mass matrix.  Model (IV) has non-degenerate
SUSY-breaking masses for both the left- and right-handed down-type
squarks in the first-two generations, so that  $K^0$-$\bar{K}^0$
oscillation receives contributions from off-diagonal elements of both
the left- and right-handed squark mass matrices.  As a result, Model
(IV) is more severely constrained from  $K^0$-$\bar{K}^0$ mixing than 
Models (II, III), as will be discussed later.

We now discuss the other SUSY-breaking parameters.
Since the gaugino masses are constrained by naturalness argument, we
consider them to be of the order of the weak scale.
Indeed, in the anomalous U(1) SUSY-breaking model given in
Appendix~\ref{Ap_model}, the gaugino masses arise from the $F$-term of
the dilaton field \cite{ADM} and their sizes can be of the order of the
weak scale \cite{BCCM}.
We treat the gaugino mass (at the GUT scale) as a free parameter in
the phenomenological analyses below.
Also, the superpotential Eq.~(\ref{hierarchy}) generates the
SUSY-breaking trilinear scalar couplings,
\begin{eqnarray}
  (A_u)_{ij} &=& (Q^X_{{\bf 10}_i}+Q^X_{{\bf 10}_j}) 
    \frac{F_{\Phi}}{\vev{\Phi}}, 
\label{A_u} \\
  (A_d)_{ij} &=& (Q^X_{{\bf 10}_i}+Q^X_{{\bf 5^\star}_j}) 
    \frac{F_{\Phi}}{\vev{\Phi}}.
\label{A_d}
\end{eqnarray}
The $(A_u)_{ij}$ and $(A_d)_{ij}$ are defined by ${\cal L}_{\rm soft} =
(y_u)_{ij}(A_u)_{ij} \tilde{\psi}_{{\bf 10}_i} \tilde{\psi}_{{\bf 10}_j}
h_u + (y_d)_{ij}(A_d)_{ij} \tilde{\psi}_{{\bf 10}_i} 
\tilde{\psi}_{{\bf 5^\star}_j} h_d$.
The $(y_u)_{ij}$ and $(y_d)_{ij}$ are the Yukawa couplings $W =
(y_u)_{ij} \Psi_{{\bf 10}_i} \Psi_{{\bf 10}_j} H_u + (y_d)_{ij}
\Psi_{{\bf 10}_i} \Psi_{{\bf 5^\star}_j} H_d$, 
and  $\tilde{\psi}$ and $h$ represent the
scalar components of the corresponding supermultiplets $\Psi$ and $H$,
respectively.
From Eq.~(\ref{m0-mD-rel}), we find that $A_u \simeq A_d \simeq m_D$ 
except for the ones which involve only the light sfermions.

The supersymmetric masses and the holomorphic SUSY-breaking masses for
the extra matters come from the U(1)$_X$ symmetry breaking.  The
supersymmetric masses should be of the same order with $m_D$
as we mentioned in Introduction.  While these features are naturally
explained in the explicit model given in Ref.~\cite{HKN} (see
Appendix~\ref{Ap_model}), we mainly adopt a phenomenological approach in
this article and take the masses as input parameters.
We parameterize the supersymmetric and holomorphic SUSY-breaking mass 
parameters as
\begin{eqnarray}
  W_{\rm ex} = 
     \mex \, \Psi_{{\bf 3}_{\rm ex}} 
    \Psi_{{\bf 3^\star}_{\rm ex}}
   + \mex' \, \Psi_{{\bf 2}_{\rm ex}} 
    \Psi_{{\bf 2^\star}_{\rm ex}},
\end{eqnarray}
\begin{eqnarray}
  {\cal L}_{\rm ex,\,soft} = - (F_\psi\, \tilde{\psi}_{{\bf 3}_{\rm ex}} 
    \tilde{\psi}_{{\bf 3^\star}_{\rm ex}}
   + F'_\psi\, \tilde{\psi}_{{\bf 2}_{\rm ex}} 
    \tilde{\psi}_{{\bf 2^\star}_{\rm ex}} + {\rm h.c.}).
\end{eqnarray}
Here, the triplet extra matters $\Psi_{{\bf 3}_{\rm ex}}$ and
$\Psi_{{\bf 3^\star}_{\rm ex}}$ and the doublet extra matters
$\Psi_{{\bf 2}_{\rm ex}}$ and $\Psi_{{\bf 2^\star}_{\rm ex}}$ are
embedded in the $\Psi_{{\bf 5}_{\rm ex}}$ and 
$\Psi_{{\bf 5^\star}_{\rm ex}}$.
In this article, we impose the SU(5)$_{\rm GUT}$ relations on the
supersymmetric and the holomorphic SUSY-breaking masses, $\mex=\mex'$
and $F_\psi = F'_\psi$, at the GUT scale unless otherwise stated.
Then, the masses for the heavy sfermions and extra matters are
completely determined by three free parameters $m_D$, $\mex$ and
$F_\psi$ once a U(1)$_X$ charge assignment is specified.

\section{Corrections to the Light Sfermions}

In the NESSM, the two-loop RG contributions to the light sfermions are
canceled between the heavy sfermions and the extra-matter
multiplets. However, there are still negative finite corrections. 
In this section, we numerically estimate
the effects of the finite corrections in various cases, using their
explicit form which is completely calculated at the two-loop level in
Appendix~\ref{Ap_finite}. We conclude this section by showing how
hierarchical the sfermion mass spectrum can be in the NESSM
without breaking color and charge, comparing the NESSM with the
original effective SUSY in which there is no extra-matter multiplet.

Now, let us numerically estimate the finite corrections in the models
discussed in the previous section.
We first consider Models (I, II). In these models, the
U(1)$_X$ charge assignments are consistent with the SU(5)$_{\rm GUT}$
and the extra-matter multiplets have an invariance under the parity
$\Psi_{{\bf 5}_{\rm ex}} \leftrightarrow \Psi_{{\bf 5^\star}_{\rm
ex}}$, so that the finite corrections to the light sfermions,
$m_{\tilde{f},\,{\rm finite}}^2$, take relatively simple forms.  The
explicit expressions for $m_{\tilde{f},\,{\rm finite}}^2$ are given in
Eqs.~(\ref{Ap_m2_direct_fin_simple}, \ref{Ap_m2_FI-1loop_fin_simple},
\ref{Ap_m2_FI-2loop_fin_simple}) in Appendix~\ref{Ap_finite}.  These
expressions can be further simplified in the case of one-pair
extra-matter multiplets as follows:
\begin{eqnarray}
m_{\tilde{f},\,{\rm finite}}^2
  &=&
    4 \sum_A \left( \frac{\alpha_A}{4\pi} \right)^2
    C_{\tilde{f}}^A 
\Biggl\{
      -\Tr_{\HS} \Biggl[ T_{\HS}^A\, 
    m_{\HS}^2 \biggl( \log \frac{\mex^2}{m_{\HS}^2} 
    +\frac{1}{3}\pi^2 \biggr) \Biggr] 
    +\frac{1}{2}\mex^2\,G(y_1,y_2) \Biggr\}
     \nonumber\\
  && \nonumber\\
  && 
    +4 \sum_A \left( \frac{\alpha_Y}{4\pi} \right) 
    \left( \frac{\alpha_A}{4\pi} \right) Y_{\tilde{f}} 
    \Tr_{\HS} \Biggl[ Y_{\HS} C_{\HS}^A\, m_{\HS}^2 
    \log m_{\HS}^2 \Biggr],
\label{finite}
\end{eqnarray}
where
\begin{eqnarray}
  G(y_1,y_2)&=&
  \left(y_1 \log y_1-2 y_1 {\rm Li}_2 (1-\frac{1}{y_1}) 
  +\frac{1}{2} y_1 {\rm Li}_2(1-\frac{y_2}{y_1})\right)
  +(y_1 \leftrightarrow y_2),\\
  &&
  \hspace{2cm}
  y_1\equiv\frac{m_1^2}{\mex^2},
  \hspace{0.5cm}
  y_2\equiv\frac{m_2^2}{\mex^2},
\end{eqnarray}
where $\HS$ denotes the heavy sfermions; $m_1$ and $m_2$ are the
mass eigenvalues for the extra scalars (see Appendix~\ref{Ap_finite}). 
Here, we have neglected the difference between the supersymmetric masses 
(and the holomorphic SUSY-breaking masses) of the triplet and doublet
extra matters for demonstrational purpose, though it will be included in
later numerical calculations.

The first term of the first line in Eq.~(\ref{finite}) gives dominant
contributions to the light-sfermion mass squareds.  They are negative
and could cause color and charge breaking since $\mex>m_{\HS}$.
The second term has the same form as the scalar mass squared generated
by integrating out the messenger fields in the gauge-mediated SUSY
breaking mechanism \cite{Martin}. 
Therefore, it gives the positive contributions
almost proportional to the SUSY-breaking bilinear coupling $F_\psi$ of
the extra scalars. Even if $F_\psi=0$, however, $G(y_1,y_2)$ remains
positive, since $m_1^2$ and $m_2^2$ are smaller than $\mex^2$ by the
U(1)$_X$ $D$-term. The last term which is generated through U(1)$_Y$
$D$-term does not give significant contributions due to the smallness
of the hypercharge gauge coupling. 
The renormalization-point dependence of the term is canceled out due to 
Eq.~(\ref{U1X_u1d2}).

In order to see the dependence of the finite corrections on $\mex$,
$m_D$ and $F_\psi$, we fix the heavy-sfermion mass scale
$m_D=10~\TEV$ as an overall scale and take
$\mex/m_D$ and $F_\psi/(\mex m_D)$ as remaining two dimensionless free
parameters. The ratio $\mex/m_D$ must be of order unity in the NESSM,
since otherwise the heavy sfermions generate the large radiative
corrections below the scale $\mex$ where the extra-matter multiplets
decouple.  Then, ${F_\psi}/(\mex m_D)$ should not be much larger than
one to avoid the negative mass squared for the extra scalar.
Indeed, ${F_\psi}/(\mex m_D)$ is of order unity in the explicit
example constructed.

\begin{figure}
\centerline{\epsfxsize=17cm \epsfbox{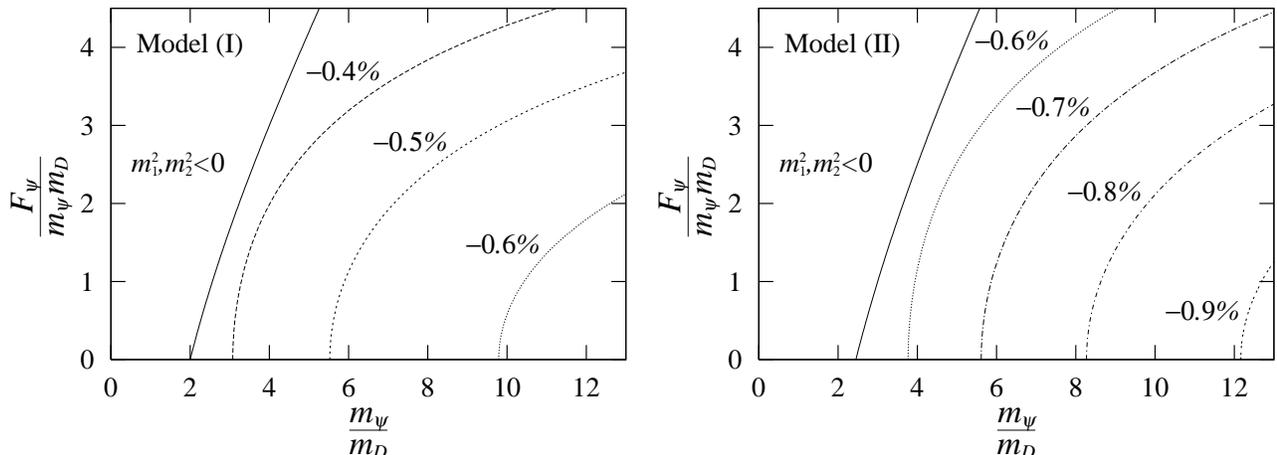}}
\vspace*{-5mm}
\caption{Finite corrections to the light squarks through the strong
 interaction. The percentages on the figures represent the ratio of
 the finite corrections to the heavy-sfermion mass scale,
 $m_{\tilde{q},\,{\rm finite}}^2/m_D^2$. The left figure is in Model
 (I) and the right is in Model (II). The left side of the solid line
 is excluded since the mass squared for the extra scalar is
 negative and color is broken. Here, we have taken $m_D=10~\TEV$ and 
 evaluated all the parameters at $\mu=10~\TEV$.}
\label{fig_correct}
\end{figure}
Fig.~\ref{fig_correct} illustrates the dependence of the finite
corrections on these two parameters.  As an example, we consider the
corrections to the light squarks in Models (I, II) and plot the ratio
of the finite corrections to the heavy sfermion mass-scale squared
$m_D^2$. Here, we only take into account the dominant two-loop
contribution through the strong interaction, that is we drop the last
term and set $A=3$ and $C_{\tilde{f}}^{(3)}=4/3$ in Eq.~(\ref{finite}).
The minimum of the ratio is $-0.36\%$ ($\mex/m_D=2.2$) and
$-0.54\%$ ($\mex/m_D=2.7$) in Models (I) and (II), respectively, 
when $F_\psi=0$. The difference between Models (I) and (II) in
Fig.~\ref{fig_correct} comes mainly from the group-theoretical factor in
Eq.~(\ref{finite}), $\Tr_{\HS} \left[T_{\HS}^A\, m_{\HS}^2\right] /
m_D^2 = \Tr_{\HS} \left[T_{\HS}^A\, Q^X_{\HS}\right]$, which is $4$ in
Model (I) and $6$ in Model (II). Thus, the corrections in Model (II) are
larger than in Model (I) approximately by a factor of $1.5$.

The ratio given in Fig.~\ref{fig_correct} determines how hierarchical
the sfermion mass spectrum can be without breaking color and charge. 
Since the bare mass $m_{\tilde{q},0}$ for the light squarks has to be
larger than these negative corrections, it gives a lower bound on 
$m_{\tilde{q},0}$. In Model (I), for example, we obtain 
\begin{eqnarray}
 m_{\tilde{q},0}^2 
  \ge -m_{\tilde{q},\,{\rm finite}}^2 \simeq 0.4\% \times m_D^2,
\label{ratio_NESSM}
\end{eqnarray}
when the supersymmetric mass for the extra matter, $\mex$, is
not much larger than $m_D$ as discussed in the previous
section.\footnote{
In the region where $\mex/m_D \gsim O(10)$ and ${F_\psi}/(\mex m_D) =
O(10)$, the finite corrections can be much smaller due to the accidental 
cancellation between the negative contribution from the heavy-sfermion
loops and the gauge-mediated contribution from the extra-matter
multiplets.
However, this does not necessarily mean that more hierarchical
superparticle mass spectrum can be realized, since the gluino mass
receives the correction from the extra-matter loops 
($m_{\tilde{g}} \simeq (\alpha_3/4\pi) (F_\psi/ \mex) \sim 0.1 m_D$).}
Since $m_{\tilde{q},0}$ set the scale for the light sfermion masses, we
find that the splitting between the heavy and light sfermions cannot far
exceed an order of magnitude.  In the case of the original effective
SUSY models (models without extra-matter multiplets), the RG
contributions to the light sfermions, $m_{\tilde{f},\,{\rm log}}^2$, are
written by solving the RG equation Eq.~(\ref{rge2sf}) as follows:
\begin{eqnarray}
m_{\tilde{f},\,{\rm log}}^2(\mu)
 &=& 4\sum_A \frac{C^A_{\tilde{f}}}{b_A} 
     \Tr_{\HS} \Biggl[T_{\HS}^A\, m_{\HS}^2\Biggr]
     \left(\frac{\alpha_A(\mu)}{4\pi}-\frac{\alpha_{\rm GUT}}{4\pi}
         \right),
\end{eqnarray}
where $b_A$ is the coefficient for the one-loop beta function of the SM
gauge coupling and $\alpha_{\rm GUT}$ is the gauge coupling constant 
at the GUT scale. Here, the contributions from the U(1)$_Y$ $D$-term are
neglected. In Model (I), for example, radiative corrections to the light
squarks through the strong interaction are estimated as
\begin{eqnarray}
m_{\tilde{q},\,{\rm log}}^2(\mu=10~\TEV)
  \simeq -2.3\% \times m_D^2,
\label{ratio_ESSM}
\end{eqnarray}
so that the bare light-squark mass squared $m_{\tilde{q},0}^2$ should be
larger than about 
$-m_{\tilde{q},\,{\rm log}}^2(\mu=10~\TEV) \simeq 2.3\% \times m_D^2$.
The rough estimations in Eqs.~(\ref{ratio_NESSM}, \ref{ratio_ESSM})
show that the splitting between the heavy and light sfermion masses in 
the NESSM can be more than twice larger than in the original effective
SUSY. This conclusion remains true even if full radiative corrections
are taken into account, as will be shown below.

We here comment on the effect of introducing more than one pairs of
extra-matter multiplets.
If we introduce more extra-matter pairs, the extra matters could have
smaller U(1)$_X$ charges satisfying Eqs.~(\ref{U1X_af}, \ref{U1X_u1d2})
and thus smaller supersymmetric masses.
Thus, we can reduce the finite corrections to the light-sfermion mass
squareds by introducing more extra matters, since it is the
supersymmetric mass for the extra matters that determines the size of
the negative finite corrections.
For example, the minimum size of the finite corrections with $F_\psi=0$
in Model (I) are reduced to $-0.36\%$, $-0.30\%$, $-0.26\%$, and
$-0.24\%$ by introducing one, two, three and four pairs of 
$\Psi_{{\bf 5}_{\rm ex}}$ and $\Psi_{{\bf 5^\star}_{\rm ex}}$, 
respectively.
However, introducing extra matters changes the beta functions, so that
the behavior of the gauge couplings becomes worse at high energy in this 
case.  In this article, we limit ourselves to the case of one pair of
$\Psi_{{\bf 5}_{\rm ex}}$ and $\Psi_{{\bf 5^\star}_{\rm ex}}$,
hereafter. 

We now include the effects from the gauginos.  Since the gauginos give
positive RG contributions at the one-loop level, they relax the lower
bounds on the light sfermion masses.  The lower bounds on the
light sfermion masses can be determined as follows.  We first set the
boundary conditions at the GUT scale $2 \times 10^{16}~\GEV$: we, for
simplicity, take the universal gaugino mass $m_{1/2}$ and give the
bare mass $m_0$ to the light sfermions at the GUT scale.  Then, we run
all the soft SUSY-breaking masses using two-loop RG equations.  We
neglect the effects of the Yukawa couplings, since they are rather
model-dependent without the assumption of universal scalar
masses. The effects will be discussed in Section 5.  
The heavy sfermions are
decoupled at the scale $m_D=10~\TEV$, where the full finite
corrections are added to the light sfermions.  The positive
contributions from the gauginos are included from the GUT scale to the
scale where the gluino decouples.  After these steps, if any light
sfermions have negative mass squareds at the weak scale, the parameter
region is excluded and we need larger value of $m_0$ at the
GUT scale.

\begin{figure}[t]
\centerline{\epsfxsize=14cm \epsfbox{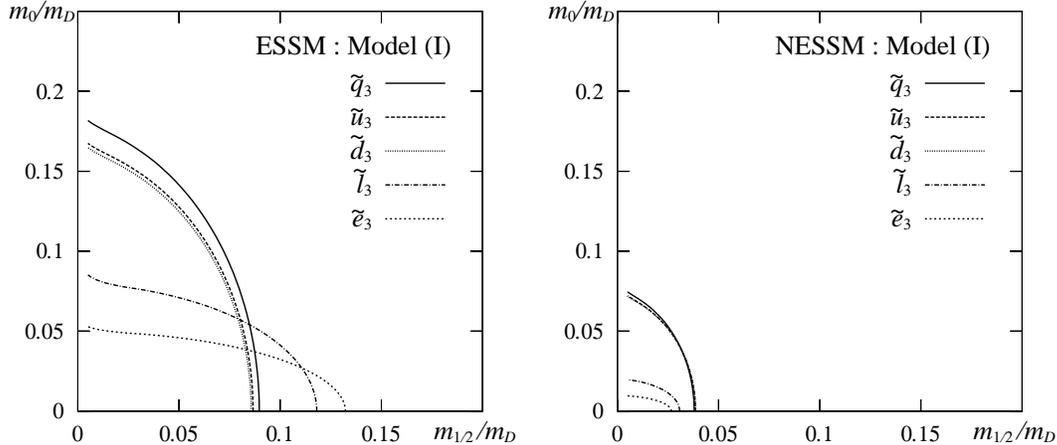}}
\caption{Lower bounds on the mass ratios of the light sfermions to the
 heavy sfermions in Model (I). $m_0$ and $m_{1/2}$ represent
 the light sfermion and gaugino masses at the GUT scale, respectively.
 The regions below the curves are excluded, since the light sfermions
 indicated by various lines have negative mass squareds at the weak
 scale, breaking color and charge.  
 Here, we have set $m_D=10~\TEV$ and $\mex=2.0\times
 m_D$ at the GUT scale, and neglected the Yukawa couplings.}
\label{fig_ratio_1}
\end{figure}
\begin{figure}[t]
\centerline{\epsfxsize=14cm  \epsfbox{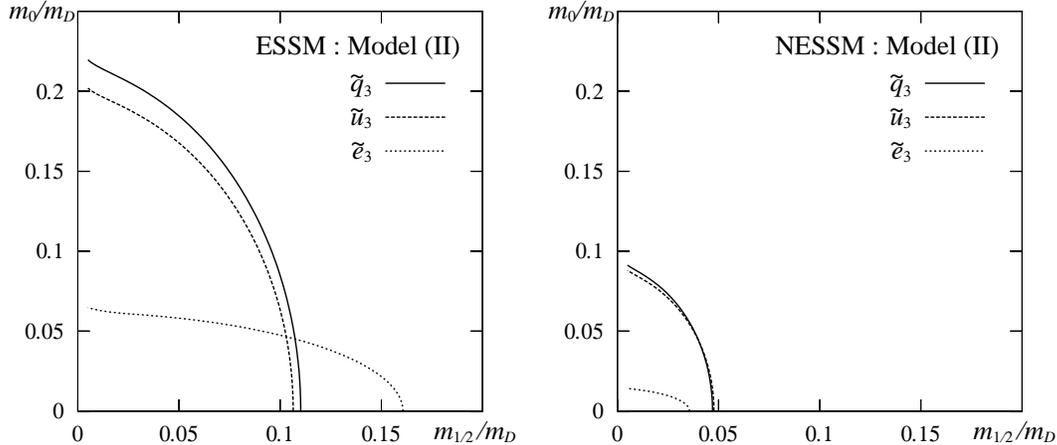}}
\caption{Lower bounds on the mass ratios of the light sfermions to the
 heavy sfermions in Model (II).  We have set $m_D=10~\TEV$ and 
 $\mex=2.5\times m_D$ at the GUT scale.}
\label{fig_ratio_2}
\end{figure}
In Fig.~\ref{fig_ratio_1} and Fig.~\ref{fig_ratio_2}, we have plotted 
the lower bounds on the light sfermion masses $m_0$ as a
function of the gaugino mass $m_{1/2}$ in Models (I) and (II).
The $m_0$ and $m_{1/2}$ are the GUT-scale values and normalized by the
heavy-sfermion mass scale $m_D$.
In each figure, the left graph (ESSM) is in the case of the original 
Effective SUSY SM (without extra-matter multiplets) and the right one
(NESSM) is of the Natural Effective SUSY SM. 
We have taken the supersymmetric mass $\mex$ for the extra-matter
multiplets as $\mex=2.0\times m_D$ in Model (I) and $\mex=2.5\times m_D$
in Model (II) at the GUT scale.
The various curves represent the lower bounds on the bare masses
for the indicated light sfermions, below which they have negative mass 
squareds at the weak scale.
From Figs.~\ref{fig_ratio_1} and \ref{fig_ratio_2} we find that in the 
NESSM the light sfermion masses and/or the gaugino masses can take 
values less than half in the ESSM without breaking color and charge.
This in turn means that we can obtain more hierarchical mass spectrum 
in the NESSM than in the ESSM.

Note that we have assumed that the triplet and doublet extra matters
have the same supersymmetric masses at the GUT scale.  Then, at low
energy the supersymmetric mass for the triplet extra matter becomes
about twice as large as that for the doublet one due to the RG
evolution.  As a result, we cannot take the supersymmetric mass which
minimizes the finite corrections in Fig.~\ref{fig_correct}, since then
the doublet extra scalar has negative mass squared.  Thus, if we allow
different supersymmetric masses for the triplet and doublet extra
matters at the GUT scale, we can further reduce the size of finite
corrections.\footnote{The different supersymmetric
masses for the triplet and doublet extra matters do not necessarily
contradict with the GUT, since we can make them split using
GUT-breaking VEV such as $\vev{\Sigma({\bf 24})}$.}  
In this case, however, the running masses for the triplet extra scalars
take negative values at the high-energy scale, which means that the
scalar potential has another minimum at the nonzero values of the
triplet extra scalars.

\begin{figure}[t]
\centerline{\epsfxsize=14cm \epsfbox{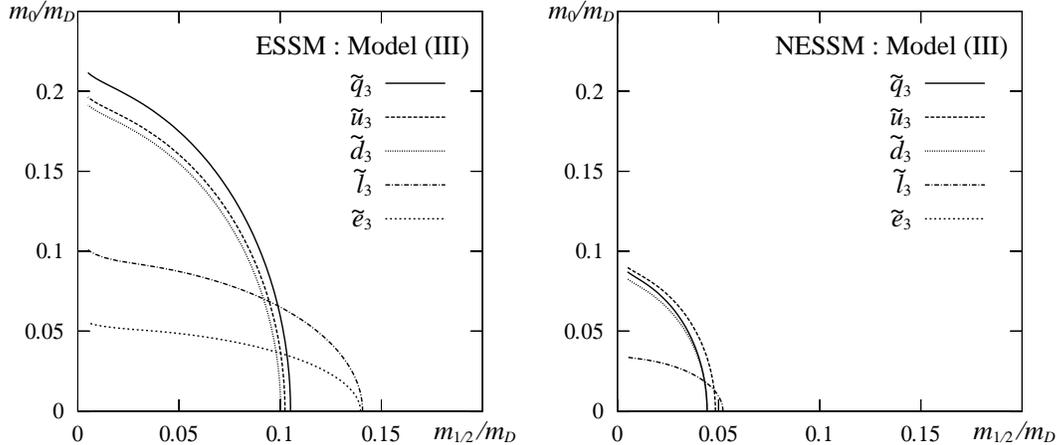}}
\caption{Lower bounds on the mass ratios of the light sfermions to the
 heavy sfermions in Model (III). We have set $m_D=10~\TEV$ and 
 $\mex=2.5\times m_D$ at the GUT scale.}
\label{fig_ratio_3}
\end{figure}
\begin{figure}[t]
\centerline{\epsfxsize=14cm \epsfbox{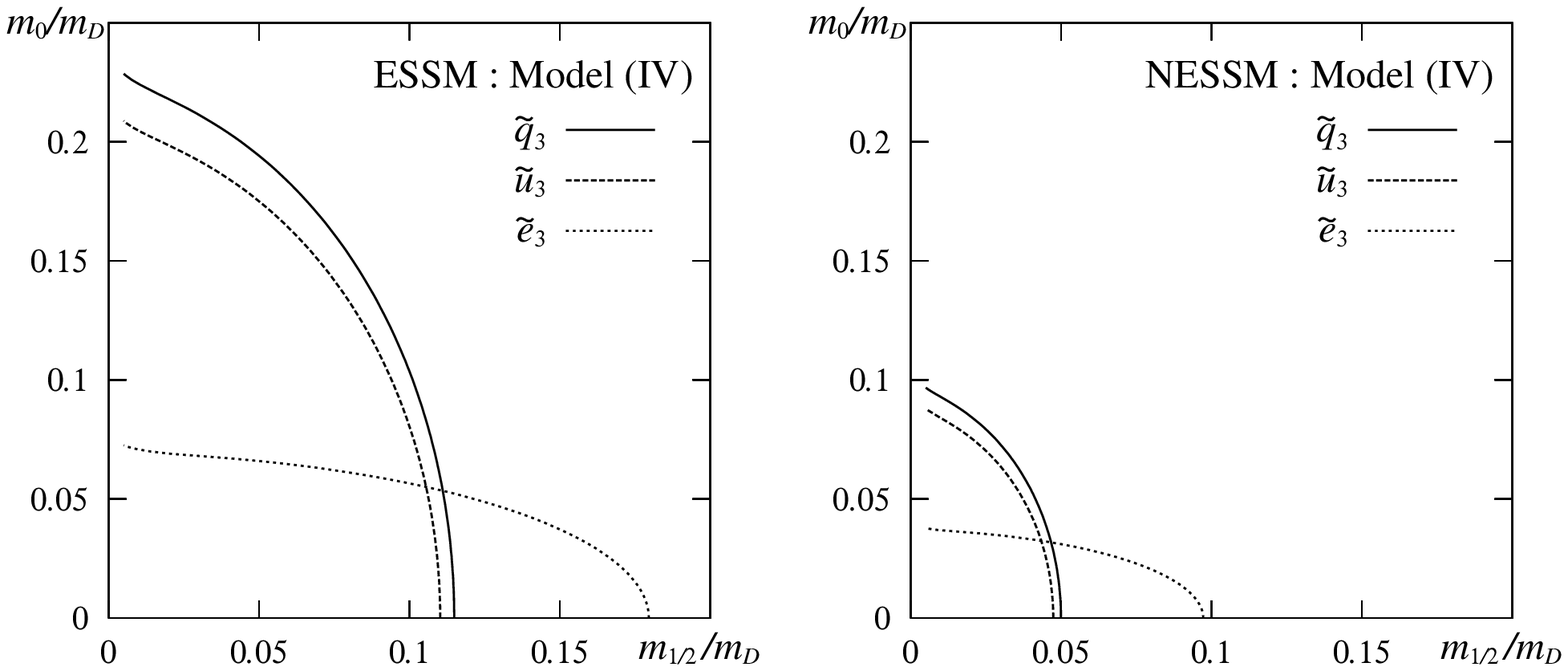}}
\caption{Lower bounds on the mass ratios of the light sfermions to the
 heavy sfermions in Model (IV). We have set $m_D=10~\TEV$ and 
 $\mex=2.7\times m_D$ at the GUT scale.} 
\label{fig_ratio_4}
\end{figure}
In the rest of this section, we consider Models (III, IV). 
The finite corrections in these models are given by 
Eqs.~(\ref{Ap_m2_direct_fin}, \ref{Ap_m2_FI-1loop_fin}, 
\ref{Ap_m2_FI-2loop_fin}) in Appendix~\ref{Ap_finite}.
They take less simple forms due to the absence of the parity between 
the extra-matter multiplets, $\Psi_{{\bf 5}_{\rm ex}} \leftrightarrow 
\Psi_{{\bf 5^\star}_{\rm ex}}$.
In Fig.~\ref{fig_ratio_3} and Fig.~\ref{fig_ratio_4}, we have plotted 
the lower bounds on $m_0$ in Models (III) and (IV), respectively. 
We have taken the supersymmetric mass for the extra-matter multiplet as 
$\mex=2.5\times m_D$ in Model (III) and $\mex=2.7\times m_D$ in Model
(IV) at the GUT scale.
All the other parameters have been taken the same as in Models (I, II). 

The main difference of Models (III, IV) from Models (I, II) is the 
existence of the one-loop U(1)$_Y$ $D$-term,
\begin{eqnarray}
m_{\tilde{f},\,{\rm FI-1loop}}^2 
= \left( \frac{\alpha_Y}{4\pi} \right) Y_{\tilde{f}}
    \Tr_{\EX} \Biggl[ Y_{\EX} \cos 2\theta 
    \biggl( m_1^2 \log m_1^2 - m_2^2 \log m_2^2 \biggr) \Biggr],
\label{1loop-U1Y-FI}
\end{eqnarray}
where $\EX$ denotes the extra matters.  This term arises because the RG
evolution from the GUT to low-energy scale splits the supersymmetric
masses for the triplet and doublet extra matters.  However, the
corrections Eq.~(\ref{1loop-U1Y-FI}) do not dominate the two-loop
contributions, since the hypercharge gauge coupling is much smaller than
the strong coupling and there is no enhancement by factors such as 4 or
$\Tr_{\HS} \left[T_{\HS}^A\, m_{\HS}^2 \right]$ in
Eq.~(\ref{finite}). As a consequence, Figs.~\ref{fig_ratio_3} and
\ref{fig_ratio_4} are roughly the same as Figs.~\ref{fig_ratio_1} and
\ref{fig_ratio_2}.  Nonetheless, we can still find the small effects
of the one-loop U(1)$_Y$ $D$-term. In Model (III), the lower bound
on the mass for $\tilde{u}_3$ is tighter than that for $\tilde{q}_3$,
since a positive one-loop U(1)$_Y$ $D$-term is generated.  In Model
(IV), the lower bound on the mass for $\tilde{e}_3$ is much larger
than that in Model (II) because of a negative one-loop U(1)$_Y$
$D$-term.

\section{Constraints from $K^0$-$\bar{K}^0$ Mixing}

$K^0$-$\bar{K}^0$ mixing gives the severest constraints on the
masses for the first-two generation sfermions in flavor-changing
processes. In this section, we calculate the lower bounds on the
heavy-sfermion mass scale $m_D$ from the $K_L$-$K_S$ mass difference
$\Delta m_K$ and the CP-violating parameter $\epsilon_K$.  
Combining these bounds on $m_D$ with the bounds on the mass ratio
$m_0/m_D$ given in the previous section, 
we obtain the lower bounds on the light sfermion
masses $m_0$ at the GUT scale. It turns out that no severe fine tuning
is needed in the NESSM, compared with the original Effective SUSY SM
(ESSM).

We first discuss the structure of the mass matrix for the down-type
squarks, since it induces a dominant flavor violation in 
$K^0$-$\bar{K}^0$ mixing through the gluino-squark box diagram.  We
restrict our attention to the contribution from the heavy first-two
generation squarks below.  The mass matrix for the down-type squarks
in the U(1)$_X$ gauge basis is given by
\begin{eqnarray}
  M^2 &=&
  \left(
  \begin{array}{cc}
  M_{LL}^2 & {\bf 0}\\ 
  {\bf 0} & M_{RR}^2
  \end{array} 
  \right),
\end{eqnarray}
where we have set the left-right mixing mass terms zero 
due to the smallness of the corresponding Yukawa couplings.
The left- and right-handed squark mass matrices
$M_{LL}^2$ and $M_{RR}^2$ consist of two parts as follows: 
\begin{eqnarray}
  M_{LL}^2 = M_{LL,D}^2 +M_{LL,0}^2,
  \hspace{5mm}
  M_{RR}^2 = M_{RR,D}^2 +M_{RR,0}^2.
\end{eqnarray}
$M_{LL,D}^2$ and $M_{RR,D}^2$ come from the 
U(1)$_X$ $D$-term and are proportional to their U(1)$_X$ charges,
\begin{eqnarray}
  M_{LL,D}^2=
  \left(
  \begin{array}{cc}
  Q^X_{{\bf 10}_1}  & 0\\ 
  0 & Q^X_{{\bf 10}_2} 
  \end{array} 
  \right) m_D^2,
  \hspace{5mm}
  M_{RR,D}^2=
  \left(
  \begin{array}{cc}
  Q^X_{{\bf 5^\star}_1}  & 0\\ 
  0 & Q^X_{{\bf 5^\star}_2} 
  \end{array} 
  \right) m_D^2.
\end{eqnarray}
$M_{LL,0}^2$ and $M_{RR,0}^2$ are contributions from supergravity,
whose components are of the order of the light-sfermion mass squared
$m^2_0$ and induce flavor violation,\footnote{
If the U(1)$_X$ charges for the first and second generations are
different, the off-diagonal elements of $M_{LL,0}^2$ and $M_{RR,0}^2$
are suppressed by 
$(\vev{\Phi}/M_{\rm pl})^{|Q^X_{{\bf 10}_1}-Q^X_{{\bf 10}_2}|}$ and 
$(\vev{\Phi}/M_{\rm pl})^{|Q^X_{{\bf 5^\star}_1} - 
Q^X_{{\bf 5^\star}_2}|}$, respectively.}
\begin{eqnarray}
  M_{LL,0}^2 = \left(
  \begin{array}{cc}
  O(m_0^2)  & O(m_0^2)\\ 
  O(m_0^2) & O(m_0^2) 
  \end{array} 
  \right) 
  , \hspace{5mm} 
  M_{RR,0}^2 = \left(
  \begin{array}{cc}
  O(m_0^2)  & O(m_0^2)\\ 
  O(m_0^2) & O(m_0^2) 
  \end{array} 
  \right).
\end{eqnarray}

The SUSY contribution to  $K^0$-$\bar{K}^0$ mixing
is controlled by two parameters qualitatively, as shown in
Appendix~\ref{Ap_KK}:
the averaged squark masses $\bar{m}_{LL,\ RR}^2$ and the off-diagonal
elements $\delta_{LL,\ RR}$.\footnote{In the following figures, we use
the exact formula for the $K^0$-$\bar{K}^0$ mixing given in
Appendix~\ref{Ap_KK}.}
The averaged squark masses $\bar{m}_{LL}^2$ and $\bar{m}_{RR}^2$
are given by 
\begin{eqnarray}
  \bar{m}_{LL}^2=\sqrt{Q^X_{{\bf 10}_1}Q^X_{{\bf 10}_2}} m_D^2,
  \hspace{5mm}
  \bar{m}_{RR}^2=\sqrt{Q^X_{{\bf 5^\star}_1}Q^X_{{\bf 5^\star}_2}}m_D^2,
\end{eqnarray}
where we have neglected $O(m_0^2)$ contributions. 
On the other hand, $\delta_{LL}$
and $\delta_{RR}$ are the off-diagonal elements of the squark mass
matrices normalized with the averaged squark masses, in a basis where
the quark mass matrix is diagonal (see also Appendix~\ref{Ap_KK}).
In order to change the U(1)$_X$ gauge basis to this basis, we introduce
unitary matrices $V^L$ and $V^R$ which diagonalize the down-type quark 
Yukawa matrix given in the U(1)$_X$ gauge basis.
We parameterize them as
\begin{eqnarray}
  V^L=
  \left(
  \begin{array}{cc}
  \cos\theta_L & -e^{i\alpha_L} \sin\theta_L \\
  e^{i\alpha_L}\sin\theta_L & \cos\theta_L
  \end{array}
  \right),
\end{eqnarray}
and $V^R$ with the replacement $L\rightarrow R$ in $V^L$.
Then, the squark mass matrices, ${\cal M}_{LL}^2$ and ${\cal M}_{RR}^2$,
in a basis where the quark mass matrix is diagonal are given by
\begin{eqnarray}
   {\cal M}_{LL}^2=V^{L\dagger}M_{LL}^2V^{L},
  \hspace{5mm}
   {\cal M}_{RR}^2=V^{R\dagger}M_{RR}^2V^{R}.
\end{eqnarray}
Consequently, the off-diagonal element $\delta_{LL}$ is represented 
as follows:
\begin{eqnarray}
  \delta_{LL}=
  -e^{i\alpha_L}\sin\theta_L \cos\theta_L 
  {(Q^X_{{\bf 10}_1}-Q^X_{{\bf 10}_2})}\frac{m_D^2}{\bar{m}_{LL}^2}
  + O\left(\frac{m_0^2}{\bar{m}_{LL}^2}\right),
\end{eqnarray}
and $\delta_{RR}$ is obtained with the replacement $L\rightarrow R$
and $Q^X_{{\bf 10}_i} \rightarrow Q^X_{{\bf 5^\star}_i}$.  If
$Q^X_{{\bf 10}_1}\neq Q^X_{{\bf 10}_2}$, $\delta_{LL}$ is dominated by
the U(1)$_X$ $D$-term contribution and is given definitely up to the
mixing angle $\sin\theta_L$. On the other hand, if $Q^X_{{\bf 10}_1} =
Q^X_{{\bf 10}_2}$, the U(1)$_X$ $D$-term contribution vanishes and the
second term dominates $\delta_{LL}$.
In this case, the off-diagonal element has a large model dependence 
since we cannot calculate the supergravity contributions.

The contribution to  $K^0$-$\bar{K}^0$ mixing from the
gluino-squark box diagram is calculated in Appendix~\ref{Ap_KK}.
We have included the leading-order QCD corrections \cite{LO, AG} and
used the bag parameters obtained by lattice calculations 
\cite{NLO, Con_Sci}.
The details are given in Appendix~\ref{Ap_KK}.  
The constraints from $\Delta m_K$ are summarized as follows:
\begin{eqnarray}
  \delta_{LL,\ RR} &<& \frac{\bar{m}_{LL,\ RR}}{(25 \sim 35)~\TEV},
  \label{KK_LL}\\
  (\delta_{LL}\delta_{RR})^{1/2} 
   &<& 
  \frac{(\bar{m}_{LL}\bar{m}_{RR})^{1/2}}{(150 \sim 250)~\TEV}.
\label{KK_LR}
\end{eqnarray}
This shows that if $\delta_{LL}$ and $\delta_{RR}$ are of the same
order, Eq.~(\ref{KK_LR}) gives about ten times as severe bounds on the
heavy squark masses as Eq.~(\ref{KK_LL}).  
Furthermore, if $\delta_{LL}$ and/or $\delta_{RR}$ have CP-violating
phases of order unity, the constraints from $\epsilon_K$ give twelve
times severer bounds,
\begin{eqnarray}
  \delta_{LL,\ RR} &<& \frac{\bar{m}_{LL,\ RR}}{(310 \sim 430)~\TEV},
\label{KK_LLep}\\
  (\delta_{LL}\delta_{RR})^{1/2} 
   &<& 
  \frac{(\bar{m}_{LL}\bar{m}_{RR})^{1/2}}{(1900 \sim 3100)~\TEV},
\label{KK_LRep}
\end{eqnarray}
as discussed in Appendix~\ref{Ap_KK}.

Now, let us discuss Models (I-IV) in turn.
In Model (I), the averaged squark masses are
\begin{eqnarray}
 \bar{m}_{LL}^2=\bar{m}_{RR}^2=m_D^2.
\end{eqnarray}
Since the U(1)$_X$ charges for the first-two generations are the same,
the $D$-term contribution does not induce any flavor violation. 
Thus, Model (I) is the hybrid scenario of the decoupling and degeneracy
in a sense.  Then, the flavor violation comes from the supergravity
contributions.  Although we cannot calculate the off-diagonal elements
in this case, we expect $\delta_{LL,RR}$ to be of order $(0.1 \sim 1)
{m_0^2}/{\bar{m}_{LL,\ RR}}$.  For simplicity, we here assume
\begin{eqnarray}
  \delta_{LL}=\frac{m_0^2}{m_D^2},
  \hspace{5mm}
  \delta_{RR}=\frac{m_0^2}{m_D^2}.
\label{delta_1}
\end{eqnarray}
Then, we obtain the bound on the light sfermion mass $m_0$ from
Eq.~(\ref{KK_LR}) as
\begin{eqnarray}
  m_0 &>& (150 \sim 250)~\left(\frac{m_0}{m_D}\right)^{3}~\TEV, 
\label{bound_1} 
\end{eqnarray}
if $\delta_{LL}$ and $\delta_{RR}$ do not have CP-violating phases.
The ratio $m_0/m_D$ is bounded from below in Fig.~\ref{fig_ratio_1} in
the previous section, so that the lower bounds on $m_0$ can be estimated
as follows:
\begin{eqnarray}
\begin{array}{cc}
   {\rm ESSM} & {\rm NESSM} \\
  m_0/m_D \ge 0.18 & m_0/m_D \ge 0.07  \\
  m_0 > (0.9 \sim 1.5)~\TEV  &  m_0 > (60 \sim 100)~\GEV 
\label{bound_1_table}
\end{array}
\end{eqnarray}
Here, the bounds on $m_0/m_D$ are those for zero gaugino mass. Note
that the lower bound on $m_0$ depends on the third power of the ratio
$m_0/m_D$ in Eq.~(\ref{bound_1}). As a result, although the bound on
$m_0/m_D$ in the NESSM is only $1/2.5$ of that in the ESSM, the bound on
the light sfermion mass $m_0$ in the NESSM becomes much smaller than in
the ESSM by a factor of $(1/2.5)^3 \sim 1/15$.

\begin{figure}[t]
\centerline{\epsfxsize=15cm \epsfbox{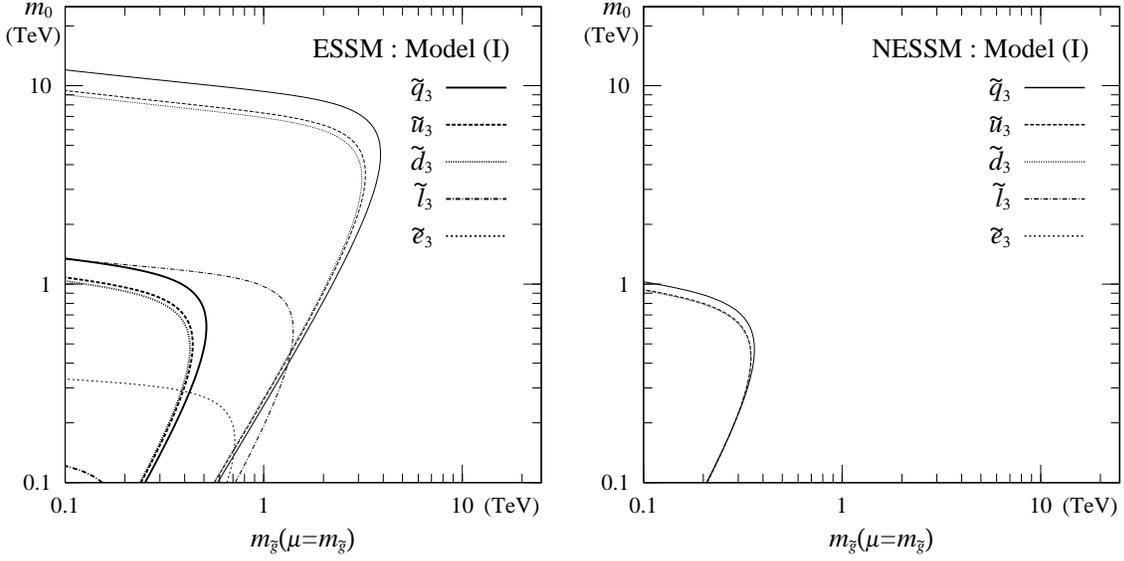}}
\caption{Lower bounds on the bare masses for the light sfermions in
 Model (I). The horizontal axis represents the gluino running mass
 evaluated at the gluino decoupling scale, and the vertical axis the
 indicated light sfermion mass at the GUT scale. 
 The regions below the curves are excluded due to the negative mass
 squareds for the corresponding sfermions.  
 Bold lines represent the constraints 
 from $\Delta m_K$.  Fine lines represent the constraints from
 $\epsilon_K$ with $O(1)$ CP phases.} 
\label{fig_KK_1}
\end{figure}

If $\delta_{LL}$ and $\delta_{RR}$ have CP phases of order unity,
the constraint from $\epsilon_K$ gives twelve times larger bounds on
$m_0$ than that from $\Delta m_K$ as,
\begin{eqnarray}
\begin{array}{cc}
  {\rm ESSM} & {\rm NESSM} \\
  m_0/m_D \ge 0.18 & m_0/m_D \ge 0.07  \\
  m_0 > (11 \sim 18)~\TEV  &  m_0 > (0.8 \sim 1.3)~\TEV 
  \label{bound_1_table1}
\end{array}
\end{eqnarray}
Thus, it seems to require some tuning of the phases or electroweak
symmetry breaking even in the NESSM.
In order to reduce the bound by a factor of $3$, for example, the phase
of $\delta_{LL}\delta_{RR}$ must be tuned to be $1/3^2 \sim 0.1$. 
However, including the gaugino contributions reduces the lower bound on
$m_0$, so that we can, in fact, realize the hierarchical spectrum
without tuning as shown in Fig.~\ref{fig_KK_1}.

In Fig.~\ref{fig_KK_1}, we have plotted the lower bounds on $m_0$, the
masses for the various light squarks and sleptons at the GUT scale,
including the gaugino contributions. The horizontal axis represents
the running gluino mass evaluated at the gluino decoupling scale, which
is twice as large as the GUT-scale gaugino mass $m_{1/2}$ in the
previous section. The boundary
conditions at the GUT scale, such as $\mex=2.0\times m_D$, are the same
as in the previous section. 
The only difference from the previous section is that $m_D$ is not
fixed to $10~\TEV$ but varies with the constraints from 
$K^0$-$\bar{K}^0$ mixing.

Fig.~\ref{fig_KK_1} shows that there is no constraints from $\Delta
m_K$ in the NESSM. Moreover, the NESSM solves the CP problem in 
$K^0$-$\bar{K}^0$ mixing without tuning of the phases or electroweak
symmetry breaking.  Note that the shape of the excluded region in
Fig.~\ref{fig_KK_1} is very different from that in
Fig.~\ref{fig_ratio_1}.  This is because $m_D$ can take a smaller
value as $m_0$ decreases,
\begin{eqnarray}
  m_D > {\rm const.} \times (m_0)^{2/3},
\end{eqnarray}    
as can be seen from Eq.~(\ref{bound_1}).  This is peculiar to the case
where the flavor violation comes from the supergravity contributions. 
In contrast, in the case where the U(1)$_X$ $D$-term contributes to the
flavor violation such as in Models (II-IV), the lower bound on $m_D$
does not change with $m_0$, so that the shape of the excluded region of
$m_0$ is similar to that of $m_0/m_D$ in the previous section. 

Now, we turn to the other models.
In Model (II), the averaged squark masses are given by
\begin{eqnarray}
  \bar{m}_{LL}^2=\sqrt{2}m_D^2,
  \hspace{5mm}
  \bar{m}_{RR}^2=m_D^2.
\end{eqnarray}
The left-handed down-type squarks of the first and second generations
have different U(1)$_X$ charges, so that the U(1)$_X$
$D$-term contribution induces flavor violation. 
In this case, the mixing angle $\theta_L$ is necessary to determine
$\delta_{LL}$.  Since the product of $V^L$ and the diagonalizing matrix
for the left-handed up-type quark gives the Cabbibo angle 
$\sin\theta_C=0.22$, it is natural to take $\sin\theta_L \sim
\sin\theta_C$.  Thus, we here take $\sin\theta_L =0.22$ and set the
off-diagonal element $\delta_{LL}$ as
\begin{eqnarray}
  \delta_{LL}=
  \frac{\sin\theta_L\cos\theta_L}{\sqrt{2}}\sim 0.15.
\end{eqnarray}
On the other hand, $\delta_{RR}$ is determined from the supergravity
contributions and is more model-dependent than $\delta_{LL}$. 
Therefore, we here restrict our attention to the flavor violation from
the U(1)$_X$ $D$-term contribution and give the lower bounds on the
light sfermion masses using only the constraint on $\delta_{LL}$.  
The supergravity contributions are discussed later.

Then, we obtain the following bound from Eq.~(\ref{KK_LL}):
\begin{eqnarray}
  m_0 &>&  (3.2\sim 4.4)~\left(\frac{m_0}{m_D}\right)~\TEV.
\end{eqnarray}
Since the bounds on the ratio $m_0/m_D$ are given in
Fig.~\ref{fig_ratio_2} in the previous section, we can estimate the
lower bounds on $m_0$ from $\Delta m_K$ as
\begin{eqnarray}
\begin{array}{cc}
  {\rm ESSM} & {\rm NESSM}\\
  m_0/m_D \ge 0.22 & m_0/m_D \ge 0.09 \\
  m_0 > (700 \sim 970)~\GEV & m_0 > (290 \sim 400)~\GEV 
\end{array}
\end{eqnarray}
\begin{figure}[t]
\centerline{\epsfxsize=15cm \epsfbox{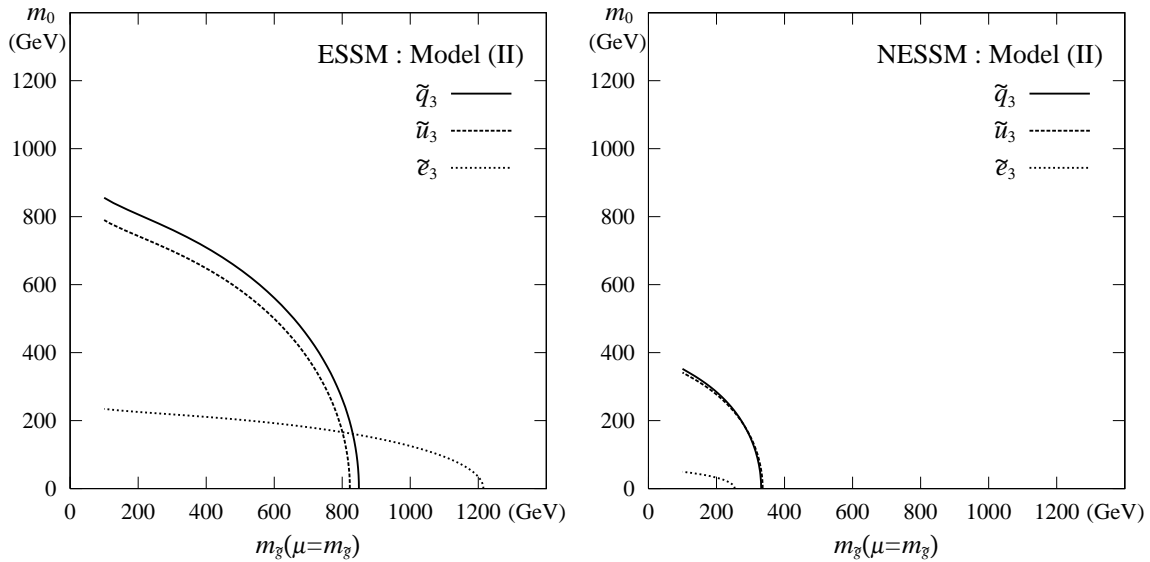}}
\caption{Lower bounds on the bare masses for the light sfermions in
 Model (II). The constraints from $\Delta m_K$ (Eq.~(\ref{KK_LL})) are
 plotted.}
\label{fig_KK_2}
\end{figure}
In Fig.~\ref{fig_KK_2}, we have plotted the lower bounds on $m_0$ in
Model (II) including the effect of the gaugino masses. 
It shows that we can take more natural mass scale for the light
sfermions in the NESSM than in the ESSM.
If $\delta_{LL}$ has a phase of order unity, however, the lower bounds
on the light sfermion masses become higher by a factor of $12$.
Therefore, the phase is necessary to be less than $10^{-2}$ to avoid 
extreme fine tuning of the electroweak symmetry breaking.

Before turning to Model (III), we comment on the 
supergravity contributions.
If we take $\delta_{RR}=m_0^2/m_D^2$ as we did in Model (I), we obtain
the following lower bounds on $m_0$ from Eq.~(\ref{KK_LR}): 
$m_0 > (2.6 \sim 4.3)~\TEV$ in the ESSM and $m_0 > (430 \sim 720)~\GEV$
in the NESSM.  Thus, it seems to require a slight tuning of the
electroweak symmetry breaking even in the NESSM. 
However, the size of $\delta_{RR}$ is strongly model-dependent as we
emphasized, and if we take $\delta_{RR}=(1/4)\times m_0^2/m_D^2$, for
example, the bounds on $m_0$ become half.
Moreover, the bounds are greatly lowered by including the effect of the
gaugino masses as in Fig.~\ref{fig_KK_1}. Therefore, the supergravity
contributions are less important to restrict the parameters in Model
(II).

\begin{figure}[t]
\centerline{\epsfxsize=15cm \epsfbox{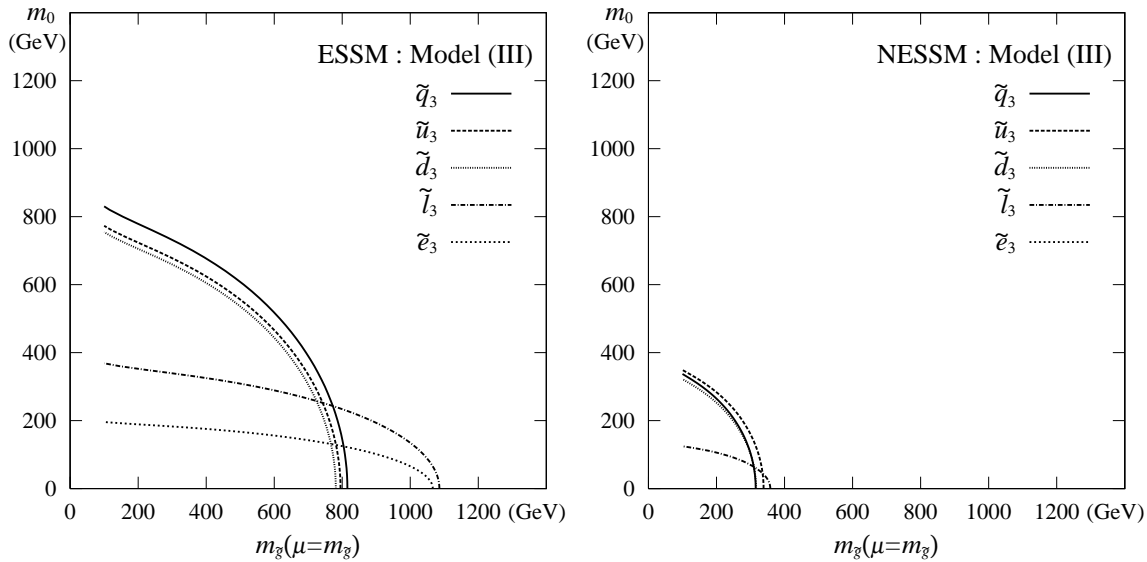}}
\caption{Lower bounds on the bare masses for the light sfermions in
 Model (III). The constraints from $\Delta m_K$ (Eq.~(\ref{KK_LL}))
 are plotted.}
\label{fig_KK_3}
\end{figure}
In Fig.~\ref{fig_KK_3}, we have plotted the lower bounds on $m_0$ in
Model (III).
Here, we have taken $\sin\theta_L=0.22$ as in Model (II).
Since the U(1)$_X$ charge assignment for the first-two generations in
Model (III) is the same as in Models (II),
$K^0$-$\bar{K}^0$ mixing gives the same constraints on both models. 
The lower bounds on the mass ratios $m_0/m_D$ are also almost the same
between Models (II) and (III) as shown in the previous section.
Therefore, the bounds on $m_0$ in Fig.~\ref{fig_KK_3} are similar
to those in Fig.~\ref{fig_KK_2}.

In Model (IV), both $\delta_{LL}$ and $\delta_{RR}$ come from the large
U(1)$_X$ $D$-term contribution, so that the constraint is the severest
among Models (I-IV).  The averaged squark masses are given by
\begin{eqnarray}
  \bar{m}_{LL}^2=\sqrt{2}m_D^2,
  \hspace{5mm}
  \bar{m}_{RR}^2=\sqrt{2}m_D^2. 
\end{eqnarray}
If we set both angles $\theta_L$ and $\theta_R$ equal to the Cabbibo
angle for simplicity, then the $\delta_{LL}$ and $\delta_{RR}$ are given 
by
\begin{eqnarray}
  \delta_{LL}=
  \frac{\sin\theta_L\cos\theta_L}{\sqrt{2}}\sim 0.15,
  \hspace{5mm}
  \delta_{RR}=
  \frac{\sin\theta_R\cos\theta_R}{\sqrt{2}}\sim 0.15.
\end{eqnarray}
Substituting these values into Eq.~(\ref{KK_LR}), we obtain the
following bound:
\begin{eqnarray}
  m_0 > (19 \sim 32)~\left(\frac{m_0}{m_D}\right)~\TEV.
\end{eqnarray}
The bounds on the ratio $m_0/m_D$ are given in Fig.~\ref{fig_ratio_4},
so that the lower bounds on $m_0$ can be estimated as
\begin{eqnarray}
\begin{array}{cc}
  {\rm ESSM} & {\rm NESSM} \\
  m_0/m_D \ge 0.23 & m_0/m_D \ge 0.10\\
  m_0 > (4.4 \sim 7.4)~\TEV & m_0 > (1.9 \sim 3.2)~\TEV
\end{array}
\end{eqnarray}
\begin{figure}[t]
\centerline{\epsfxsize=15cm \epsfbox{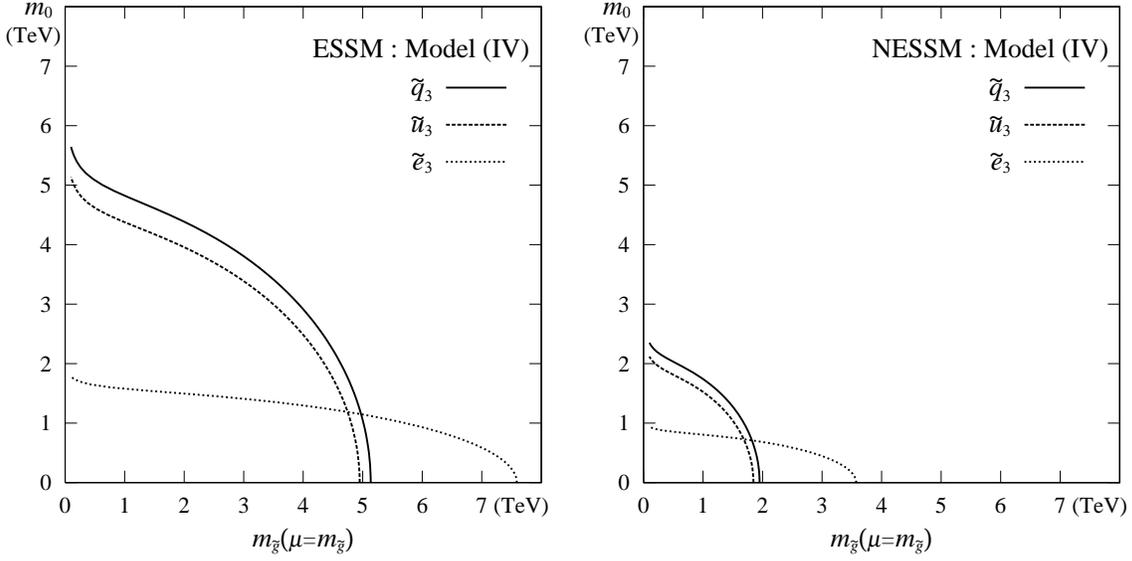}}
\caption{Lower bounds on the bare masses for the light sfermions in
 Model (IV). The constraints from $\Delta m_K$ (Eq.~(\ref{KK_LR})) are
 plotted.}
\label{fig_KK_4}
\end{figure}
In Fig~\ref{fig_KK_4}, we have plotted the lower bounds on $m_0$ in
Model (IV) including the effect of the gaugino mass.
It shows that an extreme fine tuning is required in Model (IV). 
We can reduce the degree of fine tuning by taking smaller values
for the mixing angles $\sin\theta_L$ and $\sin\theta_R$.
However, in order to reduce the bounds on the masses for both 
the light sfermions and the gluino to $500~\GEV$, we must take
$\sin\theta_R$ less than $0.03$ and it causes another tuning problem.
In consequence, it seems difficult to realize Model (IV) without
fine tuning in a framework of the NESSM, 
even if there is no CP phase.

\section{Other Renormalization-Group Effects}

In this section, we discuss other RG effects on the light sfermion
masses which we have not considered in the previous sections. 
There are three types of contributions which potentially affect the
previous analyses: the three-loop contribution through the gauge
couplings, the contribution from the bottom and tau Yukawa couplings,
and the contribution from the top Yukawa coupling.

First, we consider the correction to the light sfermions from the RG
equations at the three-loop level. 
We here adopt the $\overline{\rm SDR}$ scheme defined by ``the analytic
continuation into superspace'' \cite{Anal_Cont}.  The RG equations for
the SUSY-breaking scalar mass squareds in this scheme coincide with
those in the $\overline{\rm DR}^{\prime}$ scheme at the two-loop level,
so that $\overline{\rm SDR}$ scheme is considered to be all-order
definition of the $\overline{\rm DR}^{\prime}$ scheme.

When the SUSY-breaking scalar mass squareds are regarded as a $D$-term
of an external U(1)$_X$ gauge multiplet, the RG contributions from the
gauge interactions can be divided into two classes in the limit of
vanishing gaugino mass in this scheme.  One is the contribution from
mixed anomalies between the external U(1)$_X$ and the internal gauge
symmetries.  The other is the contribution from the kinetic-term mixing
between U(1) symmetries, which exists only if there is an internal U(1)
gauge symmetry. 
Since we have imposed Eqs.~(\ref{mass_af}, \ref{mass_u1d1},
\ref{mass_u1d2}) in the NESSM, the RG contributions to the light
sfermion masses from the heavy sfermions and extra matters only come
from the U(1)$_Y$ and U(1)$_X$ mixing contribution at the three-loop
level \cite{HKN, AR}.
Then, the contributions are suppressed by small $\alpha_Y$ and
negligible compared with the two-loop finite correction.  
Furthermore, in models where the extra matters are the fundamental
representation of SU(3)$_C$ and SU(2)$_L$, the contributions from the
heavy sfermions and extra matters in the three-loop RG equations are
proportional to $\alpha_3 \alpha_2 \alpha_Y$ or $\alpha^3_Y$, since the
terms proportional to $\alpha_3^2 \alpha_Y$ and $\alpha_2^2 \alpha_Y$
vanish due to Eqs.~(\ref{mass_u1d1}, \ref{mass_u1d2}) in this case.
Thus, the three-loop RG contributions are more strongly suppressed in
such models.  

Next, we consider the effect of the bottom and tau Yukawa couplings. In
Models (II, IV) the right-handed bottom squark and the left-handed
slepton of the third generation obtain masses from the U(1)$_X$ $D$-term 
due to $Q^{X}_{{\bf 5^\star}_3}\ne 0$.  Then, the bottom and tau Yukawa
couplings may drive the mass squareds for the doublet squark of the
third generation and the right-handed stau to be negative at the
one-loop level. The RG contributions to the doublet squark of the third
generation and the right-handed stau, $m_{\tilde{q}_3,\,{\rm yukawa}}^2$
and $m_{\tilde{e}_3,\,{\rm yukawa}}^2$, are given as
\begin{eqnarray}
  m_{\tilde{q}_3,\,{\rm yukawa}}^2 &\simeq& 
    - 0.003\% \times (1+\tan^2\beta) (m_D^2+|A_b|^2), \\
  m_{\tilde{e}_3,\,{\rm yukawa}}^2 &\simeq& 
    - 0.005\% \times (1+\tan^2\beta) (m_D^2+|A_\tau|^2).
\end{eqnarray}
Here, we have assumed that $\tan\beta$ is not so large, and $A_b$ and
$A_\tau$ are the soft SUSY-breaking trilinear scalar couplings
associated with the bottom and tau Yukawa couplings.
Then, the RG contribution to the doublet quark of the third generation
is negligible if $\tan\beta < 10$, compared with the finite correction.
On the other hand, the RG contribution to the right-handed stau can be
larger than the finite correction.
However, the bound on the bare mass given by these contributions is
still looser than the bounds on the light squarks from the finite
correction, as long as $\tan\beta$ is sufficiently small.
Since Models (II, IV) predict the small bottom and tau Yukawa coupling
constants due to $Q^{X}_{{\bf 5^\star}_3} \ne 0$ and thus require small
$\tan\beta$, these contributions are sufficiently suppressed.

Finally, we discuss the effect of the top Yukawa coupling. In the
previous sections, we have neglected it, since the effect of the top
Yukawa coupling depends on the supergravity contributions, which are
model-dependent.
There are two negative contributions to the top-squark mass squared
through the top Yukawa coupling.  One is the RG contribution which
mainly comes from the up-type Higgs mass $m_{h_u}$.  The other is the
contribution from the left-right mixing term, which gives the negative
contribution in diagonalizing the top-squark mass matrix.
The left-right mixing term for top squark is given by
\begin{eqnarray}
  (M_{\tilde{t}}^2)_{LR}=m_t (A_t + \mu\cot\beta),
\end{eqnarray}
where $m_t$ is the top quark mass, $A_t$ the soft SUSY-breaking 
trilinear coupling associated with the top Yukawa coupling and 
$\mu$ the supersymmetric mass for the Higgs doublets.
Thus, it turns out that $\mu$ and $A_t$ play an important role in
calculating the negative contributions. However, $A_t$ at the weak
scale is almost saturated by the radiative correction from the gluino,
so that we take $A_t=0$ at the GUT scale for simplicity in the
following calculations. The size of $\mu$ is determined by
the electroweak symmetry breaking.

\begin{figure}[t]
\centerline{\epsfxsize=15cm \epsfbox{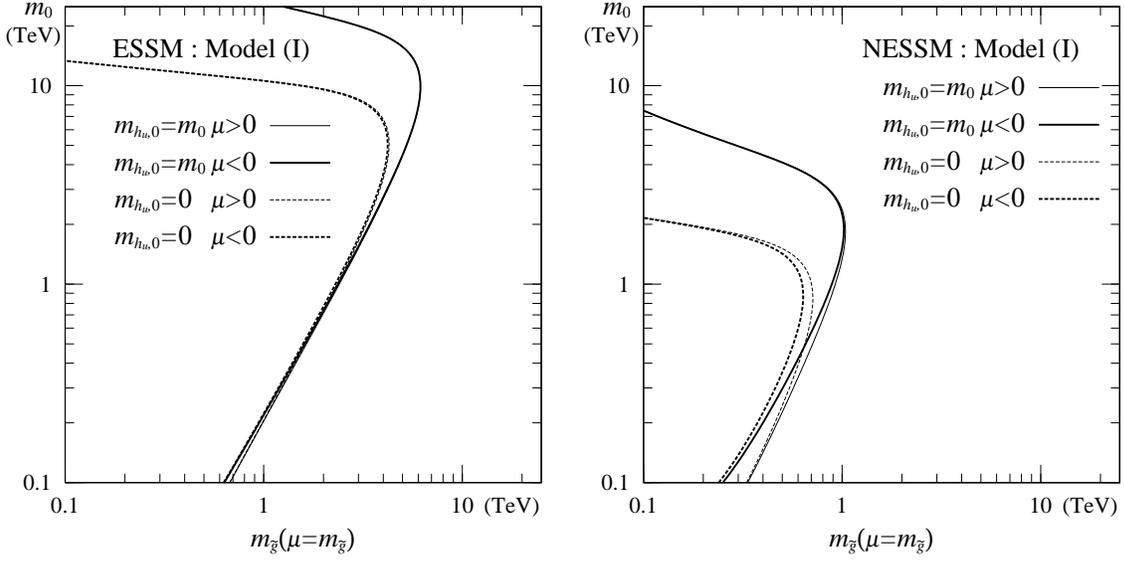}}
\caption{Lower bounds on the bare masses for the light sfermions in
 Model (I) are plotted by combining the constraint from $\epsilon_K$
 (Eq.~(\ref{KK_LRep})) with the condition that $m_{\tilde{t}_1}, 
 m_{\tilde{t}_2}>0$, where $\tilde{t}_1$ and $\tilde{t}_2$ represent the 
 mass eigenstates for top squarks.
 The regions below the curves are excluded due to 
 the negative mass-squared eigenvalues for top squarks.
 Here, we have set $\tan\beta=3.0$.}
\label{fig_stop_1}
\end{figure}
\begin{figure}[t]
\centerline{\epsfxsize=15cm \epsfbox{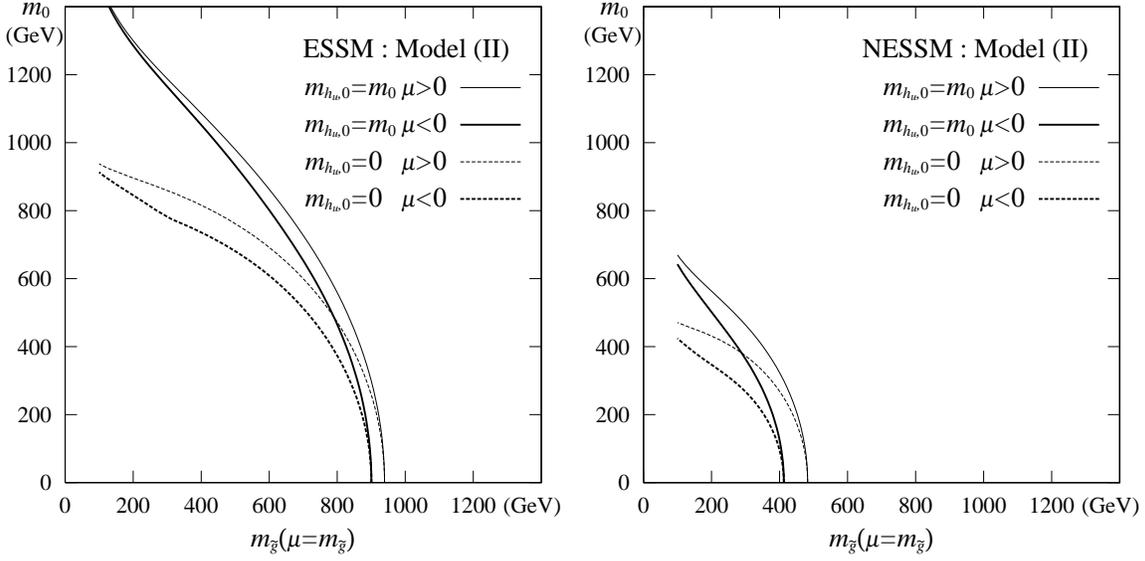}}
\caption{Lower bounds on the bare masses for the light sfermions in
 Model (II) are plotted by combining the constraint
 from $\Delta m_K$ (Eq.~(\ref{KK_LL})) with
 the condition that $m_{\tilde{t}_1}^2, m_{\tilde{t}_2}^2 >0$.
 The regions below the curves are excluded due to 
 the negative mass-squared eigenvalues for top squarks. 
 Here, we have set $\tan\beta=3.0$.}
\label{fig_stop_2}
\end{figure}
In Figs.~\ref{fig_stop_1} and \ref{fig_stop_2}, we have plotted the
lower bounds on the bare masses for top squarks.  Here, we have assumed
the universal scalar mass at the GUT scale except for the up-type Higgs
mass, and taken $A_t=0$ at the GUT scale for simplicity.
In the figures, we have shown two extreme cases for the up-type Higgs
mass in each choice for the sign of the $\mu$ parameter: one is the
universal case $m_{h_u}=m_0$ and the other is $m_{h_u}=0$.
If $\mu$ is positive, the $A_t$ and $\mu$ are added up constructively
in the left-right mixing term, since $A_t$ receive the positive
contribution from the gaugino mass.
Thus, the lower bound on the top squark mass is somewhat severe in this
case.  On the other hand, if $\mu$ is negative, the cancellation between 
two contributions from $A_t$ and $\mu$ occurs, so that the bound is
weaker than the case with $\mu > 0$.

In the region where $m_0$ is much larger than the gaugino masses,
the up-type Higgs mass $m_{h_u}$ gives the dominant effect.
Thus, the effect of the top Yukawa coupling is much model-dependent in
this region, since we cannot predict the up-type Higgs mass at the GUT
scale.
For instance, if the up-type Higgs mass is much smaller than the
top squark mass at the GUT scale, the effect is negligible as shown in
Figs.~\ref{fig_stop_1} and \ref{fig_stop_2}.
On the other hand, in the region where $m_0$ is small enough, the
dominant effect of the top Yukawa coupling comes from the left-right
mixing term through diagonalization of the top-squark mass matrix.
Since $A_t$ term is generated radiatively by the gluino as 
$A_t \sim m_{\tilde{g}}$, it gives the negative contribution to the
top-squark mass eigenvalue of order $-m_t A_t\sim -m_t m_{\tilde{g}}$.
Thus, in order to obtain the positive top-squark mass squared at the
weak scale, we have to take a larger mass for the gluino than the case
without top-Yukawa contributions.
The required increase $\delta m_{\tilde{g}}$ of the gluino mass is
determined by the following inequality: 
$2 m_{\tilde{g}} \delta m_{\tilde{g}} \gsim m_t m_{\tilde{g}}$, since
the positive contribution from the gluino is estimated as of order 
$m_{\tilde{g}}^2$.
That is, we have to take a larger mass for the gluino by the order of
$100~\GEV$ compared with the previous case as shown in
Figs.~\ref{fig_stop_1} and \ref{fig_stop_2}.

To summarize, the lower bounds on the light sfermion masses become
somewhat severer by including the effects of the top Yukawa coupling.
However, we have found that we can still take the light sfermion and
gluino masses as small as $\sim 400~\GEV$ in the natural effective SUSY, 
while we have to take them larger than $\sim 1~\TEV$ in the original
effective SUSY.

\section{Conclusions and Discussion}

In this article, we have calculated the finite corrections to the light
sfermion masses from the heavy sfermions and extra matters at the
two-loop level in the NESSM, assuming that all fields are embedded in
the SU(5)$_{\rm GUT}$ multiplets at the GUT scale.  
While the sfermion mass squareds in the NESSM are RG invariant
at the two-loop level in the limit of vanishing gaugino-mass and
Yukawa-coupling contributions, the finite corrections may still drive
the light-sfermion mass squareds to be negative.  The finite
corrections increase (decrease) if the supersymmetric masses (the
holomorphic SUSY-breaking masses) of the extra matters become large. 
We have found that the corrections can be less than $(0.3-1)\%$ of the
heavy-squark mass squareds.

We have also discussed whether the models in which the NESSM is realized
by the U(1)$_X$ gauge interaction at the GUT scale are viable or not in
the light of the experimental constraints from $\Delta m_K$ and
$\epsilon_K$.  On this setup, the supergravity contributions to the
SUSY-breaking terms, which are generically non-universal in flavor
space, are suppressed compared with the U(1)$_X$ $D$-term contribution.
We have found that, when both the left- and right-handed down-type
squarks of the first-two generations have common U(1)$_X$ charges, the
supersymmetric contributions to $\Delta m_K$ and $\epsilon_K$ are
sufficiently suppressed without breaking naturalness, even if the
flavor-violating supergravity contributions to the sfermion mass
matrices are included.  On the other hand, when only the right-handed
squarks of the first and second generations have a common U(1)$_X$
charge, we can still satisfy the constraint from $\Delta m_K$ naturally,
but evading the bound from $\epsilon_K$ requires somewhat small CP phase
of order $10^{-2}$.

The formulae we have given in Appendix \ref{Ap_finite} are applicable to
any models which have hierarchical mass spectrum for the sfermions such
as effective SUSY.
In particular, the formulae for the finite corrections can also be
applied to the models where the U(1)$_X$ $D$-term is generated at much
lower energy than the GUT scale \cite{U(1)'}, since they do not have
explicit renormalization-point dependence.
Then, it gives the lower bounds on the light-sfermion bare masses
similar to the ones we have derived in this article.
Thus, we believe that our result gives the most conservative bound for
the naturalness in the effective SUSY models.

\section*{Acknowledgments}

We would like to thank Hitoshi Murayama and Youichi Yamada for useful
discussions.  K.K. and Y.N. thank the Japan Society for the Promotion of 
Science for financial support.  This work was supported in part by the
Grant-in-Aid for Scientific Research from the Ministry of Education,
Science, Sports and Culture of Japan, on Priority Area 707
``Supersymmetry and Unified Theory of Elementary Particles" and by the
Grant-in-Aid No.10740133 (J.H.).

\newpage
\appendix

\section{Anomalous U(1) Model}
\label{Ap_model}

In this appendix, we briefly review the model given in Ref.~\cite{HKN}
and show that it naturally realizes the setup of the NESSM. The model
is based on the anomalous U(1) SUSY-breaking model \cite{anom-U1_1}.

The anomalous U(1) gauge symmetry frequently appears in low-energy
effective theories of string theory.
The matter content is anomalous under this U(1) symmetry, but its
anomalies are canceled by a nonlinear transformation of the dilaton
chiral multiplet \cite{GS}.
This leads to the generation of a nonzero Fayet-Iliopoulos (FI) $D$-term
$\xi$ which is smaller than the reduced Planck scale $M_{\rm pl}$ by a
one-loop factor, $\xi^2 \simeq (g^2/192\pi^2){\rm Tr}Q M_{\rm pl}^2$.
We parameterize it as $\xi \equiv \epsilon M_{\rm pl}$ with
$\epsilon=O(0.1)$, and take the sign convention of U(1) charges such
that $\xi^2>0$.

We identify this anomalous U(1) gauge symmetry with U(1)$_X$ given in
the text.
Thus, we assign the U(1)$_X$ charges $Q^{X}_{{\bf 10}_i}$ and 
$Q^{X}_{{\bf 5^\star}_i}$ to the quark and lepton chiral multiplets
$\Psi_{{\bf 10}_i}$ and $\Psi_{{\bf 5^\star}_i}$, respectively
($Q^{X}_{{\bf 10}_i}, Q^{X}_{{\bf 5^\star}_i} \geq 0$).
We also introduce extra-matter chiral multiplets 
$\Psi_{{\bf 5}_{\rm ex}}$ and $\Psi_{{\bf 5^\star}_{\rm ex}}$ with
U(1)$_X$ charges $Q^{X}_{{\bf 5}_{\rm ex}}$ and 
$Q^{X}_{{\bf 5^\star}_{\rm ex}}$ ($Q^{X}_{{\bf 5}_{\rm ex}}, 
Q^{X}_{{\bf 5^\star}_{\rm ex}} < 0$), in order to satisfy 
Eqs.~(\ref{U1X_af}, \ref{U1X_u1d2}).
Then, the SUSY-breaking model is constructed as follows.

We consider the SU($N_c$) SUSY gauge theory with $N_f$ flavors 
${\cal Q}^a$ and $\bar{\cal Q}_{\bar{a}}$ ($N_c/2 < N_f < N_c$), and
introduce two singlet chiral superfields $\Phi_1$ and $\Phi_2$.
Here, $a, \bar{a} = 1,\cdots,N_f$ represent flavor indices.
We assign the U(1)$_X$ charges as ${\cal Q}^a 
(-(Q^{X}_{{\bf 5}_{\rm ex}}+Q^{X}_{{\bf 5^\star}_{\rm ex}})/2)$,  
$\bar{\cal Q}_{\bar{a}} 
(-(Q^{X}_{{\bf 5}_{\rm ex}}+Q^{X}_{{\bf 5^\star}_{\rm ex}})/2)$, 
$\Phi_1(-1)$ and 
$\Phi_2(1+Q^{X}_{{\bf 5}_{\rm ex}}+Q^{X}_{{\bf 5^\star}_{\rm ex}})$.
The tree-level superpotential of the model is given by
\begin{eqnarray}
  W_{\rm tree} = \frac{{\cal Q}^a \bar{\cal Q}_{\bar{a}}}{M_{\rm pl}} 
    \left( (f_{\phi}){}^{\bar{a}}_a \, \Phi_1 \Phi_2 
    + (f_{\psi}){}^{\bar{a}}_a \, \Psi_{{\bf 5}_{\rm ex}} 
    \Psi_{{\bf 5^\star}_{\rm ex}} \right).
\end{eqnarray}
Then, the dynamical superpotential is generated by nonperturbative
effects of the SU($N_c$) gauge interaction, and the effective
superpotential of the model is exactly given by
\begin{eqnarray}
  W_{\rm eff} &=& \frac{1}{M_{\rm pl}} 
    \Bigl( \Phi_1 \Phi_2 {\rm Tr}(f_{\phi} M) 
    + \Psi_{{\bf 5}_{\rm ex}} \Psi_{{\bf 5^\star}_{\rm ex}}
    {\rm Tr}(f_{\psi} M) \Bigr) \nonumber\\
  && + (N_c-N_f)\left(\frac{\Lambda^{3N_c-N_f}}{\det M} 
    \right)^{\frac{1}{N_c-N_f}},
\end{eqnarray}
in terms of gauge-invariant composite fields 
$M^a_{\bar{a}} = {\cal Q}^a \bar{\cal Q}_{\bar{a}}$ \cite{ADS}.
Here, $\Lambda$ is the dynamical scale of the SU($N_c$) gauge theory.
The $D$-term potential for U(1)$_X$ is given by 
\begin{eqnarray}
  V_D &=& \frac{g_X^2}{2} 
    \Bigl( \xi^2 - 
    (Q^{X}_{{\bf 5}_{\rm ex}}+Q^{X}_{{\bf 5^\star}_{\rm ex}})
    {\rm Tr}\sqrt{M^{\dagger}M} \nonumber\\
  && - |\Phi_1|^2 
    + (1+Q^{X}_{{\bf 5}_{\rm ex}}+Q^{X}_{{\bf 5^\star}_{\rm ex}})|\Phi_2|^2 
    + Q^{X}_{{\bf 5}_{\rm ex}}|\Psi_{{\bf 5}_{\rm ex}}|^2
    + Q^{X}_{{\bf 5^\star}_{\rm ex}}|\Psi_{{\bf 5^\star}_{\rm ex}}|^2
    \Bigr)^2.
\end{eqnarray}
In this appendix, we use the same letter for the chiral superfield and
its scalar component.

The dynamics of the model can be understood as follows.  First,
nonperturbatively generated superpotential forces $M$ to have VEV's,
which gives supersymmetric mass terms for the singlet fields and the
extra matter fields.  The large FI $D$-term, $\xi$, is
absorbed by the shift of the singlet fields $\Phi_1$ and $\Phi_2$.
The point is that if $f_{\psi}$ is larger than $f_{\phi}$, only the
singlet fields shift to absorb the FI $D$-term and the
extra-matter fields do not develop VEV's, avoiding phenomenologically
disastrous large breaking of the SM gauge groups.  We assume that this
condition, $f_{\psi} > f_{\phi}$, is satisfied.  Then, since the
singlet fields have the supersymmetric mass induced by the VEV's of
$M$, they cannot absorb $\xi$ completely and nonzero U(1)$_X$ $D$-term
remains.  This nonvanishing $D$-term gives the MSSM sfermions and extra
scalars the SUSY-breaking mass squareds proportional to their U(1)$_X$
charges.

Now, let us minimize the potential explicitly.  The minimum of the
potential can be obtained by making an expansion in the small
parameter $(\Lambda/M_{\rm pl})^{(3N_c-N_f)/N_c} \ll 1$.  To the
leading order, the VEV's of the fields $\Phi_1$, $\Phi_2$ and $M$ at
the minimum are given by solving the following equations:
\begin{eqnarray}
  && |\Phi_1|^2 - (1+Q^{X}_{{\bf 5}_{\rm ex}}+Q^{X}_{{\bf 5^\star}_{\rm ex}}) 
    |\Phi_2|^2 = \xi^2, 
\label{imp_phi1} \\
  && |\Phi_1|^2 \left( (N_f-N_c)|\Phi_1|^2 + N_f|\Phi_2|^2 \right) 
  \nonumber\\
  && \qquad\qquad\qquad = 
    -(1+Q^{X}_{{\bf 5}_{\rm ex}}+Q^{X}_{{\bf 5^\star}_{\rm ex}})
    |\Phi_2|^2 \left( (N_f-N_c)|\Phi_2|^2 + N_f|\Phi_1|^2 \right), 
\label{imp_phi2} \\
  && (M^{-1}){}^{\bar{a}}_a 
    \left(\frac{\Lambda^{3N_c-N_f}}{\det M}\right)^{\frac{1}{N_c-N_f}} = 
    \frac{(f_{\phi}){}^{\bar{a}}_a}{M_{\rm pl}} \Phi_1 \Phi_2.
\label{imp_M}
\end{eqnarray}
From Eqs.~(\ref{imp_phi1}, \ref{imp_phi2}), we find that both the 
$\Phi_1$ and $\Phi_2$ fields have nonvanishing VEV's of order $\xi$,
and the VEV of the $M$ field is given by Eq.~(\ref{imp_M}) as
\begin{eqnarray}
  \vev{\Phi_1} \simeq \vev{\Phi_2} \simeq \xi, \qquad
  \vev{M} \simeq \Lambda^{\frac{3N_c-N_f}{N_c}} 
    \left( \frac{M_{\rm pl}}{\xi^2} \right)^{\frac{N_c-N_f}{N_c}}.
\label{imp_VEVap}
\end{eqnarray}
Then, the Yukawa matrices for the quarks and leptons are generated
through the VEV of the $\Phi_1$ field suppressed by suitable powers of
$\vev{\Phi_1}/M_{\rm pl} \simeq \xi/M_{\rm pl} = \epsilon$ (see
Eq.~(\ref{hierarchy})).

Calculating higher order in $(\Lambda/M_{\rm pl})^{(3N_c-N_f)/N_c}$, we
find that the VEV's of the auxiliary fields can be written in terms of
the VEV's of the scalar fields as
\begin{eqnarray}
  -(\bar{F}_{M^{\dagger}}){}^{\bar{a}}_a &=& 
    - \frac{N_c-N_f}{N_c N_f} \frac{1}{M_{\rm pl}}
    \frac{|\Phi_1|^2+|\Phi_2|^2}{\Phi_1^* \Phi_2^*}
    |{\rm Tr}(f_{\phi} M)|^2 (f_{\phi}^{* -1}){}^{\bar{a}}_a,
\label{imp_fVEV_M} \\
  -\bar{F}_{\Phi_1^{\dagger}} &=& 
    \left( \frac{\Phi_2}{M_{\rm pl}} \right) {\rm Tr}(f_{\phi} M),
\label{imp_fVEV_phi1} \\
  -\bar{F}_{\Phi_2^{\dagger}} &=& 
    \left( \frac{\Phi_1}{M_{\rm pl}} \right) {\rm Tr}(f_{\phi} M),
\label{imp_fVEV_phi2} \\
  -D_X &=& 
    \frac{2N_f-N_c}{N_c} \frac{1}{M_{\rm pl}^2}
    \frac{|\Phi_1|^2+|\Phi_2|^2}{|\Phi_1|^2-(1+Q^{X}_{{\bf 5}_{\rm ex}}+
    Q^{X}_{{\bf 5^\star}_{\rm ex}})|\Phi_2|^2}
    |{\rm Tr}(f_{\phi} M)|^2.
\label{imp_DVEV}
\end{eqnarray}
Substituting Eq.~(\ref{imp_VEVap}) into
Eqs.~(\ref{imp_fVEV_M} - \ref{imp_DVEV}), we obtain 
\begin{eqnarray}
  && -\bar{F}_{M^{\dagger}} \simeq
    \left( \frac{\Lambda^{2(3N_c-N_f)}}
    {\xi^{4(N_c-N_f)}M_{\rm pl}^{2N_f-N_c}} 
    \right)^{\frac{1}{N_c}},
\label{imp_fVEV_M_Ap}\\
  && -\bar{F}_{\Phi_1^{\dagger}} \simeq -\bar{F}_{\Phi_2^{\dagger}} \simeq
    \left( \frac{\Lambda^{3N_c-N_f}}
    {\xi^{2(N_c-N_f)}M_{\rm pl}^{N_f}} 
    \right)^{\frac{1}{N_c}} \xi,
\label{imp_fVEV_phi_Ap}\\
  && -D_X \simeq
    \left( \frac{\Lambda^{3N_c-N_f}}
    {\xi^{2(N_c-N_f)}M_{\rm pl}^{N_f}} 
    \right)^{\frac{2}{N_c}}.
\label{imp_DVEV_Ap}
\end{eqnarray}
The dynamical scale $\Lambda$ is determined so that $\sqrt{|-D_X|}
\simeq (1 \sim 10)~\TEV$ to give the heavy sfermions multi-TeV masses.
From Eqs.~(\ref{imp_fVEV_M_Ap}, \ref{imp_fVEV_phi_Ap},
\ref{imp_DVEV_Ap}), we find the following useful relation:
\begin{eqnarray}
  \left| -D_X \right| \simeq 
  \left| \frac{-F_{\Phi_1}}{\vev{\Phi_1}} \right|^2 \simeq
  \left| \frac{-F_{\Phi_2}}{\vev{\Phi_2}} \right|^2 \simeq
  \left| \frac{-F_M}{\vev{M}} \right|^2.
\end{eqnarray}
This relation ensures that flavor-breaking supergravity contributions
are an order of magnitude smaller than the contribution from
the U(1)$_X$ $D$-term, since their sizes are estimated as\footnote{
Since the gravitino mass is of the order of the weak scale, we can
induce $\mu$-term (supersymmetric mass term for the Higgs field, 
$W = \mu H_u H_d$) of the desired size (the weak scale) by introducing a 
holomorphic term $K = H_u H_d$ in the K\"ahler potential \cite{GM}.}
\begin{eqnarray}
  \left| \frac{-F_{\Phi_1}}{M_{\rm pl}} \right|
  \simeq \frac{\xi}{M_{\rm pl}}\left|\frac{-F_{\Phi_1}}{\vev{\Phi_1}}\right|
  \simeq \epsilon \sqrt{|-D_X|}
  \simeq (10^2 \sim 10^3)~\GEV.
\end{eqnarray}
From Eqs.~(\ref{imp_VEVap}, \ref{imp_DVEV_Ap}), we also find that
$\sqrt{|-D_X|} \simeq \vev{M}/M_{\rm pl}$.
Thus, the supersymmetric mass for the extra matters is actually the same
order with the SUSY-breaking masses for the heavy sfermions.

The gaugino masses arise from the $F$-term of the dilaton field
\cite{ADM}.
Their sizes can be of the order of the weak scale \cite{BCCM}, and then
it does not much affect the preceding analysis of the dynamics.
The SUSY-breaking trilinear scalar couplings of order $\sqrt{|-D_X|}$
are also generated by the superpotential which generates Yukawa matrices
for the quarks and leptons, except for the ones which involve only the
light sfermions (see Eqs.~(\ref{A_u}, \ref{A_d})).

Finally, we comment on the anomaly.
We have identified the U(1)$_X$ in the text with the anomalous U(1)
gauge symmetry.
Then, Eq.~(\ref{U1X_af}) might seem contradicted by the fact that the
anomalous U(1) gauge symmetry has mixed anomalies for all the other
gauge groups including the SM ones.
However, it is not a contradiction.
Since U(1)$_X$ is broken down at very high energy scale of order
$\xi$, it does not necessarily mean that the matter content is anomalous
below $\xi$ scale. 
That is, if we introduce fields $\Psi_{{\bf 5}_{\rm anom}}$ and
$\Psi_{{\bf 5^\star}_{\rm anom}}$ of masses of order $\xi$ with the
superpotential 
\begin{equation}
  W \sim \vev{\Phi_1} \Psi_{{\bf 5}_{\rm anom}}
    \Psi_{{\bf 5^\star}_{\rm anom}},
\end{equation}
we can match the anomalies as required by the anomalous U(1) symmetry,
keeping Eqs.~(\ref{U1X_af}, \ref{U1X_u1d2}) satisfied between two scales
$\sqrt{|-D_X|}$ and $\xi$.
Then, the large two-loop RG contributions are absent below the $\xi$
scale as we have explained above.\footnote{
The light-sfermion mass squareds may receive negative RG and finite
contributions above and at the decoupling scale of 
$\Psi_{{\bf 5}_{\rm anom}}$ and $\Psi_{{\bf 5^\star}_{\rm anom}}$.
However, we can make these negative contributions smaller than the
supergravity contributions by choosing their group-theoretical factors
to be small, since there is no large log-factor in the contributions.
We include these contributions in the supergravity contributions when we 
make phenomenological analyses in the text.}

\section{Finite Corrections}
\label{Ap_finite}

In this appendix, we calculate the contributions to the gaugino and the
light sfermion masses arising from loops of the extra-matter multiplets
and the heavy sfermions.
We use the $\overline{\rm DR}^{\prime}$ scheme \cite{DR_Prime} to 
regularize the theory, since we have adopted the $\overline{\rm SDR}$
scheme \cite{Anal_Cont} which is all-order definition of the 
$\overline{\rm DR}^{\prime}$ scheme in the text (see Section 5).
In this scheme, the $\epsilon$-scalar mass $m_{\tilde{A}}$ is zero at
the tree level and it does not appear in the relation between physical
quantities.
However, if the supertrace of the matter fields is nonzero,
$m_{\tilde{A}}$ receives divergent radiative correction at one loop
through loops of the sfermions, so that the counterterm is needed to
cancel this divergence.
The insertion of this counterterm gives divergent contribution to the
sfermion masses at one loop.
Thus, we have to carefully treat the $\epsilon$-scalar in order to
obtain the two-loop contribution to the light sfermion masses when the
supertrace is nonvanishing.

First, we explain our notations.
The superpotential of the vector-like extra matters, $\Psi_{\rm ex}$ and
$\bar{\Psi}_{\rm ex}$, is given by
\begin{eqnarray}
  W_{\rm ex} = \mex\, \Psi_{\rm ex} \bar{\Psi}_{\rm ex}.
\end{eqnarray}
We denote the scalar and fermion components of $\Psi_{\rm ex}$ 
($\bar{\Psi}_{\rm ex}$) as $\tilde{\psi}_{\rm ex}$ and $\psi_{\rm ex}$
($\tilde{\bar{\psi}}_{\rm ex}$ and $\bar{\psi}_{\rm ex}$),
respectively.
In addition, the extra scalars have the SUSY-breaking mass terms
\begin{eqnarray}
  {\cal L}_{\rm ex,\,soft} = -(F_\psi\, \tilde{\psi}_{\rm ex} 
    \tilde{\bar{\psi}}_{\rm ex} + {\rm h.c.})
    - \tilde{m}^2 \tilde{\psi}_{\rm ex}^* \tilde{\psi}_{\rm ex}
    - \tilde{\bar{m}}^2 \tilde{\bar{\psi}}_{\rm ex}^* 
      \tilde{\bar{\psi}}_{\rm ex}.
\end{eqnarray}
Then, the mass terms for the extra scalars are written as
\begin{eqnarray}
&& {\cal L}_{\rm ex} = 
    - (\tilde{\psi}_{\rm ex}^*, \tilde{\bar{\psi}}_{\rm ex})
    \tilde{M}_{\rm ex}^2
    \left(
      \begin{array}{c}
        \tilde{\psi}_{\rm ex} \\ \tilde{\bar{\psi}}_{\rm ex}^*
      \end{array}
    \right); \\
&& \tilde{M}_{\rm ex}^2 = \left(
    \begin{array}{cc}
      |\mex|^2 + \tilde{m}^2 &        F_\psi^*               \\
              F_\psi         & |\mex|^2 + \tilde{\bar{m}}^2 
    \end{array}
    \right).
\end{eqnarray}
The mass matrix $\tilde{M}_{\rm ex}^2$ can be diagonalized by the
unitary matrix $V$ as
\begin{eqnarray}
  {\rm diag} \left( m_1^2, m_2^2 \right)=
  V \tilde{M}_{\rm ex}^2 V^{\dagger}.
\end{eqnarray}
We parameterize $V$ as
\begin{eqnarray}
  V = \left(
  \begin{array}{cc}
    \cos\theta  &  -{\rm e}^{i\alpha} \sin\theta  \\
    {\rm e}^{-i\alpha} \sin\theta  &  \cos\theta
  \end{array}
  \right),
\end{eqnarray}
and define
\begin{eqnarray}
  y_1 \equiv \frac{m_1^2}{\mex^2}, \qquad
  y_2 \equiv \frac{m_2^2}{\mex^2}.
\end{eqnarray}
We calculate the gaugino and the light sfermion masses arising from 
loops of the extra-matter multiplets in terms of these parameters.
Then, the contribution from the heavy sfermions are obtained by taking
the limit $\mex \rightarrow 0$ and $\theta \rightarrow 0$.

\subsection{The gaugino masses}

\begin{figure}
\begin{center} 
\begin{picture}(100,80)(150,130)
  \Line(100,150)(150,150) \Text(125,160)[b]{$\tilde{g}$}
  \Photon(100,150)(150,150){3}{5} \Line(150,150)(250,150)
  \Text(200,140)[t]{$\psi_{\rm ex}$} \DashCArc(200,150)(50,0,180){3}
  \Line(300,150)(250,150) \Text(275,160)[b]{$\tilde{g}$}
  \Photon(300,150)(250,150){3}{5} \Vertex(150,150){2}
  \Vertex(250,150){2}
  \Text(200,215)[b]{$\tilde{\psi}_{\rm ex}$}
\end{picture}
\caption{Diagram contributing to the gaugino masses} 
\label{Fig_gaugino}
\end{center}
\end{figure}
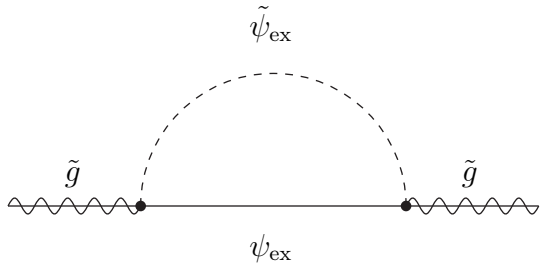

The gauginos acquire their masses through the one-loop diagram of the
extra-matter multiplet shown in Fig.~\ref{Fig_gaugino}.
If there is a pair of vector-like extra matters, it is given by 
\begin{eqnarray}
  m_{\tilde{g}_A} = \frac{g_A^2}{4\pi^2} T^A \mex
    \sum_{\alpha=1,2} V_{2 \alpha}^{\dagger} V_{\alpha 1}
    \frac{m_{\alpha}^2}{m_{\alpha}^2 - \mex^2} 
    \log\left( \frac{m_{\alpha}^2}{\mex^2} \right),
\end{eqnarray}
where $A=1-3$ represents the standard-model gauge groups, and we have
adopted the SU(5) GUT normalization for the U(1)$_Y$ gauge coupling 
($g_1 \equiv \sqrt{5/3}\, g_Y$).
Here, $T^A$ is the Dynkin index of the extra-matter representation, in a
normalization $T^A=1/2$ for a fundamental of SU($N$) and $T^1=(3/5)Y^2$
for U(1)$_Y$.
Note that $T^A$ is the Dynkin index of the extra matter and not the sum
of the index of the extra matter and anti-matter.
That is, we use $T^A = 1/2$ for a vector-like pair of fundamental extra
matters.

Then, the gaugino masses are given by
\begin{eqnarray}
  m_{\tilde{g}_A} = 2 \left( \frac{\alpha_A}{4\pi} \right)
    \Tr_\EX \Biggl[ T_\EX^A\, \mex\, {\rm e}^{-i\alpha}
    \sin 2\theta\, \frac{y_1 \log y_1 - y_2 \log y_2 
    - y_1 y_2 \log(y_1/y_2)}{(y_1-1)(y_2-1)} \Biggr],
\label{Ap_gaugino_mass}
\end{eqnarray}
where the trace is taken over pairs of the extra matters 
$\EX$. (Note that $\mex, y_1, y_2, \alpha$ and $\theta$ depend
on $\EX$.)
This is in agreement with the result given in Ref.~\cite{Pop_Tri}.

Obviously, the heavy sfermions do not contribute to the gaugino masses.
It can also be seen by taking the limit $\mex \rightarrow 0$ and $\theta
\rightarrow 0$ in Eq.~(\ref{Ap_gaugino_mass}).

\subsection{The sfermion masses}

We now calculate the light sfermion masses at the two-loop level.
There are two types of contributions which generate the light sfermion
masses.
One is the contribution directly arising from the loop diagrams: the
one-loop graph involving the $\epsilon$-scalar and the two-loop graphs
involving heavy particles, both of which are of order $\alpha_A^2$.
The other contribution is that through the generation of U(1)$_Y$
FI $D$-term at one loop and two loops, which are of
order $\alpha_Y$ and $\alpha_Y \alpha_A$, respectively.

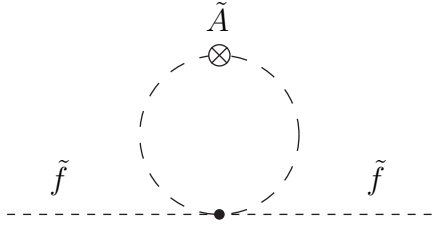
\begin{figure}
\begin{center} 
\begin{picture}(100,80)(150,140)
  \DashLine(120,150)(280,150){3} \Vertex(200,150){2}
  \Text(140,160)[b]{$\tilde{f}$} \Text(260,160)[b]{$\tilde{f}$}
  \DashCArc(200,180)(30,0,360){6} \Text(200,220)[b]{$\tilde{A}$}
  \BCirc(200,210){4} \Line(197,207)(203,213) \Line(197,213)(203,207)
\end{picture}
\caption{One-loop diagram contributing to the light sfermion masses
 which involves the $\epsilon$-scalar $\tilde{A}$.} 
\label{Fig_sferm_epsilon}
\end{center}
\end{figure}

We first consider the direct contribution from the extra-matter
multiplets. 
One diagram contributing to the light sfermion masses is the one-loop
$\epsilon$-scalar graph shown in Fig.~\ref{Fig_sferm_epsilon}.
It contains the counterterm $\delta_{m_{\tilde{A}}^2}$ for the
$\epsilon$-scalar mass, ${\cal L} = -(1/2)\delta_{m_{\tilde{A}}^2}
\tilde{A}_{\tilde{\mu}}^a \tilde{A}_{\tilde{\mu}}^a$, and gives the
light-sfermion mass squared,
\begin{eqnarray}
  m_{\tilde{f},\,\tilde{A}}^2 
    = \frac{2\epsilon\,\Gamma(\epsilon)}{(4\pi)^{2-\epsilon}}
    (\Lambda_{\rm IR}^2)^{-\epsilon} \mu^{2\epsilon} \sum_{A}g_A^2 
    C_{\tilde{f}}^A \delta_{m_{\tilde{A}}^2},
\label{Ap_m2_epsilon}
\end{eqnarray}
in $D = 4-2\epsilon$ dimensions.
Here, $C_{\tilde{f}}^A$ is the quadratic Casimir coefficient for the
light sfermion $\tilde{f}$, $\mu$ is the renormalization scale, and
$\Lambda_{\rm IR}$ is the infra-red cutoff.
The counterterm $\delta_{m_{\tilde{A}}^2}$ is determined to cancel the
divergence of the $\epsilon$-scalar mass arising from one loop of the
extra-matter multiplets,
\begin{eqnarray}
\begin{picture}(380,50)(0,0)
  \DashLine(0,20)(80,20){6} \DashCArc(40,35)(15,0,360){3} \Vertex(40,20){2}
  \Text(100,20)[]{$+$}
  \DashLine(120,20)(140,20){6} \DashLine(180,20)(200,20){6}
  \CArc(160,20)(20,0,360) \Vertex(140,20){2} \Vertex(180,20){2}
  \Text(220,20)[]{$+$}
  \DashLine(240,20)(320,20){6}
  \BCirc(280,20){4} \Line(277,17)(283,23) \Line(277,23)(283,17)
  \Text(340,20)[l]{$=\quad {\rm finite},$}
\end{picture}
\end{eqnarray}
as
\begin{eqnarray}
  \delta_{m_{\tilde{A}}^2} = -\frac{2g_A^2}{(4\pi)^2}
    \Tr_\EX \Biggl[ T_\EX^A\, (m_1^2+m_2^2-2\mex^2) \Biggr]
    \left( \frac{1}{\epsilon} \right).
\label{Ap_count_epsilon}
\end{eqnarray}
Substituting Eq.~(\ref{Ap_count_epsilon}) into
Eq.~(\ref{Ap_m2_epsilon}), we obtain
\begin{eqnarray}
  m_{\tilde{f},\,\tilde{A}}^2 
    = -4 \sum_A \left( \frac{\alpha_A}{4\pi} \right)^2
    C_{\tilde{f}}^A \Tr_\EX 
       \Biggl[ T_\EX^A\, \mex^2\, (y_1+y_2-2) \Biggr]
    \left( \frac{1}{\epsilon} - \log\frac{\Lambda_{\rm IR}^2}{\mu^2} \right).
\label{Ap_m2_A}
\end{eqnarray}
Here, the combination $- \gamma + \log(4\pi)$ has been absorbed by an
appropriate redefinition of the renormalization scale, 
$\mu \rightarrow \mu \sqrt{{\rm e}^{\gamma}/4\pi}$ (where $\gamma$ is
the Euler number).
The result has $1/\epsilon$ pole of order $\alpha_A^2$, so that it
contributes to the two-loop RG equations for the
light sfermion masses.
We will omit $\mu$, hereafter.

The remaining diagram consists of two-loop graphs involving the
extra-matter multiplets given in Fig.~\ref{Fig_sferm_2loop}.
These graphs are identical to those considered in Ref.~\cite{Martin} in 
the case of vanishing supertrace.
Their contribution is ultra-violet finite even in the case of
nonvanishing supertrace.
Together with Eq.~(\ref{Ap_m2_A}), we obtain the light sfermion masses
induced directly by loops of the extra-matter multiplets as
\begin{eqnarray}
  m_{\tilde{f},\,{\rm direct}}^2 |_{\EX}
    &=& -4 \sum_A \left( \frac{\alpha_A}{4\pi} \right)^2
    C_{\tilde{f}}^A \Tr_\EX 
     \Biggl[ T_\EX^A\, \mex^2\, (y_1+y_2-2) \Biggr]
    \left( \frac{1}{\epsilon} \right) \nonumber\\
  && +4 \sum_A \left( \frac{\alpha_A}{4\pi} \right)^2
    C_{\tilde{f}}^A \Tr_\EX 
     \Biggl[ T_\EX^A\, \mex^2 \biggl\{ (y_1+y_2-2) 
    (\log \mex^2 + 2) \nonumber\\
  && \qquad + (y_1 \log y_1 + y_2 \log y_2)
     -2 \biggl( y_1 {\rm Li}_2(1-\frac{1}{y_1}) 
     + y_2 {\rm Li}_2(1-\frac{1}{y_2}) \biggr) \nonumber\\
  && \qquad + \frac{1}{2}\sin^2 2\theta 
     \biggl( y_1 {\rm Li}_2(1-\frac{y_2}{y_1})
     + y_2 {\rm Li}_2(1-\frac{y_1}{y_2}) \biggr) 
     \biggr\} \Biggr],
\end{eqnarray}
which is in agreement with the result given in Ref.~\cite{Pop_Tri}.
Here, ${\rm Li}_2(x) = -\int_0^1 dt\,\log(1-xt)/t$ is the dilogarithm
function.
Note that $\Lambda_{\rm IR}^2$ is canceled between the diagrams of
Fig.~\ref{Fig_sferm_epsilon} and Fig.~\ref{Fig_sferm_2loop}.
Taking the limit $\mex \rightarrow 0$ and $\theta \rightarrow 0$ and
regarding $m_1^2$ as the heavy sfermion masses $m_{\HS}^2$, we
obtain the light sfermion masses induced by loops of the heavy
sfermions as
\begin{eqnarray}
  m_{\tilde{f},\,{\rm direct}}^2 |_{\HS}
    &=& -4 \sum_A \left( \frac{\alpha_A}{4\pi} \right)^2
    C_{\tilde{f}}^A \Tr_{\HS} 
    \Biggl[ T_{\HS}^A\, m_{\HS}^2 \Biggr] 
    \left( \frac{1}{\epsilon} \right) \nonumber\\
  && +4 \sum_A \left( \frac{\alpha_A}{4\pi} \right)^2
    C_{\tilde{f}}^A \Tr_{\HS} \Biggl[ T_{\HS}^A\, 
    m_{\HS}^2 \biggl( \log m_{\HS}^2 + 2 
    - \frac{1}{3}\pi^2 \biggr) \Biggr],
\end{eqnarray}
where $\HS$ denotes the heavy sfermions.
The direct contribution $m_{\tilde{f},\,{\rm direct}}^2$ to the light
sfermion masses is given by summing up that from the extra-matter
multiplets and that from the heavy sfermions,
\begin{eqnarray}
  m_{\tilde{f},\,{\rm direct}}^2 
    = m_{\tilde{f},\,{\rm direct}}^2 |_{\EX} 
    + m_{\tilde{f},\,{\rm direct}}^2 |_{\HS}.
\label{Ap_m2_direct_div}
\end{eqnarray}
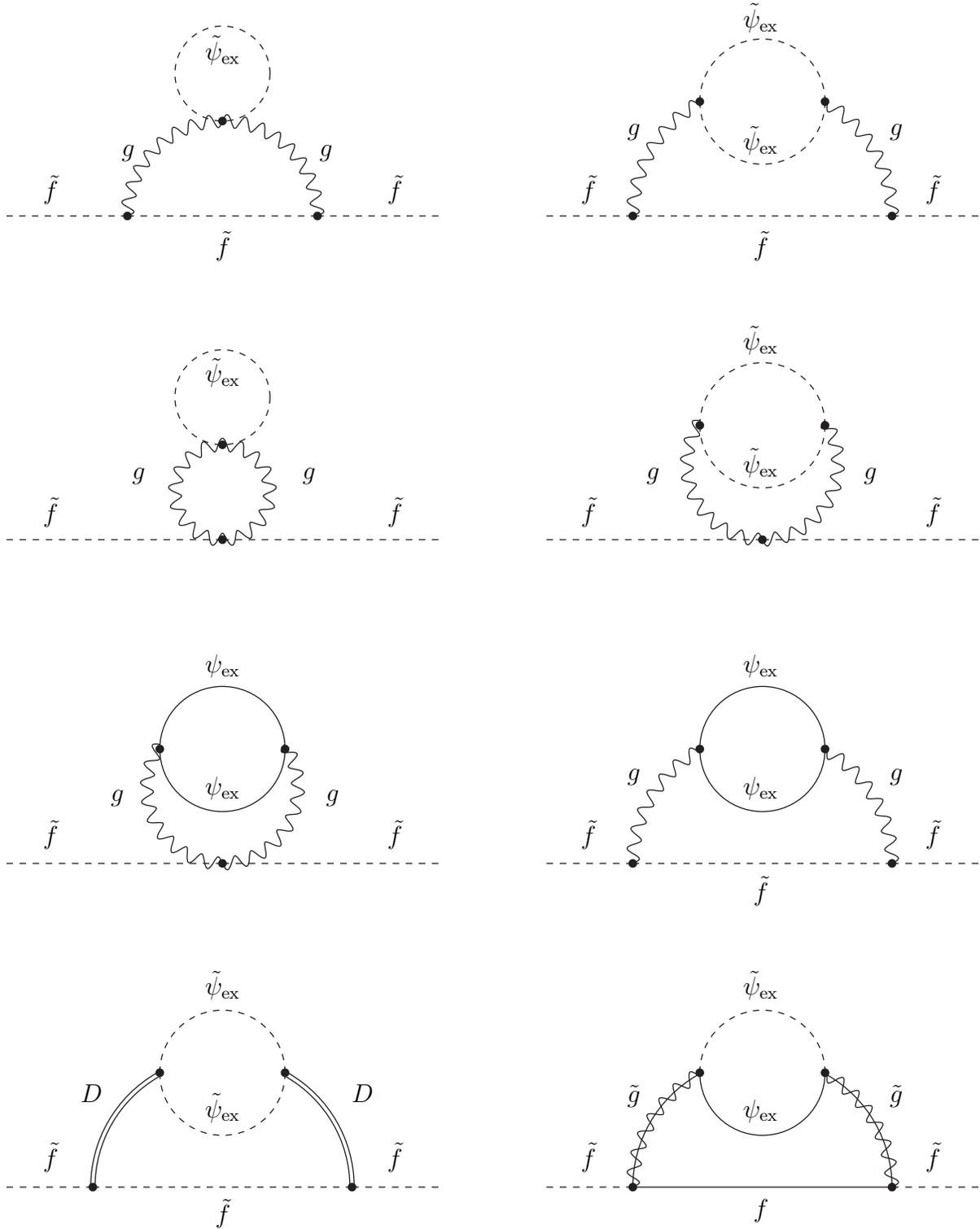
\begin{figure}
\begin{center}
\begin{picture}(450,560)(-140,120)
  \DashLine(-140,600)(-84,600){3} \Text(-120,610)[b]{$\tilde{f}$}
  \Vertex(-84,600){2} \Vertex(4,600){2}
  \Text(-81,630)[rb]{$g$} \PhotonArc(-40,600)(44,0,180){3}{17}
  \DashLine(-84,600)(4,600){3} \Text(-40,595)[t]{$\tilde{f}$}
  \Vertex(-40,644){2} \Text(5,630)[lb]{$g$}
  \DashCArc(-40,666)(22,0,360){3} 
  \Text(-40,675)[b]{$\tilde{\psi}_{\rm ex}$}
  \DashLine(4,600)(60,600){3} \Text(40,610)[b]{$\tilde{f}$}
  \DashLine(-140,450)(-40,450){3} \Text(-120,460)[b]{$\tilde{f}$}
  \Vertex(-40,450){2} \Text(-76,480)[rb]{$g$}
  \PhotonArc(-40,472)(22,0,360){3}{17} \Vertex(-40,494){2}
  \Text(-3,480)[lb]{$g$} \DashCArc(-40,516)(22,0,360){3}
  \Text(-40,525)[b]{$\tilde{\psi}_{\rm ex}$} 
  \DashLine(-40,450)(60,450){3}
  \Text(40,460)[b]{$\tilde{f}$}
  \DashLine(-140,300)(-40,300){3} \Text(-120,310)[b]{$\tilde{f}$}
  \Vertex(-40,300){2} \PhotonArc(-40,335)(35,145,35){3}{18}
  \Text(-86,330)[rb]{$g$} \Vertex(-69,353){2}
  \Text(8,330)[lb]{$g$} \Vertex(-11,353){2}
  \CArc(-40,353)(29,0,180) \Text(-40,390)[b]{$\psi_{\rm ex}$}
  \CArc(-40,353)(29,180,360) \Text(-40,333)[b]{$\psi_{\rm ex}$}
  \DashLine(-40,300)(60,300){3} \Text(40,310)[b]{$\tilde{f}$}
  \DashLine(-140,150)(-100,150){3} \Text(-120,160)[b]{$\tilde{f}$}
  \Vertex(-100,150){2} \DashLine(-100,150)(20,150){3} 
  \Text(-40,145)[t]{$\tilde{f}$}
  \CArc(-40,150)(59,120,180) \Text(-96,190)[rb]{$D$}
  \CArc(-40,150)(61,120,180) \Vertex(-69,203){2}
  \CArc(-40,150)(59,0,60) \Text(20,190)[lb]{$D$}
  \CArc(-40,150)(61,0,60) \Vertex(-11,203){2}
  \DashCArc(-40,203)(29,0,180){3} 
  \Text(-40,240)[b]{$\tilde{\psi}_{\rm ex}$}
  \DashCArc(-40,203)(29,180,360){3} 
  \Text(-40,183)[b]{$\tilde{\psi}_{\rm ex}$}
  \DashLine(20,150)(60,150){3} \Text(40,160)[b]{$\tilde{f}$}
  \Vertex(20,150){2}
  \DashLine(110,600)(150,600){3} \Text(130,610)[b]{$\tilde{f}$}
  \Vertex(150,600){2} \DashLine(150,600)(270,600){3}
  \Text(210,595)[t]{$\tilde{f}$} \Text(154,640)[rb]{$g$}
  \PhotonArc(210,600)(60,120,180){3}{7} \Vertex(181,653){2}
  \Text(270,640)[lb]{$g$} \PhotonArc(210,600)(60,0,60){3}{7}
  \Vertex(239,653){2} \DashCArc(210,653)(29,0,180){3}
  \Text(210,690)[b]{$\tilde{\psi}_{\rm ex}$}
  \DashCArc(210,653)(29,180,360){3}
  \Text(210,633)[b]{$\tilde{\psi}_{\rm ex}$} 
  \DashLine(270,600)(310,600){3}
  \Text(290,610)[b]{$\tilde{f}$} \Vertex(270,600){2}
  \DashLine(110,450)(210,450){3} \Text(130,460)[b]{$\tilde{f}$}
  \Vertex(210,450){2} \PhotonArc(210,485)(35,145,35){3}{18}
  \Text(163,480)[rb]{$g$} \Vertex(181,503){2}
  \Text(258,480)[lb]{$g$} \Vertex(239,503){2}
  \DashCArc(210,503)(29,0,180){3} 
  \Text(210,540)[b]{$\tilde{\psi}_{\rm ex}$}
  \DashCArc(210,503)(29,180,360){3}
  \Text(210,483)[b]{$\tilde{\psi}_{\rm ex}$} 
  \DashLine(210,450)(310,450){3}
  \Text(290,460)[b]{$\tilde{f}$}
  \DashLine(110,300)(150,300){3} \Text(130,310)[b]{$\tilde{f}$}
  \Vertex(150,300){2} \DashLine(150,300)(270,300){3}
  \Text(210,295)[t]{$\tilde{f}$} \Text(154,340)[rb]{$g$}
  \PhotonArc(210,300)(60,120,180){3}{7} \Vertex(181,353){2}
  \Text(270,340)[lb]{$g$} \PhotonArc(210,300)(60,0,60){3}{7}
  \Vertex(239,353){2} \CArc(210,353)(29,0,180)
  \Text(210,390)[b]{$\psi_{\rm ex}$} \CArc(210,353)(29,180,360)
  \Text(210,333)[b]{$\psi_{\rm ex}$} \DashLine(270,300)(310,300){3}
  \Text(290,310)[b]{$\tilde{f}$} \Vertex(270,300){2}
  \DashLine(110,150)(150,150){3} \Text(130,160)[b]{$\tilde{f}$}
  \Vertex(150,150){2} \Line(150,150)(270,150) \Text(210,145)[t]{$f$}
  \CArc(210,150)(60,120,180) \Text(154,190)[rb]{$\tilde{g}$}
  \PhotonArc(210,150)(60,120,180){3}{7} \Vertex(181,203){2}
  \CArc(210,150)(60,0,60) \Text(270,190)[lb]{$\tilde{g}$}
  \PhotonArc(210,150)(60,0,60){3}{7} \Vertex(239,203){2}
  \DashCArc(210,203)(29,0,180){3} 
  \Text(210,240)[b]{$\tilde{\psi}_{\rm ex}$}
  \CArc(210,203)(29,180,360) \Text(210,183)[b]{$\psi_{\rm ex}$}
  \DashLine(270,150)(310,150){3} \Text(290,160)[b]{$\tilde{f}$}
  \Vertex(270,150){2}
\end{picture}
\caption{Two-loop diagram contributing to the light sfermion masses
 which involves the extra-matter multiplets $\Psi_{\rm ex}$ and 
 $\bar{\Psi}_{\rm ex}$.}
\label{Fig_sferm_2loop}
\end{center}
\end{figure}

\begin{figure}
\begin{center} 
\begin{picture}(250,100)(-165,140)
  \DashLine(-130,150)(-40,150){3} \Text(-110,160)[b]{$\tilde{f}$}
  \Vertex(-40,150){2} \Vertex(-40,174){2}
  \Line(-39,150)(-39,174) \Line(-41,150)(-41,174) 
  \Text(-35,162)[l]{$D_Y$}
  \DashCArc(-40,203)(29,0,360){3} 
  \Text(-40,240)[b]{$\tilde{\psi}_{\rm ex}$}
  \DashLine(-40,150)(50,150){3} \Text(30,160)[b]{$\tilde{f}$}
\end{picture}
\caption{One-loop diagram contributing to the light sfermion masses
 through the generation of U(1)$_Y$ FI $D$-term.}
\label{Fig_FI_1loop}
\end{center}
\end{figure}
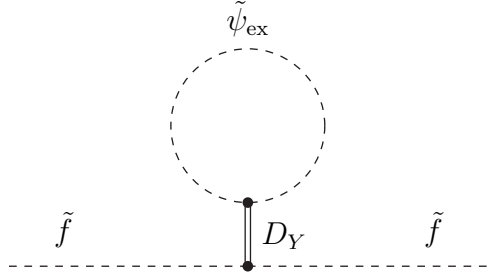

We next consider the light sfermion masses induced via the generation of
U(1)$_Y$ FI $D$-term.
At the one-loop level, the light sfermion masses generated by loops of
the extra-matter multiplets are given by (see Fig.~\ref{Fig_FI_1loop})
\begin{eqnarray}
  m_{\tilde{f},\,{\rm FI-1loop}}^2 |_{\EX} 
    &=& -\left( \frac{\alpha_Y}{4\pi} \right) Y_{\tilde{f}} 
    \Tr_{\EX} \Biggl[ Y_{\EX} \cos 2\theta\, (m_1^2-m_2^2) \Biggr]
    \left( \frac{1}{\epsilon} + 1 \right) \nonumber\\
  && + \left( \frac{\alpha_Y}{4\pi} \right) Y_{\tilde{f}}
    \Tr_{\EX} \Biggl[ Y_{\EX} \cos 2\theta 
    \biggl( m_1^2 \log m_1^2 - m_2^2 \log m_2^2 \biggr) \Biggr].
\end{eqnarray}
This contribution vanishes if the extra-matter multiplets have an
invariance under the parity 
$\Psi_{\rm ex} \leftrightarrow \bar{\Psi}_{\rm ex}$, as can be readily
seen by taking $\theta = \pm \pi/4$ in the expression.
The heavy sfermions also generate one-loop U(1)$_Y$ FI $D$-term, which
gives
\begin{eqnarray}
  m_{\tilde{f},\,{\rm FI-1loop}}^2 |_{\HS} 
    &=& -\left( \frac{\alpha_Y}{4\pi} \right) Y_{\tilde{f}} 
    \Tr_{\HS} \Biggl[ Y_{\HS}\, m_{\HS}^2 \Biggr]
    \left( \frac{1}{\epsilon} + 1 \right) \nonumber\\
  && + \left( \frac{\alpha_Y}{4\pi} \right) Y_{\tilde{f}}
    \Tr_{\HS} \Biggl[ Y_{\HS}\, 
    m_{\HS}^2 \log m_{\HS}^2 \Biggr].
\end{eqnarray}
The total contribution through the generation of U(1)$_Y$ FI $D$-term at
one loop is
\begin{eqnarray}
  m_{\tilde{f},\,{\rm FI-1loop}}^2
    = m_{\tilde{f},\,{\rm FI-1loop}}^2 |_{\EX} 
    + m_{\tilde{f},\,{\rm FI-1loop}}^2 |_{\HS}.
\label{Ap_m2_FI-1loop_div}
\end{eqnarray}

\begin{figure}
\begin{center} 
\begin{picture}(450,400)(-150,-20)
  \DashLine(-130,290)(-40,290){3} \Text(-110,300)[b]{$\tilde{f}$}
  \Vertex(-40,290){2} \Vertex(-40,314){2}
  \Line(-39,290)(-39,314) \Line(-41,290)(-41,314) 
  \Text(-35,302)[l]{$D_Y$}
  \DashCArc(-40,343)(29,0,180){3} 
  \Text(-40,380)[b]{$\tilde{\psi}_{\rm ex}$}
  \Photon(-69,343)(-11,343){3}{7} \Text(-40,352)[b]{$g$}
  \Vertex(-69,343){2} \Vertex(-11,343){2}
  \DashCArc(-40,343)(29,180,360){3} 
  \Text(-14,329)[lt]{$\tilde{\psi}_{\rm ex}$}
  \DashLine(-40,290)(50,290){3} \Text(30,300)[b]{$\tilde{f}$}
  \DashLine(-130,150)(-40,150){3} \Text(-110,160)[b]{$\tilde{f}$}
  \Vertex(-40,150){2} \Vertex(-40,174){2}
  \Line(-39,150)(-39,174) \Line(-41,150)(-41,174) 
  \Text(-35,162)[l]{$D_Y$}
  \DashCArc(-40,203)(29,0,180){3} 
  \Text(-40,240)[b]{$\tilde{\psi}_{\rm ex}$}
  \Line(-69,202)(-11,202) \Line(-69,204)(-11,204)
  \Text(-40,208)[b]{$D$} \Vertex(-69,203){2} \Vertex(-11,203){2}
  \DashCArc(-40,203)(29,180,360){3} 
  \Text(-14,189)[lt]{$\tilde{\psi}_{\rm ex}$}
  \DashLine(-40,150)(50,150){3} \Text(30,160)[b]{$\tilde{f}$}
  \DashLine(-15,10)(75,10){3} \Text(5,20)[b]{$\tilde{f}$}
  \Vertex(75,10){2} \Vertex(75,34){2}
  \Line(76,10)(76,34) \Line(74,10)(74,34) \Text(80,22)[l]{$D_Y$}
  \DashCArc(75,63)(29,0,360){3} 
  \Text(75,103)[b]{$\tilde{\psi}_{\rm ex}$}
  \BCirc(75,92){4} \Line(72,89)(78,95) \Line(72,95)(78,89)
  \DashLine(75,10)(165,10){3} \Text(145,20)[b]{$\tilde{f}$}
  \DashLine(100,290)(190,290){3} \Text(120,300)[b]{$\tilde{f}$}
  \Vertex(190,290){2} \Vertex(190,314){2}
  \Line(191,290)(191,314) \Line(189,290)(189,314) 
  \Text(195,302)[l]{$D_Y$}
  \CArc(190,343)(29,0,180) \Text(190,380)[b]{$\psi_{\rm ex}$}
  \Photon(161,343)(219,343){3}{7} \Line(161,343)(219,343)
  \Text(190,352)[b]{$\tilde{g}$}
  \Vertex(161,343){2} \Vertex(219,343){2}
  \DashCArc(190,343)(29,180,360){3} 
  \Text(216,329)[lt]{$\tilde{\psi}_{\rm ex}$}
  \DashLine(190,290)(280,290){3} \Text(260,300)[b]{$\tilde{f}$}
  \DashLine(100,150)(190,150){3} \Text(120,160)[b]{$\tilde{f}$}
  \Vertex(190,150){2} \Vertex(190,174){2}
  \Line(191,150)(191,174) \Line(189,150)(189,174) 
  \Text(195,162)[l]{$D_Y$}
  \DashCArc(190,193)(19,0,360){3} 
  \Text(215,193)[l]{$\tilde{\psi}_{\rm ex}$}
  \Vertex(190,212){2}
  \PhotonArc(190,231)(19,0,360){2.5}{16} \Text(218,231)[l]{$g$}
  \DashLine(190,150)(280,150){3} \Text(260,160)[b]{$\tilde{f}$}
\end{picture}
\caption{Two-loop diagram contributing to the light sfermion masses
 through the generation of U(1)$_Y$ FI $D$-term.}
\label{Fig_FI_2loop}
\end{center}
\end{figure}
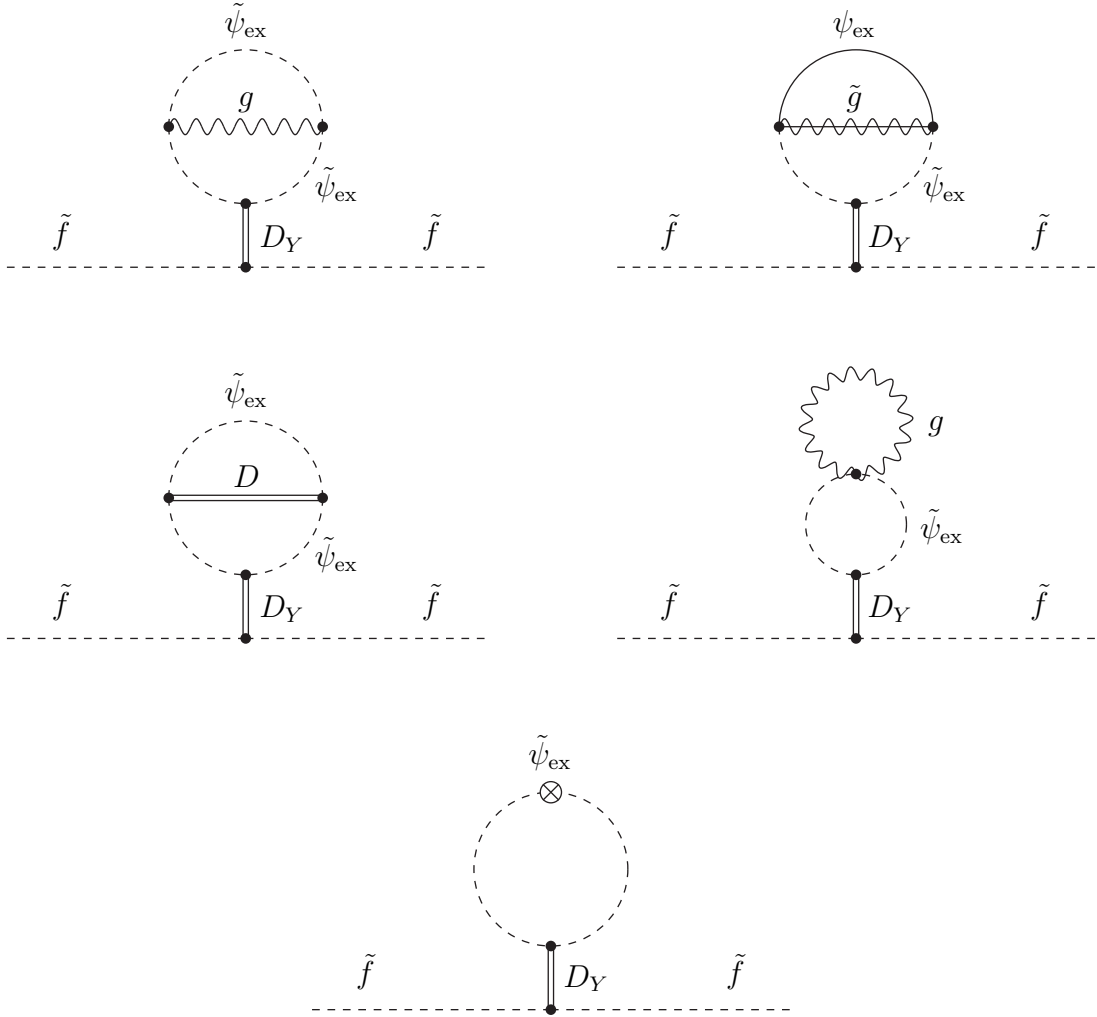

\begin{figure}
\begin{center} 
\begin{picture}(450,175)(-265,0)
  \DashLine(-265,10)(-200,10){3} \Text(-255,20)[b]{$\tilde{f}$}
  \Vertex(-200,10){2} \Vertex(-200,34){2}
  \Line(-199,10)(-199,34) \Line(-201,10)(-201,34) 
  \Text(-195,22)[l]{$D_Y$}
  \DashCArc(-200,63)(29,0,360){3}
  \Text(-165,63)[l]{$\tilde{\psi}_{\rm ex}, \tilde{\psi}_{\HS},
                     \tilde{f}$}
  \Vertex(-200,92){2} \Vertex(-200,116){2}
  \Line(-199,92)(-199,116) \Line(-201,92)(-201,116) 
  \Text(-195,104)[l]{$D_Y$}
  \DashCArc(-200,145)(29,0,360){3} 
  \Text(-200,182)[b]{$\tilde{\psi}_{\rm ex}, \tilde{\psi}_{\HS}$}
  \DashLine(-200,10)(-135,10){3} \Text(-145,20)[b]{$\tilde{f}$}
  \DashLine(-105,10)(-40,10){3} \Text(-95,20)[b]{$\tilde{f}$}
  \Vertex(-40,10){2} \Vertex(-40,34){2}
  \Line(-39,10)(-39,34) \Line(-41,10)(-41,34) \Text(-35,22)[l]{$D_Y$}
  \DashCArc(-40,63)(29,0,360){3} 
  \Text(-40,103)[b]{$\tilde{\psi}_{\rm ex}, \tilde{\psi}_{\HS},
                     \tilde{f}$}
  \BCirc(-40,92){4} \Line(-43,89)(-37,95) \Line(-43,95)(-37,89)
  \DashLine(-40,10)(25,10){3} \Text(15,20)[b]{$\tilde{f}$}
  \DashLine(55,10)(120,10){3} \Text(65,20)[b]{$\tilde{f}$}
  \DashCArc(120,39)(29,0,360){3} 
  \Text(120,76)[b]{$\tilde{\psi}_{\rm ex}, \tilde{\psi}_{\HS}$}
  \DashLine(120,10)(185,10){3} \Text(175,20)[b]{$\tilde{f}$}
  \BCirc(120,10){4} \Line(117,7)(123,13) \Line(117,13)(123,7)
\end{picture}
\caption{Two-loop diagram contributing to the light sfermion masses.}
\label{Fig_FI_1loop2}
\end{center}
\end{figure}
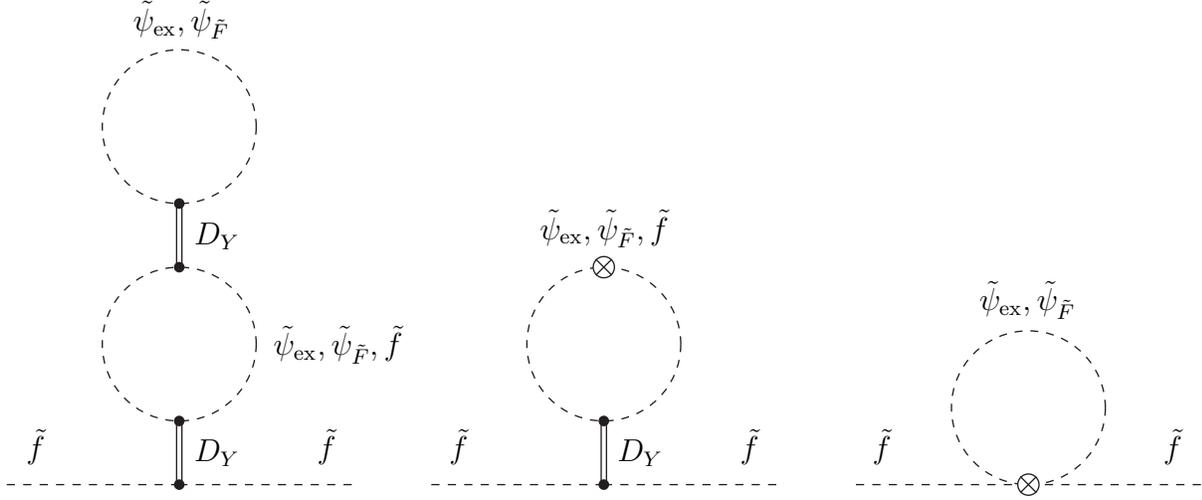

U(1)$_Y$ FI $D$-term is also generated at two loops through the
diagram shown in Fig.~\ref{Fig_FI_2loop}.
The resulting light sfermion masses from extra-matter loops are written
as 
\begin{eqnarray}
  m_{\tilde{f},\,{\rm FI-2loop}}^2 |_{\EX} 
    &=& \sum_A g_Y^2 g_A^2 Y_{\tilde{f}} \Tr_{\EX} 
    \Biggl[ Y_{\EX} C_{\EX}^A
    \biggl\{ \sum_{\alpha} (V_{\alpha 1}V_{1 \alpha}^{\dagger} - 
    V_{\alpha 2}V_{2 \alpha}^{\dagger})\, I_1(m_{\alpha}^2) \nonumber\\
  && \qquad + 4\sum_{\alpha} (V_{\alpha 1}V_{1 \alpha}^{\dagger} - 
    V_{\alpha 2}V_{2 \alpha}^{\dagger})\, I_2(m_{\alpha}^2) \nonumber\\
  && \qquad - \sum_{\alpha\beta\gamma} 
    (V_{\alpha 1}V_{1 \beta}^{\dagger} - V_{\alpha 2}V_{2 \beta}^{\dagger}) 
    (V_{\beta 1}V_{1 \gamma}^{\dagger} - V_{\beta 2}V_{2 \gamma}^{\dagger})
    \nonumber\\
  && \qquad\qquad\qquad \times
    (V_{\gamma 1}V_{1 \alpha}^{\dagger} - V_{\gamma 2}V_{2 \alpha}^{\dagger}) 
    \, I_3(m_{\alpha}^2, m_{\beta}^2, m_{\gamma}^2) \nonumber\\
  && \qquad + \sum_{\alpha\beta} (V_{\alpha 1}V_{1 \beta}^{\dagger} - 
    V_{\alpha 2}V_{2 \beta}^{\dagger}) (\delta_{m^2})_{\beta\alpha}\,
    I_c(m_{\alpha}^2, m_{\beta}^2) \biggr\} \Biggr],
\label{Ap_m2_FI2loop_Q}
\end{eqnarray}
in the Feynman gauge. Here, functions $I$'s are defined as
\begin{eqnarray}
  I_1(m_{\alpha}^2) &=&
    \int\!\!\frac{d^4 p}{(2\pi)^4} \int\!\!\frac{d^4 k}{(2\pi)^4}
    \frac{1}{(p^2-m_{\alpha}^2)^2} \frac{(2p-k)^2}{k^2}
    \frac{1}{(p-k)^2-m_{\alpha}^2}, \\
  I_2(m_{\alpha}^2) &=&
    \int\!\!\frac{d^4 p}{(2\pi)^4} \int\!\!\frac{d^4 k}{(2\pi)^4}
    \frac{1}{(p^2-m_{\alpha}^2)^2} \frac{k^2-k\cdot p}{k^2}
    \frac{1}{(p-k)^2-\mex^2}, \\
  I_3(m_{\alpha}^2, m_{\beta}^2, m_{\gamma}^2) &=&
    \int\!\!\frac{d^4 p}{(2\pi)^4} 
    \frac{1}{p^2-m_{\alpha}^2} \frac{1}{p^2-m_{\beta}^2} 
    \int\!\!\frac{d^4 q}{(2\pi)^4} 
    \frac{1}{q^2-m_{\gamma}^2}, \\
  I_c(m_{\alpha}^2, m_{\beta}^2) &=&
    i \int\!\!\frac{d^4 p}{(2\pi)^4}
    \frac{1}{p^2-m_{\alpha}^2} \frac{1}{p^2-m_{\beta}^2}.
\end{eqnarray}
The counterterm $(\delta_{m^2})_{\alpha\beta}$ for the extra-scalar
masses, ${\cal L} = -\sum_A g_A^2 C_{\EX}^A 
(\delta_{m^2})_{\alpha\beta} 
(\tilde{\psi}_{\rm ex})_{\alpha}^* (\tilde{\psi}_{\rm ex})_{\beta}$, is
determined to cancel the divergence,
\begin{eqnarray}
\begin{picture}(380,80)(0,0)
  \DashLine(0,50)(20,50){3} \DashLine(60,50)(80,50){3}
  \PhotonArc(40,50)(20,0,180){2}{10} 
  \DashLine(20,50)(60,50){3} \Vertex(20,50){2} \Vertex(60,50){2}
  \Text(100,50)[]{$+$}
  \DashLine(120,50)(140,50){3} \DashLine(180,50)(200,50){3}
  \PhotonArc(160,50)(20,0,180){2}{10} \CArc(160,50)(20,0,180) 
  \Line(140,50)(180,50) \Vertex(140,50){2} \Vertex(180,50){2}
  \Text(220,50)[]{$+$}
  \DashLine(240,50)(260,50){3} \DashLine(300,50)(320,50){3}
  \DashCArc(280,50)(20,0,180){3} 
  \Line(260,49)(300,49) \Line(260,51)(300,51) 
  \Vertex(260,50){2} \Vertex(300,50){2}
  \DashLine(120,0)(200,0){3} \Vertex(160,0){2}
  \PhotonArc(160,15)(15,0,360){1.5}{16}
  \Text(220,0)[]{$+$}
  \DashLine(240,0)(320,0){3}
  \BCirc(280,0){4} \Line(277,-3)(283,3) \Line(277,3)(283,-3)
  \Text(340,0)[l]{$=\quad {\rm finite},$}
\end{picture}
\end{eqnarray}
as
\begin{eqnarray}
  (\delta_{m^2})_{\alpha\beta} &=& \frac{1}{(4\pi)^2}
    \left( \frac{1}{\epsilon} \right) \nonumber\\
  && \times \left(
    \begin{array}{cc}
      -\sin^2 2\theta\, (m_1^2-m_2^2) - 4\mex^2  & 
      \sin 2\theta \cos 2\theta\, (m_1^2-m_2^2)\, {\rm e}^{i\alpha}  \\
      \sin 2\theta \cos 2\theta\, (m_1^2-m_2^2)\, {\rm e}^{-i\alpha}  &
      \sin^2 2\theta\, (m_1^2-m_2^2) - 4\mex^2
    \end{array}
  \right).
\label{Ap_count_mass}
\end{eqnarray}
Substituting Eq.~(\ref{Ap_count_mass}) into Eq.~(\ref{Ap_m2_FI2loop_Q}), 
we obtain
\begin{eqnarray}
  && m_{\tilde{f},\,{\rm FI-2loop}}^2 |_{\EX} = \nonumber\\
  && \qquad -2 \sum_A \left( \frac{\alpha_Y}{4\pi} \right) 
    \left( \frac{\alpha_A}{4\pi} \right) Y_{\tilde{f}} 
    \Tr_{\EX} 
    \Biggl[ Y_{\EX} C_{\EX}^A \cos 2\theta\, (m_1^2-m_2^2) \Biggr]
    \left( \frac{1}{\epsilon} + 3 \right) \nonumber\\
  && \qquad +4 \sum_A \left( \frac{\alpha_Y}{4\pi} \right) 
    \left( \frac{\alpha_A}{4\pi} \right) Y_{\tilde{f}} 
    \Tr_{\EX} \Biggl[ Y_{\EX} C_{\EX}^A \cos 2\theta
    \biggl\{ -\frac{1}{2}\mex^2 \log^2 m_1^2 
    +\frac{1}{2}\mex^2 \log^2 m_2^2 \nonumber\\
  && \qquad \qquad \qquad \qquad
    + (m_1^2+\mex^2)\log m_1^2 - (m_2^2+\mex^2)\log m_2^2 \nonumber\\
  && \qquad \qquad \qquad \qquad
    + (m_1^2-\mex^2) {\rm Li}_2(1-\frac{\mex^2}{m_1^2})
    - (m_2^2-\mex^2) {\rm Li}_2(1-\frac{\mex^2}{m_2^2})
    \biggr\} \nonumber\\
  && \qquad \qquad \qquad
    + \frac{1}{4} Y_{\EX} C_{\EX}^A \cos 2\theta \sin^2 2\theta
    \biggl\{ 2 \frac{(m_1^2 \log m_1^2 - m_2^2 \log m_2^2)^2}
    {m_1^2-m_2^2} \nonumber\\
  && \qquad \qquad \qquad \qquad
    - 4(m_1^2 \log m_1^2 - m_2^2 \log m_2^2)
    - (m_1^2 \log^2 m_1^2 - m_2^2 \log^2 m_2^2) \nonumber\\
  && \qquad \qquad \qquad \qquad
    + (m_1^2-m_2^2)(2 + \log m_1^2 + \log m_2^2 
    - \log m_1^2 \log m_2^2) \biggr\} \Biggr].
\end{eqnarray}
This contribution vanishes when $\theta = \pm \pi/4$ as it should be.
The contribution from the heavy sfermions is read off by taking the
limit $\mex \rightarrow 0$ and $\theta \rightarrow 0$ as
\begin{eqnarray}
  && m_{\tilde{f},\,{\rm FI-2loop}}^2 |_{\HS} = \nonumber\\
  && \qquad -2 \sum_A \left( \frac{\alpha_Y}{4\pi} \right) 
    \left( \frac{\alpha_A}{4\pi} \right) Y_{\tilde{f}} \Tr_{\HS} 
    \Biggl[ Y_{\HS} C_{\HS}^A\, m_{\HS}^2 \Biggr]
    \left( \frac{1}{\epsilon} + 3 \right) \nonumber\\
  && \qquad +4 \sum_A \left( \frac{\alpha_Y}{4\pi} \right) 
    \left( \frac{\alpha_A}{4\pi} \right) Y_{\tilde{f}} 
    \Tr_{\HS} \Biggl[ Y_{\HS} C_{\HS}^A\, m_{\HS}^2 
    \biggl( \log m_{\HS}^2 + \frac{1}{6}\pi^2 \biggr) \Biggr],
\end{eqnarray}
reproducing the earlier result derived in Ref.~\cite{AG}.\footnote{
Their result is different from ours by a finite part proportional to
$- \gamma + \log(4\pi)$, since they subtracted divergences not in the
$\overline{\rm MS}$ scheme.}
In addition, there is another diagram which contributes to the light
sfermion masses at the two-loop level.
The diagram is shown in Fig.~\ref{Fig_FI_1loop2} and its contribution is
\begin{eqnarray}
  m_{\tilde{f},\,{\rm FI-2(1loop)}}^2 = 
    -g_Y^4 Y_{\tilde{f}}\, J_1 J_2
    +i g_Y^4 Y_{\tilde{f}}\, \delta_{m_Y^2} J_2
    +i g_Y^4 Y_{\tilde{f}}\, J_1 \delta_{g_Y^4},
\label{Ap_m2_FI1loop2}
\end{eqnarray}
where
\begin{eqnarray}
  J_1 &=&
    \Tr_{\EX} \Biggl[ Y_{\EX} \sum_{\alpha} 
       (V_{\alpha 1}V_{1 \alpha}^{\dagger}
         -V_{\alpha 2}V_{2 \alpha}^{\dagger}) 
      \int\!\!\frac{d^4 p}{(2\pi)^4} \frac{1}{p^2-m_{\alpha}^2} \Biggr]
      \nonumber\\
  && + \Tr_{\HS} \Biggl[ Y_{\HS} 
      \int\!\!\frac{d^4 p}{(2\pi)^4} 
          \frac{1}{p^2-m_{\HS}^2} \Biggr], \\
  J_2 &=&
    \Tr_{\EX} \Biggl[ Y_{\EX}^2 \sum_{\gamma\delta} 
      (V_{\gamma 1}V_{1 \delta}^{\dagger}-V_{\gamma 2}V_{2 \delta}^{\dagger})
      (V_{\delta 1}V_{1 \gamma}^{\dagger}-V_{\delta 2}V_{2 \gamma}^{\dagger})
      \int\!\!\frac{d^4 q}{(2\pi)^4} \frac{1}{q^2-m_{\gamma}^2}
      \frac{1}{q^2-m_{\delta}^2} \Biggr] \nonumber\\
  && + \Tr_{\HS} \Biggl[ Y_{\HS}^2
      \int\!\!\frac{d^4 q}{(2\pi)^4} \frac{1}{(q^2-m_{\HS}^2)^2} \Biggr]
     + \Tr_{\tilde{f}} \Biggl[ Y_{\tilde{f}}^2
      \int\!\!\frac{d^4 q}{(2\pi)^4} \frac{1}{(q^2-m_{\tilde{f}}^2)^2} 
     \Biggr].
\end{eqnarray}
The counterterms $\delta_{m_Y^2}$ and $\delta_{g_Y^4}$ are defined by 
\begin{eqnarray}
  {\cal L} &=& -g_Y^2 \delta_{m_Y^2} \biggl\{ \sum_{\alpha\beta} 
    (V_{\alpha 1}V_{1 \beta}^{\dagger} 
     - V_{\alpha 2}V_{2 \beta}^{\dagger})
    (\tilde{\psi}_{\rm ex})_{\alpha}^* Y_{\EX} 
      (\tilde{\psi}_{\rm ex})_{\beta} +
    \tilde{\psi}_{\HS}^* Y_{\HS} \tilde{\psi}_{\HS} +
    \tilde{f}^* Y_{\tilde{f}} \tilde{f}
    \biggr\} \nonumber\\
  && - g_Y^4 \delta_{g_Y^4} 
    {\tilde{f}}^* Y_{\tilde{f}} {\tilde{f}}\,
    \biggl\{ \sum_{\alpha\beta} (V_{\alpha 1}V_{1 \beta}^{\dagger} - 
    V_{\alpha 2}V_{2 \beta}^{\dagger})
    (\tilde{\psi}_{\rm ex})_{\alpha}^* Y_{\EX}
        (\tilde{\psi}_{\rm ex})_{\beta} +
    \tilde{\psi}_{\HS}^* Y_{\HS} \tilde{\psi}_{\HS} 
    \biggr\}.
\end{eqnarray}
They are determined by the conditions,
\begin{eqnarray}
\begin{picture}(260,50)(120,0)
  \DashLine(120,10)(200,10){3} \DashCArc(160,35)(15,0,360){3}
  \Line(159,10)(159,20) \Line(161,10)(161,20)
  \Vertex(160,10){2} \Vertex(160,20){2}
  \Text(220,10)[]{$+$}
  \DashLine(240,10)(320,10){3}
  \BCirc(280,10){4} \Line(277,7)(283,13) \Line(277,13)(283,7)
  \Text(340,10)[l]{$=\quad {\rm finite},$}
\end{picture}
\end{eqnarray}
and
\begin{eqnarray}
\begin{picture}(260,50)(120,0)
  \DashLine(120,0)(160,5){3} \DashLine(120,50)(160,45){3}
  \DashLine(160,5)(200,0){3} \DashLine(160,45)(200,50){3}
  \DashCArc(160,25)(10,0,360){3}
  \Line(159,35)(159,45) \Line(161,35)(161,45)
  \Vertex(160,35){2} \Vertex(160,45){2}
  \Line(159,5)(159,15) \Line(161,5)(161,15)
  \Vertex(160,5){2} \Vertex(160,15){2}
  \Text(220,25)[]{$+$}
  \DashLine(240,0)(320,50){3} \DashLine(240,50)(320,0){3}
  \BCirc(280,25){4} \Line(277,23)(283,28) \Line(277,28)(283,23)
  \Text(340,25)[l]{$=\quad {\rm finite},$}
\end{picture}
\end{eqnarray}
respectively, which lead to
\begin{eqnarray}
  \delta_{m_Y^2} &=& \frac{1}{(4\pi)^2}
    \Biggl\{ \Tr_{\EX} 
        \Biggl[ Y_{\EX} \cos 2\theta\, (m_1^2-m_2^2) \Biggr]
    + \Tr_{\HS} \Biggl[ Y_{\HS}\, m_{\HS}^2 \Biggr] 
    \Biggr\} \left( \frac{1}{\epsilon} \right), 
\label{Ap_count_massY}\\
  \delta_{g_Y^4} &=& \frac{1}{(4\pi)^2}
    \Biggl\{ 2 \Tr_{\EX} \biggl[ Y_{\EX}^2 \biggr]
    + \Tr_{\HS} \biggl[ Y_{\HS}^2 \biggr]
    + \Tr_{\tilde{f}} \biggl[ Y_{\tilde{f}}^2 \biggr] \Biggr\}
    \left( \frac{1}{\epsilon} \right).
\label{Ap_count_gY}
\end{eqnarray}
From Eqs.~(\ref{Ap_m2_FI1loop2}, \ref{Ap_count_massY},
\ref{Ap_count_gY}), we obtain
\begin{eqnarray}
  \lefteqn{m_{\tilde{f},\,{\rm FI-2(1loop)}}^2 =} \nonumber\\
  && - \left( \frac{\alpha_Y}{4\pi} \right)^2 Y_{\tilde{f}}
    \left( \frac{1}{\epsilon} \right)^2 \nonumber\\
  && \quad \times
    \Biggl\{ \Tr_{\EX} 
       \Biggl[ Y_{\EX} \cos 2\theta\, (m_1^2-m_2^2) \Biggr]
    + \Tr_{\HS} \Biggl[ Y_{\HS}\, m_{\HS}^2 \Biggr] \Biggr\}
    \Biggl\{ 2 \Tr_{\EX} \biggl[ Y_{\EX}^2 \biggr]
    + \Tr_{\HS} \biggl[ Y_{\HS}^2 \biggr]
    + \Tr_{\tilde{f}} \biggl[ Y_{\tilde{f}}^2 \biggr] 
    \Biggr\} \nonumber\\
 && - \left( \frac{\alpha_Y}{4\pi} \right)^2 Y_{\tilde{f}}\,
    \Biggl\{ \Tr_{\EX} 
      \Biggl[ Y_{\EX} \cos 2\theta\, (m_1^2-m_2^2) \Biggr]
    + \Tr_{\HS} \Biggl[ Y_{\HS}\, m_{\HS}^2 
    \Biggr] \nonumber\\
  && \quad \qquad
    - \Tr_{\EX} \Biggl[ Y_{\EX} \cos 2\theta 
    \biggl( m_1^2 \log m_1^2 - m_2^2 \log m_2^2 \biggr) \Biggr]
    - \Tr_{\HS} \Biggl[ Y_{\HS}\, 
    m_{\HS}^2 \log m_{\HS}^2 \Biggr] \Biggr\} \nonumber\\
  && \quad \times
    \Biggl\{ \Tr_{\EX} \Biggl[ Y_{\EX}^2 \cos^2 2\theta 
    \biggl( \log m_1^2 + \log m_2^2 \biggr) 
    -2 Y_{\EX}^2 \sin^2 2\theta \biggl( 1 - 
    \frac{m_1^2 \log m_1^2 - m_2^2 \log m_2^2}{m_1^2-m_2^2} \biggr) 
    \Biggr] \nonumber\\
  && \quad \qquad
    + \Tr_{\HS} 
          \Biggl[ Y_{\HS}^2\, \log m_{\HS}^2 \Biggr] 
    + \Tr_{\tilde{f}} 
          \Biggl[ Y_{\tilde{f}}^2\, \log m_{\tilde{f}}^2 \Biggr] 
    \Biggr\}.
\end{eqnarray}
Thus, the light sfermion masses induced at two loops through the
generation of U(1)$_Y$ FI $D$-term are given by
\begin{eqnarray}
  m_{\tilde{f},\,{\rm FI-2loop}}^2
    = m_{\tilde{f},\,{\rm FI-2loop}}^2 |_{\EX} 
    + m_{\tilde{f},\,{\rm FI-2loop}}^2 |_{\HS}
    + m_{\tilde{f},\,{\rm FI-2(1loop)}}^2.
\label{Ap_m2_FI-2loop_div}
\end{eqnarray}

Altogether, the light sfermion masses at the two-loop level are given by
\begin{eqnarray}
  m_{\tilde{f},\,{\rm total}}^2
    = m_{\tilde{f},\,{\rm direct}}^2
    + m_{\tilde{f},\,{\rm FI-1loop}}^2
    + m_{\tilde{f},\,{\rm FI-2loop}}^2,
\end{eqnarray}
combining Eqs.~(\ref{Ap_m2_direct_div}, \ref{Ap_m2_FI-1loop_div},
\ref{Ap_m2_FI-2loop_div}).

\subsection{The finite case}

In the previous subsection, we have calculated the gaugino and the light
sfermion masses at the two-loop level in the 
$\overline{\rm DR}^{\prime}$ scheme.
The light sfermion masses generically have divergent contribution, so
that they receive large negative contribution from the RG evolution.
However, the divergences are canceled among loops of various heavy
particles under the conditions Eqs.~(\ref{mass_af}, \ref{mass_u1d1},
\ref{mass_u1d2}) discussed in the text.
These conditions are written as
\begin{eqnarray}
  && \Tr_{\EX} \Biggl[ T_{\EX}^A\, \mex^2\, (y_1+y_2-2) \Biggr]
    + \Tr_{\HS} \Biggl[ T_{\HS}^A\, m_{\HS}^2 \Biggr] = 0, 
\label{Ap_rel_fin_anomaly}\\
  && \Tr_{\EX} \Biggl[ Y_{\EX} \cos 2\theta\, (m_1^2-m_2^2) \Biggr]
    + \Tr_{\HS} \Biggl[ Y_{\HS}\, m_{\HS}^2  \Biggr] = 0, 
\label{Ap_rel_fin_FI-1loop}\\
  && \Tr_{\EX} 
       \Biggl[ Y_{\EX} C_{\EX}^A \cos 2\theta\, (m_1^2-m_2^2) \Biggr]
    + \Tr_{\HS} \Biggl[ Y_{\HS} C_{\HS}^A\, m_{\HS}^2 \Biggr] = 0,
\label{Ap_rel_fin_FI-2loop}
\end{eqnarray}
in the notation of this appendix.
Using these relations in Eqs.~(\ref{Ap_m2_direct_div},
\ref{Ap_m2_FI-1loop_div}, \ref{Ap_m2_FI-2loop_div}), we obtain the
finite contribution to the light sfermion masses,
\begin{eqnarray}
  m_{\tilde{f},\,{\rm total}}^2
    = m_{\tilde{f},\,{\rm direct}}^2
    + m_{\tilde{f},\,{\rm FI-1loop}}^2
    + m_{\tilde{f},\,{\rm FI-2loop}}^2;
\end{eqnarray}
\begin{eqnarray}
  \lefteqn{m_{\tilde{f},\,{\rm direct}}^2 =} \nonumber\\
  && 4 \sum_A \left( \frac{\alpha_A}{4\pi} \right)^2
    C_{\tilde{f}}^A \Tr_{\EX} 
      \Biggl[ T_{\EX}^A\, \mex^2 \biggl\{ (y_1+y_2-2) 
    \log \mex^2 + (y_1 \log y_1 + y_2 \log y_2) \nonumber\\
  && \quad
   -2 \biggl( y_1 {\rm Li}_2(1-\frac{1}{y_1}) 
   + y_2 {\rm Li}_2(1-\frac{1}{y_2}) \biggr)
   + \frac{1}{2}\sin^2 2\theta \biggl( y_1 {\rm Li}_2(1-\frac{y_2}{y_1})
   + y_2 {\rm Li}_2(1-\frac{y_1}{y_2}) \biggr) \biggr\} \Biggr] 
  \nonumber\\
  && \!\!\!\!\!
    +4 \sum_A \left( \frac{\alpha_A}{4\pi} \right)^2
    C_{\tilde{f}}^A \Tr_{\HS} \Biggl[ T_{\HS}^A\, 
    m_{\HS}^2 \biggl( \log m_{\HS}^2 
    - \frac{1}{3}\pi^2 \biggr) \Biggr],
\label{Ap_m2_direct_fin} \\
  \lefteqn{m_{\tilde{f},\,{\rm FI-1loop}}^2 =} \nonumber\\
  && \left( \frac{\alpha_Y}{4\pi} \right) Y_{\tilde{f}}
    \Tr_{\EX} \Biggl[ Y_{\EX} \cos 2\theta 
    \biggl( m_1^2 \log m_1^2 - m_2^2 \log m_2^2 \biggr) \Biggr] 
     \nonumber\\
  && \!\!\!\!\!\!
    + \left( \frac{\alpha_Y}{4\pi} \right) Y_{\tilde{f}}
    \Tr_{\HS} \Biggl[ Y_{\HS}\, 
    m_{\HS}^2 \log m_{\HS}^2 \Biggr],
\label{Ap_m2_FI-1loop_fin} \\
  \lefteqn{m_{\tilde{f},\,{\rm FI-2loop}}^2 =} \nonumber\\
  && 4 \sum_A \left( \frac{\alpha_Y}{4\pi} \right) 
    \left( \frac{\alpha_A}{4\pi} \right) Y_{\tilde{f}} 
    \Tr_{\EX} \Biggl[ Y_{\EX} C_{\EX}^A \cos 2\theta
    \biggl\{ -\frac{1}{2} \mex^2 \log^2 m_1^2 
    +\frac{1}{2} \mex^2 \log^2 m_2^2 \nonumber\\
  && \quad \qquad
    + (m_1^2+\mex^2)\log m_1^2 - (m_2^2+\mex^2)\log m_2^2 \nonumber\\
  && \quad \qquad
    + (m_1^2-\mex^2) {\rm Li}_2(1-\frac{\mex^2}{m_1^2})
    - (m_2^2-\mex^2) {\rm Li}_2(1-\frac{\mex^2}{m_2^2})
    \biggr\} \nonumber\\
  && \qquad
    + \frac{1}{4} Y_{\EX} C_{\EX}^A \cos 2\theta \sin^2 2\theta
    \biggl\{ 2 \frac{(m_1^2 \log m_1^2 - m_2^2 \log m_2^2)^2}
    {m_1^2-m_2^2} \nonumber\\
  && \quad \qquad
    - 4(m_1^2 \log m_1^2 - m_2^2 \log m_2^2)
    - (m_1^2 \log^2 m_1^2 - m_2^2 \log^2 m_2^2) \nonumber\\
  && \quad \qquad
    + (m_1^2-m_2^2)(2 + \log m_1^2 + \log m_2^2 
    - \log m_1^2 \log m_2^2) \biggr\} \Biggr] \nonumber\\
  && \!\!\!\!\!\!
    +4 \sum_A \left( \frac{\alpha_Y}{4\pi} \right) 
    \left( \frac{\alpha_A}{4\pi} \right) Y_{\tilde{f}} 
    \Tr_{\HS} \Biggl[ Y_{\HS} C_{\HS}^A\, m_{\HS}^2 
    \biggl( \log m_{\HS}^2 + \frac{1}{6}\pi^2 \biggr) 
    \Biggr] \nonumber\\
  && \!\!\!\!\!\!
    + \left( \frac{\alpha_Y}{4\pi} \right)^2 Y_{\tilde{f}}\,
    \Biggl\{ \Tr_{\EX} \Biggl[ Y_{\EX} \cos 2\theta 
    \biggl( m_1^2 \log m_1^2 - m_2^2 \log m_2^2 \biggr) \Biggr]
    + \Tr_{\HS} \Biggl[ Y_{\HS}\, 
    m_{\HS}^2 \log m_{\HS}^2 \Biggr] \Biggr\} \nonumber\\
  && \quad \times
    \Biggl\{ \Tr_{\EX} \Biggl[ Y_{\EX}^2 \cos^2 2\theta 
    \biggl( \log m_1^2 + \log m_2^2 \biggr) 
    -2 Y_{\EX}^2 \sin^2 2\theta \biggl( 1 - 
    \frac{m_1^2 \log m_1^2 - m_2^2 \log m_2^2}{m_1^2-m_2^2} \biggr) 
    \Biggr] \nonumber\\
  && \quad \qquad
    + \Tr_{\HS} \Biggl[ Y_{\HS}^2\, \log m_{\HS}^2 
    \Biggr] \Biggr\}.
\label{Ap_m2_FI-2loop_fin}
\end{eqnarray}
It should be understood that the coupling constants $\alpha_A$ and
$\alpha_Y$ are the renormalized ones at the scale $\mu$,
$\alpha_A(\mu)$ and $\alpha_Y(\mu)$, and the logarithms $\log m^2$ of
various masses are normalized at the scale $\mu$, $\log (m^2/\mu^2)$.
Here, we have dropped the term proportional to $\Tr_{\tilde{f}} [
Y_{\tilde{f}}^2\, \log m_{\tilde{f}}^2 ]$ which is the contribution
from the RG evolution below $\mu$.  Note that the contribution
$m_{\tilde{f},\,{\rm FI-1loop}}^2$ from one-loop U(1)$_Y$ FI $D$-term
is generically nonvanishing even if we assign U(1)$_X$ charges
consistent with the SU(5)$_{\rm GUT}$, since the extra scalars embedded
in a common SU(5)$_{\rm GUT}$ multiplet could have different masses
due to the difference between their supersymmetric masses caused by RG
evolution below the GUT scale.

Furthermore, the above expression is considerably simplified when the
following two conditions are satisfied:
\begin{enumerate}
\item
The U(1)$_X$ charge assignment is consistent with SU(5)$_{\rm GUT}$:
$\Tr_{\HS} \Biggl[ Y_{\HS}\, f(m_{\HS}^2) \Biggr] = 0$
\item
The extra-matter multiplets have an invariance under the parity 
$\Psi_{\rm ex} \leftrightarrow \bar{\Psi}_{\rm ex}$: 
$\theta = \pm \frac{\pi}{4}$
\end{enumerate}
These conditions are satisfied in the case of Model (I) and Model (II)
considered in the text.
Then, Eqs.~(\ref{Ap_m2_direct_fin}, \ref{Ap_m2_FI-1loop_fin},
\ref{Ap_m2_FI-2loop_fin}) are reduced to
\begin{eqnarray}
  && m_{\tilde{f},\,{\rm direct}}^2 =
    4 \sum_A \left( \frac{\alpha_A}{4\pi} \right)^2
    C_{\tilde{f}}^A \Tr_{\EX} 
       \Biggl[ T_{\EX}^A\, \mex^2 \biggl\{ (y_1+y_2-2) 
        \log \mex^2 + (y_1 \log y_1 + y_2 \log y_2) \nonumber\\
  && \qquad
   -2 \biggl( y_1 {\rm Li}_2(1-\frac{1}{y_1}) 
   + y_2 {\rm Li}_2(1-\frac{1}{y_2}) \biggr)
   + \frac{1}{2} \biggl( y_1 {\rm Li}_2(1-\frac{y_2}{y_1})
   + y_2 {\rm Li}_2(1-\frac{y_1}{y_2}) \biggr) \biggr\} \Biggr] \nonumber\\
  && \qquad
    +4 \sum_A \left( \frac{\alpha_A}{4\pi} \right)^2
    C_{\tilde{f}}^A \Tr_{\HS} \Biggl[ T_{\HS}^A\, 
    m_{\HS}^2 \biggl( \log m_{\HS}^2 
    - \frac{1}{3}\pi^2 \biggr) \Biggr], 
\label{Ap_m2_direct_fin_simple} \\
  && m_{\tilde{f},\,{\rm FI-1loop}}^2 = 0, 
\label{Ap_m2_FI-1loop_fin_simple} \\
  && m_{\tilde{f},\,{\rm FI-2loop}}^2 =
    4 \sum_A \left( \frac{\alpha_Y}{4\pi} \right) 
    \left( \frac{\alpha_A}{4\pi} \right) Y_{\tilde{f}} 
    \Tr_{\HS} \Biggl[ Y_{\HS} C_{\HS}^A\, m_{\HS}^2 
    \log m_{\HS}^2 \Biggr].
\label{Ap_m2_FI-2loop_fin_simple}
\end{eqnarray}

\section{Constraints from $\Delta m_K$ and $\epsilon_K$}
\label{Ap_KK}

In this appendix, we calculate the constraints on the first-two
generation sfermion masses from $\Delta m_K$ and $\epsilon_K$ 
in effective SUSY, where the gluino is much lighter than the
first-two generation sfermions.
In the following calculation, we only consider the gluino box diagram 
since it gives a dominant contribution.
According to Refs.~\cite{LO, NLO, Con_Sci},
we not only take into account the leading QCD corrections but also 
make use of $B$ parameters instead of the vacuum insertion
approximation.

The mass matrix for the down-type squark is relevant to the gluino box
diagram. In a basis where the down-type Yukawa matrix is diagonal, 
the mass matrix is
\begin{eqnarray}
  {\cal L}_{\rm mass} 
  &=& - \left(\tilde d^*_{Li} \;\;\tilde d^*_{Ri}\right)
  \left(
    \begin{array}{cc}
       ({\cal M}^2_{LL})_{ij} & ({\cal M}^2_{LR})_{ij} \\ 
       ({\cal M}^{2\dagger}_{LR})_{ij} & ({\cal M}^2_{RR})_{ij}
    \end{array}
  \right)
  \left(
    \begin{array}{c}
       \tilde d_{Lj} \\ 
       \tilde d_{Rj} 
    \end{array}
  \right),
\end{eqnarray}
where the subscripts $i$, $j$ are the indices of the generation. 
We here restrict the subscript $i$ to $i=1,2$,
since we calculate the constraints on the first-two generation sfermion
masses. From now we take the left-right mixing mass term 
${\cal M}_{LR}^2={\bf 0}$ because of the small Yukawa couplings. 
In consequence, ${\cal M}^2_{LL}$ and ${\cal M}^2_{RR}$ are diagonalized
separately by the sfermion mixing matrices $(U^L)_{X_L,i}$ and
$(U^R)_{X_R,i}$ as follows:
\begin{eqnarray}
  U^L {\cal M}_{LL}^2 U^{L\dagger} 
   ={\rm diag} \left(m_{X_L}^2\right),
  \hspace{5mm}
  U^R {\cal M}_{RR}^2 U^{R\dagger} 
   ={\rm diag} \left(m_{X_R}^2\right),
\end{eqnarray}
where $X_L=1_L,2_L$ and $X_R=1_R,2_R$.

The $\Delta S=2$ effective Lagrangian at the scale $m_D$, 
where the heavy sfermions decouple, is written as
\begin{eqnarray}
  \hspace*{-10mm}
  {\cal L}_{\rm eff}
   = \alpha_3^2(m_D)
      \biggl[C_1 {\cal O}_1+\tilde{C}_1 \tilde{\cal O}_1
               + C_4{\cal O}_4 +C_5 {\cal O}_5   \biggr].
\label{Lag_eff}
\end{eqnarray}
The operators ${\cal O}_{1,4,5}$ and their coefficients $C_{1,4,5}$
are defined as follows:
\begin{eqnarray}
  {\cal O}_1 &=&(\bar{d}_\alpha \gamma_\mu P_L s_\alpha)\,
                    (\bar{d}_\beta \gamma^\mu P_L s_\beta),\\
  {\cal O}_4 &=&(\bar{d}_\alpha  P_L s_\alpha)\,
                    (\bar{d}_\beta  P_R s_\beta),\\
  {\cal O}_5 &=&(\bar{d}_\alpha  P_L s_\beta)\,
                    (\bar{d}_\beta  P_R s_\alpha),
\end{eqnarray}
\begin{eqnarray}
  C_1&=& U^{L\dagger}_{1,X_L}U^L_{X_L,2}U^{L\dagger}_{1,Y_L}U^L_{Y_L,2}
    \left(\frac{1}{9}I_{X_L,Y_L}+\frac{11}{36}\tilde{I}_{X_L,Y_L}\right),
\label{Ap_coef_exact_1}\\
  C_4&=& U^{L\dagger}_{1,X_L}U^L_{X_L,2}U^{R\dagger}_{1,Y_R} U^R_{Y_R,2}
    \left(\frac{7}{3}I_{X_L,Y_R}-\frac{1}{3}\tilde{I}_{X_L,Y_R}\right),
\label{Ap_coef_exact_4}\\
  C_5&=& U^{L\dagger}_{1,X_L}U^L_{X_L,2}U^{R\dagger}_{1,Y_R} U^R_{Y_R,2}  
    \left(\frac{1}{9}I_{X_L,Y_R}+\frac{5}{9}\tilde{I}_{X_L,Y_R}\right).
\label{Ap_coef_exact_5}
\end{eqnarray}
$\tilde{\cal O}_1$ and $\tilde{C}_1$ are obtained with the replacement
$L \rightarrow R$ in ${\cal O}_1$ and $C_1$, respectively. 
The loop integrals $I_{X,Y}$ and $\tilde{I}_{X,Y}$ in the above
equations are
\begin{eqnarray}
  I_{X,Y}= 0,
  \hspace{10mm}
  \tilde{I}_{X,Y} = -\frac{\log{m_X^2}-\log{m_Y^2}}{m_X^2-m_Y^2},
\end{eqnarray}
where we have neglected the contributions dependent on the gluino mass
$m_{\tilde{g}}$, since they are suppressed by $m_{\tilde{g}}^2/m_X^2$
($\ll 1$). 

The above effective Lagrangian is valid for an arbitrary 
down-type squark mass matrix.
We here give a convenient expression by using the so-called mass
insertion method \cite{DGH}. 
We introduce two parameters essential for the mass insertion
method.
One is the averaged mass $\bar{m}_{LL}^2$ ($\bar{m}_{RR}^2$) for the
left- (right-) handed down-type squarks in the first-two generations,
which is defined by a geometric mean as
\begin{eqnarray}
  \bar{m}_{LL}^2\equiv\sqrt{m_{1_L}^2m_{2_L}^2},
  \hspace{5mm}
  \bar{m}_{RR}^2\equiv\sqrt{m_{1_R}^2m_{2_R}^2}.
\end{eqnarray}
The other is the off-diagonal element 
$\delta_{LL}$ ($\delta_{RR}$) of ${\cal M}^2_{LL}$ (${\cal M}^2_{RR}$),
which is normalized with the averaged mass as
\begin{eqnarray}
  \delta_{LL}\equiv \frac{({\cal M}_{LL}^2)_{1_L,2_L}}{\bar{m}_{LL}^2},
  \hspace{5mm}
  \delta_{RR}\equiv \frac{({\cal M}_{RR}^2)_{1_R,2_R}}{\bar{m}_{RR}^2}.
\end{eqnarray}
By taking a leading order of the mass differences,
$(m_{1_L}^2-m_{2_L}^2)$ and $(m_{1_R}^2-m_{2_R}^2)$,  
the coefficients in Eqs.~(\ref{Ap_coef_exact_1}, \ref{Ap_coef_exact_4},
\ref{Ap_coef_exact_5}) become 
\begin{eqnarray}
  C_1&=& -\frac{11}{108}\frac{\delta_{LL}^2}{\bar{m}_{LL}^2},\\
  C_4&=&\quad\frac{1}{9}
        \frac{\delta_{LL}\delta_{RR}}{\bar{m}_{LL}\bar{m}_{RR}},\\ 
  C_5&=&-\frac{5}{27}
        \frac{\delta_{LL}\delta_{RR}}{\bar{m}_{LL}\bar{m}_{RR}}.
\end{eqnarray}
$\tilde{C}_1$ is obtained with the replacement $L\rightarrow R$ 
in $C_1$. 
In Section 4, we often use the expression in the mass insertion method
in estimating the bounds on the light sfermion masses.
However, Figs.~\ref{fig_KK_1}-\ref{fig_stop_2} are obtained
by making use of the exact coefficients given in
Eqs.~(\ref{Ap_coef_exact_1}, \ref{Ap_coef_exact_4},
\ref{Ap_coef_exact_5}).

Since the above effective Lagrangian is obtained at the heavy-sfermion
mass scale $m_D$, we must evolve it using RG equations to the hadronic
scale $\mu_{\rm had}$, where hadronic matrix elements are evaluated.  
The coefficients at $\mu_{\rm had}$ are calculated at the
one-loop level as follows \cite{LO}:
\begin{eqnarray}
C_1(\mu_{\rm had})&=& \kappa_1 C_1(m_D),\\ 
\tilde{C}_1(\mu_{\rm had})&=& \kappa_1 \tilde{C}_1(m_D),\\ 
C_4(\mu_{\rm had})&=& 
  \kappa_4 C_4(m_D)+\frac{1}{3}(\kappa_4-\kappa_5)C_5(m_D),\\
C_5(\mu_{\rm had})&=& \kappa_5 C_5(m_D),
\end{eqnarray}
where
\begin{eqnarray}
\kappa_1&=&
 \left(\frac{\alpha_3(m_b)}{\alpha_3(\mu_{\rm had})}\right)^{6/25}
 \left(\frac{\alpha_3(m_t)}{\alpha_3(m_b)}\right)^{6/23}
 \left(\frac{\alpha_3(m_{\tilde{g}})}{\alpha_3(m_t)}\right)^{6/21}
 \left(\frac{\alpha_3(m_D)}{\alpha_3(m_{\tilde{g}})}\right)^{-2/b_3},\\
\kappa_4&=& \kappa_1^{-4},\\
\kappa_5&=& \kappa_1^{1/2}.
\end{eqnarray}
Here, we have taken the hadronic scale $\mu_{\rm had}=2~\GEV$ according
to Ref.~\cite{NLO} and assumed that all the light sfermions and the
gauginos decouple at the gluino mass scale for simplicity.
$b_3$ is the coefficient of the one-loop beta function
for the strong coupling between the gluino mass $m_{\tilde{g}}$ and 
the heavy-sfermion mass scale $m_D$.

Instead of using the vacuum insertion approximation, we represent 
the hadronic matrix elements of the renormalized operators 
${\cal O}(\mu)$ at the renormalization scale $\mu$ 
in terms of the corresponding $B$ parameters as follows:  
\begin{eqnarray}
\bra{K^0} {\cal O}_1(\mu) \ket{\bar{K}^0} &=&
\bra{K^0} \tilde{{\cal O}}_1(\mu) \ket{\bar{K}^0}
 =\frac{1}{3}m_K f_K^2 B_1(\mu),\\
\bra{K^0} {\cal O}_4(\mu) \ket{\bar{K}^0}
 &=& \frac{1}{4}\left(\frac{m_K}{m_s(\mu)+m_d(\mu)}\right)^2 
          m_K f_K^2 B_4(\mu),\\
\bra{K^0} {\cal O}_5(\mu) \ket{\bar{K}^0}
 &=& \frac{1}{12}\left(\frac{m_K}{m_s(\mu)+m_d(\mu)}\right)^2
         m_K f_K^2 B_5(\mu),
\end{eqnarray}
where $m_K=497.7~\MEV$, $f_K=160~\MEV$, $m_s(2~\GEV)=125~\MEV$
and $m_d(2~\GEV)=7~\MEV$.
In our calculation, we use the following values for the $B$ parameters 
obtained by lattice calculations at $\mu=2~\GEV$ \cite{Bk, B_para}:  
\begin{eqnarray}
 B_1(\mu=2~\GEV)&=&0.60(6),\\
 B_4(\mu=2~\GEV)&=&1.03(6),\\
 B_5(\mu=2~\GEV)&=&0.73(10).
\end{eqnarray}

We have now explained all the elements necessary for calculating the
SUSY contribution to the $K_L$-$K_S$ mass difference,
\begin{eqnarray}
  \Delta m_{K,{\rm SUSY}}=-2\Re\bra{K^0}{\cal L}_{\rm eff}\ket{\bar{K}^0}.
\end{eqnarray}
Using them, we obtain constraints from $\Delta m_K$ by imposing a
condition that the SUSY contribution does not saturate the experimental
value,
\begin{eqnarray}
  2|\bra{K^0}{\cal L}_{\rm eff}\ket{\bar{K}^0}|
  < \Delta m_K=3.49\times10^{-15}~\GEV,
\end{eqnarray}
in the case where the $\delta_{LL}$ and $\delta_{RR}$ have no
CP-violating phases.
The constraints can be expressed in a simple form using the mass
insertion as follows: 
\begin{eqnarray}
  \delta_{LL,\ RR} 
  &<& \frac{\bar{m}_{LL,\ RR}}{(25 \sim 35)~\TEV},
\label{Ap_KK_LL}\\
  (\delta_{LL}\delta_{RR})^{1/2} 
  &<& \frac{(\bar{m}_{LL}\bar{m}_{RR})^{1/2}}{(150 \sim 250)~\TEV}.
\label{Ap_KK_LR}
\end{eqnarray} 

If $\delta_{LL}$ and/or $\delta_{RR}$ have CP-violating phases,
there is another constraint from $\epsilon_K$,
\begin{eqnarray}
  \frac{1}{\sqrt{2}\Delta m_K} 
     |\Im \bra{K^0} {\cal L}_{\rm eff} \ket{\bar{K}^0}| <  \epsilon_K 
  =2.3\times 10^{-3}.
\label{epsilon_K}
\end{eqnarray}
The constraint is the severest when 
$\bra{K^0} {\cal L}_{\rm eff} \ket{\bar{K}^0}$ is pure imaginary, and
then the above constraint Eq.~(\ref{epsilon_K}) is rewritten as 
\begin{eqnarray}
  2|\bra{K^0}{\cal L}_{\rm eff}\ket{\bar{K}^0}|
  < 2\sqrt{2} \epsilon_K \Delta m_K =6.5\times 10^{-3} \Delta m_K.
\end{eqnarray}
Thus, the constraints from $\epsilon_K$ can be severer than those from
$\Delta m_K$ by a factor of $(2\sqrt{2}\epsilon_K)^{-1/2} \sim 12.4$.

\newpage
%
%
%
\newcommand{\Journal}[4]{{\sl #1} {\bf #2} {(#3)} {#4}}
\newcommand{\PL}{\sl Phys. Lett.}
\newcommand{\PR}{\sl Phys. Rev.}
\newcommand{\PRL}{\sl Phys. Rev. Lett.}
\newcommand{\NP}{\sl Nucl. Phys.}
\newcommand{\ZP}{\sl Z. Phys.}
\newcommand{\PTP}{\sl Prog. Theor. Phys.}
\newcommand{\NC}{\sl Nuovo Cimento}
\newcommand{\MPL}{\sl Mod. Phys. Lett.}
\newcommand{\PRep}{\sl Phys. Rep.}

%
\end{document}